\documentclass[longauth]{aa}  

\usepackage{graphicx}
\usepackage{gensymb}

\usepackage{txfonts}
\usepackage[flushleft]{threeparttable}
\renewcommand{\arraystretch}{1.1}
\usepackage[hidelinks,colorlinks=true,linkcolor=blue,citecolor=blue]{hyperref}
\usepackage{xcolor}
\usepackage{longtable}
\usepackage{lipsum}
\usepackage{adjustbox}
\usepackage{stfloats}
\usepackage{scrextend}
\usepackage{amssymb}
\usepackage{amsmath}
\usepackage{mathrsfs}
\usepackage{multicol}
\usepackage{booktabs}
\usepackage{moresize}
\usepackage{orcidlink}
\usepackage{xspace}

\newcommand{\feh}{\mbox{[Fe/H]}\xspace}

\newcommand{\mh}{\mbox{[M/H]}\xspace}

\newcommand{\teff}{\ensuremath{T_{\rm eff}}}

\newcommand{\gccc}{\mbox{g\,cm$^{-3}$}\xspace}

\newcommand{\mjup}{\mbox{$\it{M_{\rm \mathrm{Jup}}}$}\xspace}

\newcommand{\me}{\mbox{$\it{M_{\rm \mathrm{\oplus}}}$}\xspace}
\newcommand{\re}{\mbox{$\it{R_{\rm \mathrm{\oplus}}}$}\xspace}

\newcommand{\msol}{\mbox{$\it{M_\mathrm{\odot}}$}\xspace}
\newcommand{\rsol}{\mbox{$\it{R_\mathrm{\odot}}$}\xspace}
\newcommand{\se}{\mbox{$\it{S_{\rm \mathrm{\oplus}}}$}\xspace}

\newcommand{\lsol}{\mbox{$\it{L_\odot}$}\xspace}
\newcommand{\denssol}{\mbox{$\it{\rho_\mathrm{\odot}}$}\xspace}

\begin{document}

\title{NIRPS and TESS reveal a peculiar system around the M dwarf TOI-756: A transiting sub-Neptune and a cold eccentric giant}

\titlerunning{The peculiar TOI-756 system}
\authorrunning{L. Parc, et al.}

\author{
L\'ena Parc\inst{1,*}, 
Fran\c{c}ois Bouchy\inst{1}, 
Neil J. Cook\inst{2}, 
Nolan Grieves\inst{1}, 
\'Etienne Artigau\inst{2,3}, 
Alexandrine L'Heureux\inst{2}, 
Ren\'e Doyon\inst{2,3}, 
Yuri S. Messias\inst{2,4}, 
Fr\'ed\'erique Baron\inst{2,3}, 
Susana C. C. Barros\inst{5,6}
% ...
}

\institute{
\inst{1}Observatoire de Gen\`eve, D\'epartement d’Astronomie, Universit\'e de Gen\`eve, Chemin Pegasi 51, 1290 Versoix, Switzerland\\
\inst{2}Institut Trottier de recherche sur les exoplan\`etes, D\'epartement de Physique, Universit\'e de Montr\'eal, Montr\'eal, Qu\'ebec, Canada\\
\inst{3}Observatoire du Mont-M\'egantic, Qu\'ebec, Canada\\
\inst{4}Departamento de F\'isica Te\'orica e Experimental, Universidade Federal do Rio Grande do Norte, Campus Universit\'ario, Natal, RN, 59072-970, Brazil\\
\inst{5}Instituto de Astrof\'isica e Ci\^encias do Espa\c{c}o, Universidade do Porto, CAUP, Rua das Estrelas, 4150-762 Porto, Portugal\\
\inst{6}Departamento de F\'isica e Astronomia, Faculdade de Ci\^encias, Universidade do Porto, Rua do Campo Alegre, 4169-007 Porto, Portugal\\
\inst{7}Univ. Grenoble Alpes, CNRS, IPAG, F-38000 Grenoble, France\\
\inst{8}Department of Physics, University of Toronto, Toronto, ON M5S 3H4, Canada\\
\inst{9}Department of Physics \& Astronomy, McMaster University, 1280 Main St W, Hamilton, ON, L8S 4L8, Canada\\
\inst{10}Department of Physics, McGill University, 3600 rue University, Montr\'eal, QC, H3A 2T8, Canada\\
\inst{11}Department of Earth \& Planetary Sciences, McGill University, 3450 rue University, Montr\'eal, QC, H3A 0E8, Canada\\
\inst{12}Departamento de F\'isica, Universidade Federal do Cear\'a, Caixa Postal 6030, Campus do Pici, Fortaleza, Brazil\\
\inst{13}Centro de Astrobiolog\'ia (CAB), CSIC-INTA, Camino Bajo del Castillo s/n, 28692, Villanueva de la Ca\~nada (Madrid), Spain\\
\inst{14}Centre Vie dans l’Univers, Facult\'e des sciences de l’Universit\'e de Gen\`eve, Quai Ernest-Ansermet 30, 1205 Geneva, Switzerland\\
\inst{15}Instituto de Astrof\'isica de Canarias (IAC), Calle V\'ia L\'actea s/n, 38205 La Laguna, Tenerife, Spain\\
\inst{16}Departamento de Astrof\'isica, Universidad de La Laguna (ULL), 38206 La Laguna, Tenerife, Spain\\
\inst{17}European Southern Observatory (ESO), Karl-Schwarzschild-Str. 2, 85748 Garching bei M\"unchen, Germany\\
\inst{18}Space Research and Planetary Sciences, Physics Institute, University of Bern, Gesellschaftsstrasse 6, 3012 Bern, Switzerland\\
\inst{19}Consejo Superior de Investigaciones Cient\'ificas (CSIC), E-28006 Madrid, Spain\\
\inst{20}Bishop's Univeristy, Dept of Physics and Astronomy, Johnson-104E, 2600 College Street, Sherbrooke, QC, Canada, J1M 1Z7\\
\inst{21}Department of Physics and Space Science, Royal Military College of Canada, PO Box 17000, Station Forces, Kingston, ON, Canada\\
\inst{22}Instituto de Astrof\'isica e Ci\^encias do Espa\c{c}o, Faculdade de Ci\^encias da Universidade de Lisboa, Campo Grande, 1749-016 Lisboa, Portugal\\
\inst{23}Departamento de F\'isica da Faculdade de Ci\^encias da Universidade de Lisboa, Edif\'icio C8, 1749-016 Lisboa, Portugal\\
\inst{24}Centre of Optics, Photonics and Lasers, Universit\'e Laval, Qu\'ebec, Canada\\
\inst{25}Herzberg Astronomy and Astrophysics Research Centre, National Research Council of Canada\\
\inst{26}Aix Marseille Univ, CNRS, CNES, LAM, Marseille, France\\
\inst{27}Center for Space and Habitability, University of Bern, Gesellschaftsstrasse 6, 3012 Bern, Switzerland\\
\inst{28}Center for astrophysics $\vert$ Harvard \& Smithsonian, 60 Garden Street, Cambridge, MA 02138, United States\\
\inst{29}NASA Exoplanet Science Institute, IPAC, California Institute of Technology, Pasadena, CA 91125 USA\\
\inst{30}George Mason University, 4400 University Drive, Fairfax, VA, 22030 USA\\
\inst{31}European Southern Observatory (ESO), Av. Alonso de Cordova 3107,  Casilla 19001, Santiago de Chile, Chile\\
\inst{32}Plan\'etarium de Montr\'eal, Espace pour la Vie, 4801 av. Pierre-de Coubertin, Montr\'eal, Qu\'ebec, Canada\\
\inst{33}Lund Observatory, Division of Astrophysics, Department of Physics, Lund University, Box 118, 221 00 Lund, Sweden\\
\inst{34}SETI Institute, Mountain View, CA 94043, USA NASA Ames Research Center, Moffett Field, CA 94035, USA\\
\inst{35}York University, 4700 Keele St, North York, ON M3J 1P3\\
\inst{36}Max-Planck-Institut f\"ur Astronomie, K\"onigstuhl 17, D-69117 Heidelberg, Germany\\
\inst{37}University of British Columbia, 2329 West Mall, Vancouver, BC, Canada, V6T 1Z4\\
\inst{38}Western University, Department of Physics \& Astronomy and Institute for Earth and Space Exploration, 1151 Richmond Street, London, ON N6A 3K7, Canada\\
\inst{39}Light Bridges S.L., Observatorio del Teide, Carretera del Observatorio, s/n Guimar, 38500, Tenerife, Canarias, Spain\\
\inst{40}University Observatory, Faculty of Physics, Ludwig-Maximilians-Universit\"at M\"unchen, Scheinerstr. 1, 81679 Munich, Germany\\
\inst{41}Hamburger Sternwarte, Gojenbergsweg 112, D-21029 Hamburg, Germany\\
\inst{42}Subaru Telescope, National Astronomical Observatory of Japan (NAOJ), 650 N Aohoku Place, Hilo, HI 96720, USA\\
\inst{43}Department of Astronomy \& Astrophysics, University of Chicago, 5640 South Ellis Avenue, Chicago, IL 60637, USA\\
\inst{44}Laboratoire Lagrange, Observatoire de la C\^ote d’Azur, CNRS, Universit\'e C\^ote d’Azur, Nice, France\\
\inst{*}\email{lena.parc@unige.ch}
}

\date{Received 27 May 2025 / Accepted 25 July 2025}

\abstract{The Near InfraRed Planet Searcher (NIRPS) joined HARPS on the 3.6-m ESO telescope at La Silla Observatory in April 2023, dedicating part of its Guaranteed Time Observations (GTO) program to the radial velocity follow-up of TESS planet candidates to confirm and characterize transiting planets around M dwarfs.}{We present the "Sub-Neptunes" subprogram of the NIRPS-GTO, aimed at investigating the composition and formation of sub-Neptunes orbiting M dwarfs. We report the first results of this program with the characterization of the TOI-756 system, which consists of TOI-756 b, a transiting sub-Neptune candidate detected by TESS, as well as TOI-756 c, an additional non-transiting planet discovered by NIRPS and HARPS.}{We analyzed TESS and ground-based photometry, high-resolution imaging, and high-precision radial velocities (RVs) from NIRPS and HARPS to characterize the two newly discovered planets orbiting TOI-756, as well as to derive the fundamental properties of the host star. A dedicated approach was employed for the NIRPS RV extraction to mitigate telluric contamination, particularly when the star’s systemic velocity was shown to overlap with the barycentric Earth radial velocity.}{TOI-756 is a M1V-type star with an effective temperature of $T_\text{eff}\sim3657$ K and a super-solar metallicity ([Fe/H]) of 0.20$\pm$0.03 dex. TOI-756 b is a 1.24-day period sub-Neptune with a radius of 2.81 $\pm$ 0.10~\re and a mass of 9.8$^{+1.8}_{-1.6}$~\me. TOI-756 c is a cold eccentric (e$_c$ = 0.45 $\pm$ 0.01) giant planet orbiting with a period of 149.6 days around its star with a minimum mass of 4.05 $\pm$ 0.11~\mjup. Additionally, a linear trend of 146$~\mathrm{m\,s}^{-1}\,\mathrm{yr}^{-1}$ is visible in the radial velocities, hinting at a third component, possibly in the planetary or brown dwarf regime.}{We present the discovery and characterization of the transiting sub-Neptune TOI-756 b and the non-transiting eccentric cold giant TOI-756 c. This system is unique in the exoplanet landscape, standing as the first confirmed example of such a planetary architecture around an M dwarf. With a density of 2.42 $\pm$ 0.49 g~cm$^{-3}$, the inner planet, TOI-756 b, is a volatile-rich sub-Neptune. Assuming a pure H/He envelope, we inferred an atmospheric mass fraction of 0.023 and a core mass fraction of 0.27, which is well constrained by stellar refractory abundances derived from NIRPS spectra. It falls within the still poorly explored radius cliff and at the lower boundary of the Neptune desert, making it a prime target for a future atmospheric characterization with JWST to improve our understanding of this population.}

\keywords{techniques: photometric -- techniques: radial velocities -- planets and satellites: detection -- planets and satellites: composition -- planets and satellites: formation -- stars: low-mass}

\maketitle

\section{Introduction}\label{sect:intro}

Since the discovery of a giant exoplanet orbiting 51 Pegasi \citep{Mayor1995}, nearly 5900 exoplanets have been detected\footnote{\url{https://exoplanetarchive.ipac.caltech.edu/}}, showcasing an incredible variety of planetary systems and greatly enhancing our understanding of planet formation and evolution. Notably, space-based missions such as Kepler \citep{Borucki2010} and TESS \citep{Ricker2014} have revealed the prevalence of a population of exoplanets with sizes between Earth and Neptune, known as super-Earths and sub-Neptunes. This group of smaller planets (with radii between 1 and 4~\re) is not present in our Solar System, yet more than half of all Sun-like stars in the Galaxy are believed to host a sub-Neptune within 1 AU \citep[e.g.][]{Batalha2013,Petigura2013,Marcy2014}. 

M dwarfs are the most abundant stars in our Galaxy \citep{Henry2006,Winters2015,Reyle2021} and the search for exoplanets around these low-mass stars has gained significant attention in recent years. Indeed, they appear to have a high occurrence rate of planets, particularly of rocky planets and sub-Neptunes \citep{Dressing2013,Bonfils2013,Dressing2015,Mulders2015,Gaidos2016, Mignon2025}. In addition, exoplanets that transit M dwarfs are of particular interest, as their small size and low irradiation levels allow for easier detection of smaller and cooler planets, such as those located within the habitable zone of their star, than around larger and hotter stars. TESS was specially designed to be sensitive to these redder, cooler stars, but the relative faintness of M dwarfs in the visible spectrum has hindered a comprehensive characterization of the planetary systems via ground-based follow-ups. Indeed, the empirical population of known planets hosted by low-mass stars later than mid-K spectral type is smaller by nearly an order of magnitude than planets around Sun-like stars \citep{Cloutier2020}. In particular, the PlanetS catalog\footnote{\url{https://dace.unige.ch/exoplanets/}} \citep{Parc2024,Otegi2020} of well-characterized planets (with precisions $\sigma$ on inferred masses $M$ and radius $R$ of $\sigma_M/M < 25\%$ ; $\sigma_R/R <8\%$) only contains 80 planets around M dwarfs, compared to 745 around earlier-type stars. However, the recent development of high-resolution near-infrared spectrographs such as the Near InfraRed Planet Searcher \citep[NIRPS ;][]{Bouchy2017,Bouchy2025} has enabled efficient radial velocity (RV) follow-up of these transiting exoplanets. NIRPS represents a breakthrough in this respect, allowing for precise measurements of RVs of M dwarfs too faint for HARPS, breaking the meter-per-second barrier in the infrared \citep{Suarez2025}.

The precise characterization of the radius and mass of these planets is crucial for deriving the bulk density. This, in turn, allows us to study the planet’s internal structure and composition, offering insights into the relative mass fractions of its components, such as the iron core, mantle, atmosphere, and total mass fraction of water \citep{Dorn2015, Brugger2017, Plotnykov2024}. It is also necessary for atmospheric characterization via transmission spectroscopy, as the scale height of atmospheres is inversely proportional to surface gravity \citep{Batalha2019}. Understanding the compositional differences of planets hosted by M dwarfs is crucial for comprehending their distinct planet formation environments, as M-type stars have a longer hot protostellar phases \citep{Baraffe1998,Baraffe2015}, lower protoplanetary disk masses \citep{Pascucci2016}, and higher and longer activity at young ages compared to FGK-type stars \citep{Ribas2005}. 

While studies show that low-mass planets appear to be more numerous around close-in M-dwarf systems than Solar-type stars,  the differences among how M-dwarf environments influence the composition of planets of a given radius remain uncertain. \citet{Cloutier2020} found an increase in the frequency of close-in rocky planets around increasingly lower-mass stars and that the relative occurrence rate of rocky to non-rocky planets increases $\sim$6–30 times around mid-M dwarfs compared to mid-K dwarfs. However, they did not firmly identify the physical cause of this trend. Furthermore, \citet{Kubyshkina2021} modeled the evolution of a wide range of sub-Neptune-like planets orbiting stars of different masses and evolutionary histories. They found that atmospheric escape of planets with the same equilibrium temperature ranges occurs more efficiently around lower mass stars, indirectly supporting the findings of \citet{Cloutier2020}. A key question that remains is whether M-dwarf planets primarily form as bare rocky planets, or if they form with an a envelope that they subsequently lose. Despite the evidence suggesting that M dwarfs tend to form more rocky planets, \citet{Parc2024} identified statistical evidence for small well-characterized sub-Neptunes (1.8~\re < $R_p$ < 2.8~\re) being less dense around M dwarfs than around FGK dwarfs, hinting that these planets are ice-rich and would, hence, be likely migrated objects that accreted most of their solids beyond the iceline \citep[e.g.,][]{Alibert2017,Venturini2020,Venturini2024,Burn2021,Burn2024}. However, the sample of these sub-Neptunes orbiting M dwarfs is still small and more well-characterized planets are needed in this parameter space to determine whether this low-density trend is consistent.

On the other hand, giant planets with masses comparable to Jupiter are very infrequent around M dwarfs. Recent simulations of planet formation suggest that their occurrence rate declines sharply with decreasing stellar mass, potentially reaching zero for the lowest-mass stars \citep{Burn2021}. Nevertheless, such planets do exist, although they appear to be significantly less common than around FGK-type stars \citep[e.g.,][]{Bonfils2013,Bryant2023,Pass2023,Mignon2025}. Unlike small exoplanets, giant planets are expected to form at larger orbital separations from their host star, where more material is available \citep{Alexander2012,Bitsch2015}. As a result, this population is more affected by the observational biases of the transit method, as they orbit farther from small stars, making their detection more challenging. The RV method is less sensitive to this bias given the large RV signal induced by massive planets, even at longer periods, but the monitoring over long baselines to see these giant planet signals is costly and not often done. Finally, increasing this sample will provide crucial constraints on the formation and evolution of giant planets around M dwarfs.

This paper is structured as follows: in Sect.~\ref{sect:wp2subprog}, we provide a description of the "Sub-Neptunes" NIRPS-GTO SP2 subprogram. In Sect.~\ref{sect:observations}, we present the space- and ground-based observations taken by TESS, LCO-CTIO, and ExTrA, as well as the NIRPS+HARPS RV observations. Sect.~\ref{Sect:stellar_charac} details how we determined the host star parameters from both NIRPS and HARPS high-resolution spectra and photometric observations. In Sect.~\ref{sect:photo_rv_analysis}, we present the global photometric and RV analysis and its results. Finally, in Sect.~\ref{sect:discussion}, we discuss the system. We present our conclusions in Sect.~\ref{sect:ccl}.

\section{Exploring the composition and formation of sub-Neptunes orbiting M dwarfs with NIRPS}\label{sect:wp2subprog}

NIRPS began operations in April 2023, initiating its five-year Guaranteed Time Observations (GTO) program, which spans 725 nights. The NIRPS GTO is structured into three primary scientific subprograms (SPs), each allocated 225 nights, along with smaller "Other Sciences" (OS) programs totaling 50 nights, as described by \citet{Bouchy2025}. The second major work package (SP2) focuses on the characterization of the mass and bulk density of exoplanets transiting M dwarfs. SP2 aims to constrain the internal composition of these exoplanets, including their iron, rock, and water fractions, as well as to investigate how their properties vary with stellar irradiation, stellar mass, planetary architecture, and stellar composition. The objective is to provide critical insights into the formation and evolutionary pathways of exoplanet systems around M dwarfs. One particular subprogram of the SP2 of NIRPS has been dedicated to exploring the composition and formation of sub-Neptune sized planets orbiting M dwarfs (the "sub-Neptunes" subprogram).

As discussed in Sect.~\ref{sect:intro}, this program aims to increase the sample of sub-Neptunes with precise masses. This will help elucidate whether these planets are ice-rich and, hence, whether they are likely to be objects that accreted most of their solids beyond the iceline and migrated in \citep[e.g.,][]{Alibert2017,Venturini2020,Burn2021} or whether they are water-poor and formed inside the water ice line \citep[e.g.,][]{Owen2017,Lopez2018,Cloutier2020}. The initial target list, established in 2022, focused on TESS Objects of Interest (TOIs) with radii between 2 and 3~\re\ orbiting M dwarfs, with all but one target receiving an insolation of $<30~\se$. We selected targets that are observable with NIRPS, installed at La Silla Observatory, meaning they have a declination of less than $+$20 degrees. Additionally, we prioritized targets with a \textit{J}-band magnitude lower than 11.5, ensuring that the small semi-amplitudes required for precise mass measurements with NIRPS could be accurately detected.

Over time, these selection criteria evolved, particularly following studies such as that of \citet{Parc2024}, which statistically confirmed the initially observed low-density trend and highlighted the need to explore the sparsely populated 3–4~\re range. This same study suggests that the transition between super-Earths and sub-Neptunes is dependent on stellar type and appears significantly less pronounced around M dwarfs compared to FGK dwarfs, an interesting demographic feature we decided to explore also with NIRPS and this subprogram. Initially, 12 targets were included in this program, but some were removed due to the challenging semi-amplitudes expected around very faint stars. A few others were added with the extension of the science case and the new TESS candidates. We currently have 22 targets in our GTO protected target list for this subprogram in Period 116 (1 October 2025-30 April 2026), some of which are shared with other SP2 subprograms. This list is updated each semester based on observational results and information from other facilities involved in the mass characterization of these objects. All protected targets are published on the ESO website and all initiated targets are recorded in the SG2/SG4 TESS FOllow-up Program (TFOP) spreadsheet. With \textit{V}-band magnitudes ranging from 12.2 to 15.4, NIRPS extends the traditional RV follow-up limits imposed by HARPS and 4-meter-class telescopes, enabling the study of fainter targets that would otherwise be challenging to monitor with optical spectrographs.

This study presents the first results of the NIRPS-GTO SP2 program, including the confirmation and characterization of a TESS candidate orbiting the M1V star TOI-756, as well as the discovery of TOI-756~c, the first planet detected with NIRPS.

\begin{figure}[t]
  \centering
    \includegraphics[width=0.45\textwidth]{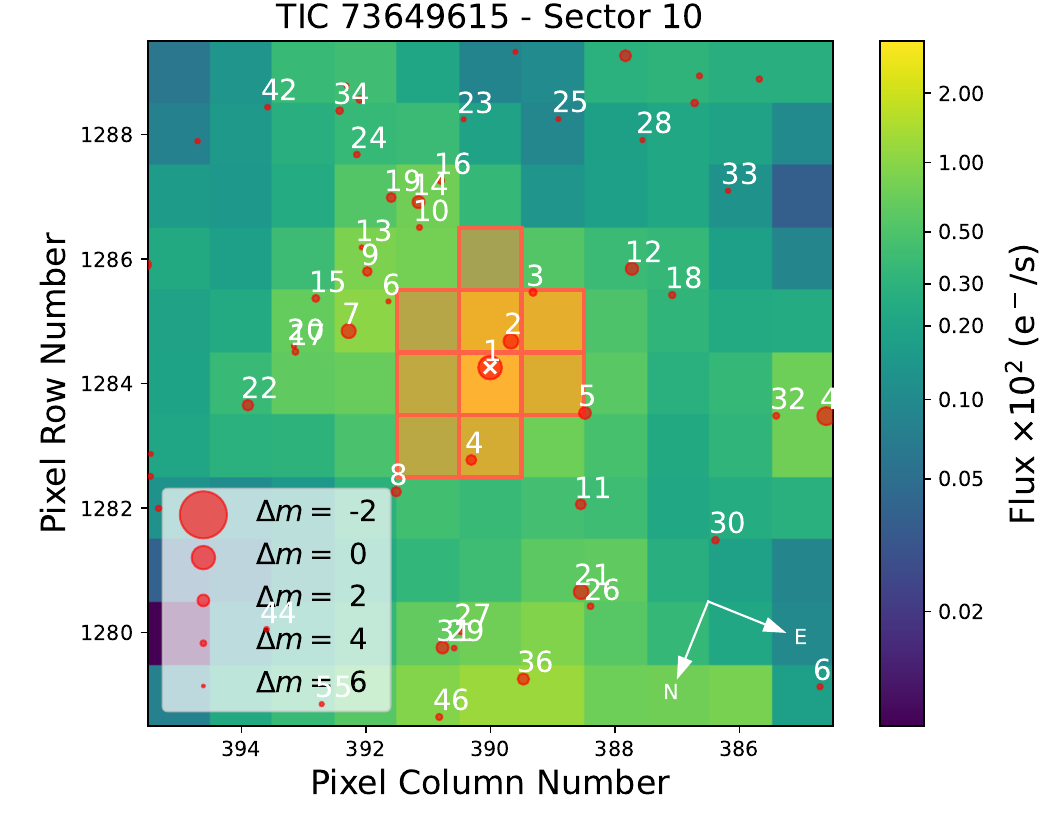}
  \caption{TESS TPF of TOI-756 created with \texttt{tpfplotter} (\citealt{Aller2020}). The orange pixels define the aperture mask used for extracting the photometry. Additionally, the red circles indicate neighboring objects from the Gaia DR3 catalog, with the circle size corresponding to the brightness difference compared to the target (as indicated in the legend). Our target is marked with a white cross. Pixel scale is 21$\arcsec$/pixel. The co-moving companion of TOI-756 corresponds to the star labeled "2."} 
  \label{fig:TPF}
\end{figure}

\section{Observations}\label{sect:observations}

\begin{table}[t]
\small
\caption{Stellar parameters for TOI-756.}
\centering
\renewcommand{\arraystretch}{1.2}
\begin{tabular}{lcc}
\hline\hline
  & \textbf{TOI-756}  & Source \\
    \hline
    \textbf{Identifiers}& &\\
    TIC ID  & 73649615 & TICv8\\
    2MASS ID & J12482523-4528140 &  2MASS\\
    Gaia ID & 6129327525817451648 & Gaia DR3\\
    WT &  351 & WT\\
    \hline
    \textbf{Astrometric parameters} & &\\
    Right ascension (J2016),  $\alpha$ & 12$^{\mathrm{h}}$ 48$^{\mathrm{m}}$ 25.21$^{\mathrm{s}}$ &Gaia DR3\\
    Declination (J2016), $\delta$ & -45$^{\circ}$ 28$^{\prime}$ 14.15$^{\prime \prime}$&Gaia DR3\\
    Parallax (mas) & 11.61 $\pm$ 0.02 &Gaia DR3\\
    Distance (pc) & 86.45$^{+1.18}_{-0.22}$ &Gaia DR3\\
    $\mu_{\rm{R.A.}}$ (mas yr$^{-1}$) &-216.502 $\pm$ 0.016 &Gaia DR3\\
    $\mu_{\rm{Dec}}$ (mas yr$^{-1}$) &29.197 $\pm$ 0.013 &Gaia DR3\\
    $V_{syst}$ (km s$^{-1}$) &15.36 $\pm$ 1.20 &Gaia DR3\\
    $U_{LSR}$ & -57.56 $\pm$ 0.64 &This work\\
    $V_{LSR}$ & -44.82 $\pm$ 0.97 &This work\\
    $W_{LSR}$ & 22.18 $\pm$ 0.36 &This work\\
    \hline
    \textbf{Photometric parameters} & &\\
    TESS (mag) & 12.5554 $\pm$ 0.007 & TICv8\\
    \textit{B} (mag) &16.102 $\pm$ 0.057 & TICv8\\
    \textit{V} (mag) &14.607 $\pm$ 0.018& TICv8\\
    \textit{G} (mag) &13.677 $\pm$ 0.003& Gaia DR3\\
    \textit{J} (mag) &11.138 $\pm$ 0.026 & 2MASS\\
    \textit{H} (mag) &10.517 $\pm$ 0.025 & 2MASS\\
    $K_s$ (mag) &10.274 $\pm$ 0.021 & 2MASS\\
    \hline
    \textbf{Bulk parameters} & &\\
    Spectral type & M1V & This work \\
    \teff\,(K) &  3657 $\pm$ 72 & This work \\
    $R_\star$  (\rsol)  &  0.505 $\pm$ 0.015 & This work \\
    $M_\star$  (\msol)  &  0.505 $\pm$ 0.019  & This work \\
    $\rho_\star$  (\gccc)  &  5.52$^{+0.56}_{-0.54}$  & This work \\
    $L_\star$  (\lsol)  &  0.041 $\pm$ 0.004  &This work\\
    \feh (dex) &  0.196 $\pm$ 0.029$\dagger$ &This work\\
    \mh (dex) &  0.17 $\pm$ 0.08 &This work\\
    $\log g_*$ (cm\,s$^{-2}$) &  4.735 $\pm$ 0.031  &This work\\
    Age (Gyr) &  $3.2^{+5.5}_{-2.3}$&This work\\
   \hline
\end{tabular}
\\
\begin{tablenotes}
\item Sources: 1) TICv8 \citep{Stassun2019} 2) 2MASS \citep{2MASS2006} 3) Gaia DR3 \citep{Gaia2018} 4) WT \citep{Wroblewski1991} $\dagger$ Value derived with a fixed T$_\mathrm{eff}$, so the uncertainties are likely underestimated (see Sect.~\ref{subSect:stellar_teff}).
\end{tablenotes}
\label{tab:Stellar parameters}
\end{table}

\subsection{TOI-756 as part of a wide binary system}\label{sect:WT351companion}

TOI-756 is an M1V star with an effective temperature of $\sim3600$\,K and magnitudes in the \textit{V} and \textit{I} of 14.6 and 11.1, respectively. It was discovered by \citet{Wroblewski1991} and named WT 351 as part of a binary system with a widely separated co-moving stellar companion WT 352. This companion is a M3/4V main sequence star ($T_{\text{eff}}\sim3300$ K) located at a separation of 11.09 arcsec, corresponding to a projected separation of about $\sim$ 955 au. By comparing their parallaxes, proper motions, and radial velocities from Gaia DR3, we confirmed that the two stars share common motion and distance, consistent with a physically bound binary system, which was previously reported by \citet{Mugrauer2020} and \citet{ElBadry2021}. The main identifiers, as well as the astrometric and photometric parameters of TOI-756, are listed in Table~\ref{tab:Stellar parameters}.

\subsection{TESS photometry}
TOI-756 (TIC 73649615) was observed in TESS Sector 10 (March 26, 2019 to  April 22, 2019), Sector 11 (April 23, 2019 to May 20, 2019), Sector 37 (April 02, 2021 to  April 28, 2021) and Sector 64 (April 06, 2023 to May 04, 2023) in 2-min cadence. The target was imaged on CCD 3 of camera 2 in Sectors 10, 37, and 64 and on CCD 4 of camera 2 in Sector 11. The TESS Science Processing Operations Center (SPOC; \citealt{Jenkins2016}) at NASA Ames processed the TESS photometric data resulting in the Simple Aperture Photometry (SAP; \citealt{Twicken2010,Morris2020}) flux and the Presearch Data Conditioning Simple Aperture Photometry (PDCSAP; \citealt{Smith2012,Stumpe2012,Stumpe2014}) flux. The latter flux was corrected for dilution in the TESS aperture by known contaminating sources. Indeed, due to its large pixel size of 21" per pixel, the TESS photometry can be contaminated by nearby companions. To evaluate the possible contamination, we plotted the target pixel file (TPF, Fig.\ref{fig:TPF}) of Sector 10 along with the aperture mask used for the SAP flux using \texttt{tpfplotter} \citep{Aller2020}. The TPFs of Sector 11, 37, and 64 are plotted in Appendix~\ref{appendix:TPF}. The apertures used for extracting the light curves in all four sectors were mostly contaminated by TIC 73649613, the co-moving companion of TOI-756 (Sect.~\ref{sect:WT351companion}) with a TESS magnitude of 13.75 (corresponding to a $\Delta m$ = 1.35). \\
On 2019 June 05, the TESS data public website\footnote{\url{https://tev.mit.edu/data/}} announced the detection of a 1.24-day TOI \citep{Guerrero2021}, TOI-756.01. The SPOC detected the transit signature of TOI-756.01 in Sector 10 and in Sector 11 and the signature was fitted with an initial limb-darkened transit model \citep{Li2019} and passed all the diagnostic tests presented in the Data Validations Reports \citep{Twicken2018}. In particular, the difference image centroiding test for the multi-sector searches strongly rejected all TIC objects other than the target star as the transit source in each case.

\subsection{Ground-based photometry}

The TESS pixel scale is~21" pixel$^{-1}$ and photometric apertures typically extend out to roughly 1', generally causing multiple stars to blend in the TESS aperture. To definitely exclude the presence of another star causing the signal in the TESS data and improve the transit ephemerides, we conducted photometric ground-based follow-up observations in different bands with ExTrA and LCO-CTIO of the field around TOI-756 as part of the TESS Follow-up Observing Program (TFOP)\footnote{\url{https://tess.mit.edu/followup/}} Sub Group 1 \citep{Collins2019}. 

\subsubsection{LCO-CTIO}

We used the Las Cumbres Observatory Global Telecope (LCOGT: \citealt{Brown2013}) 1.0-m network to observe two full transits of TOI-756~b in Sloan-\textit{i'} and \textit{g'} filters. The telescopes are equipped with 4096 $\times$ 4096 SINISTRO Cameras, having an image scale of 0.389" per pixel and a field of view of 26$' \times$ 26$'$. The raw data were calibrated by the standard LCOGT \texttt{BANZAI} pipeline \citep{McCully2018} and photometric measurements were extracted using \texttt{AstroImageJ} \citep{Collins2017}. The two transits were observed at Cerro Tololo Interamerican Observatory (CTIO), the first on 2019 June 12 UT in Sloan-\textit{i'} using 4.7" target aperture and the second on  July 03, 2019 UT in Sloan-\textit{g'} using 5.1" target aperture. The data are shown in Fig.~\ref{fig:Photometry} and Fig.~\ref{fig:LCOT2_fit}.

\subsubsection{ExTrA}

ExTrA \citep{Bonfils2015} is a near-infrared (0.85 to 1.55~$\mu m$) multi-object spectrophotometer fed by three 60-cm telescopes located at La Silla Observatory in Chile. One full transit (on 2021 February 24) and three partial transits (on  March 01, 06, and 27, 2021) of TOI-756~b were observed using two ExTrA telescopes. We used 8" aperture fibers and the low-resolution mode (R $\sim$ 20) of the spectrophotometer with an exposure time of 60 seconds. Five fiber positioners are used at the focal plane of each telescope to select light from the target and four comparison stars chosen with 2MASS \textit{J}-band magnitude \citep{2MASS2006} and Gaia effective temperatures \citep{Gaia2018} similar to the target. The resulting ExTrA data were analyzed using a custom data reduction software to produce synthetic photometry in a 0.85-1.55 micron bandpass, described in more detail in \citet{Cointepas2021}. The data are shown in Fig.~\ref{fig:ExTrA_fit}.

\subsection{High-resolution imaging}\label{sect:high_res}

As part of our standard process for validating transiting exoplanets to exclude false positives and to assess the possible contamination of bound or unbound companions on the derived planetary radii \citep{Ciardi2015}, we observed TOI-756 with adaptive optics and speckle imaging at VLT, SOAR, and Gemini.

\subsubsection{VLT}

TOI-756 was imaged with the NAOS/CONICA instrument on board the Very Large Telescope (NACO/VLT) on the night of July 13, 2019 UT in NGS mode with the \textit{Ks} filter centered on 2.18~$\mu$m \citep{Lenzen2003,Rousset2003}. We took nine frames with an integration time of 14 s each and dithered between each frame. We performed a standard reduction using a custom IDL pipeline: we subtracted flats and constructed a sky background from the dithered science frames, aligned and co-added the images, and then injected fake companions to determine a 5-$\sigma$ detection threshold as a function of radius. We obtained a contrast of 4.65 mag at 1", and no companions were detected. The contrast curve is shown in the top left panel of Fig.~\ref{fig:HighRes}.

\subsubsection{SOAR}
We observed TOI-756 with speckle imaging using the High-Resolution Camera (HRCam) imager on the 4.1\,m Southern Astrophysical Research (SOAR) telescope \citep{Tokovinin2018} on  July 14, 2019 UT, observing in Cousins \textit{I}-band, a similar visible band-pass as TESS. This observation was sensitive to objects fainter by 5.2 at an angular distance of 1 arcsec from the target. More details of the observations within the SOAR TESS survey are available in \citet{Ziegler2020}. The 5-$\sigma$ detection sensitivity and speckle autocorrelation functions from the observation are shown in the top right panel Fig.~\ref{fig:HighRes}. No nearby stars were detected within 3" of TOI-756 in the SOAR observation.

\subsubsection{Gemini}

TOI-756 was observed on  March 12, 2020 and  July 05, 2023 UT using the Zorro speckle instrument on Gemini South. Zorro provides speckle imaging in two bands (562 and 832 nm) with output data products including a reconstructed image and robust contrast limits on companion detections \citep{Howell2011}. Both observations provided similar results; TOI-756 has no close companions to within the 5-$\sigma$ contrast limits obtained (4.84 to 6.1 magnitudes) at 0.5\,arcsec (Fig.~\ref{fig:HighRes}, bottom panels).

\subsection{Spectroscopic follow-up with combined NIRPS and HARPS}\label{sect:RVs_observations}

TOI-756 was observed simultaneously from April 4, 2023,
to August 23, 2024, with NIRPS (Bouchy et al. 2025) and HARPS \citep{Mayor2003} echelle spectrographs at the ESO 3.6\,m telescope at La Silla Observatory in Chile. NIRPS is a new echelle spectrograph designed for precision radial velocities covering the \textit{YJH} bands (980–1800 nm). The instrument is equipped with a high-order adaptive optics system and two observing modes: high accuracy (HA; R $\sim$ 88\,000, 0.4" fiber) and high efficiency (HE; R $\sim$ 75\,200, 0.9" fiber), which can be utilized simultaneously with HARPS. TOI-756 was observed as part of the NIRPS-GTO program, under the Follow-up of Transiting Planets subprogram (PID:111.254T.001, 112.25NS.001, 112.25NS.002; PI: Bouchy \& Doyon) in HE mode with NIRPS and in EGGS mode (high efficiency mode, R $\sim$ 80\,000, 1.4" fiber) with HARPS. We selected these modes to minimize the modal noise of NIRPS and to maximize the flux by taking the large fibers, especially for HARPS, since the target is relatively faint in the visible (\textit{V} = 14.6). We also chose to target the sky with fiber B instead of the Fabry-Perot, due to the target's faintness, to facilitate background light correction. Over 64 individual nights, we collected three spectra of TOI-756 per night with NIRPS (3 exposures of 800\,s), which we combined to obtain a median signal-to-noise ratio (S/N) of 28.7 per pixel in the middle of \textit{H} band. As NIRPS operated alone for seven nights, the HARPS dataset comprise 57 spectra (a single 2400-s exposure per night) with a median S/N of 6.5 per pixel near 550 nm. We choose to take time series of three exposures on NIRPS and then combined them because the maximum recommended exposure time is 900 s on NIRPS due to detector readout noise limitations \citep{Bouchy2025}. We removed the three last HARPS spectra since they were affected by a HARPS shutter problem (24-07-24 ; 26-07-24 ; 22-08-24). 

For HARPS, we used the extracted spectra from the HARPS-DRS \citep{Lovis2007}. For NIRPS, the observations were reduced with both the NIRPS-DRS and APERO. The NIRPS-DRS is based on and adapted from the publicly available ESPRESSO pipeline \citep{Pepe2021}. Several updates have been implemented in the ESPRESSO pipeline to enable the reduction of infrared observations, including a telluric correction following the method of \citet{Allart2022} \citep[see][]{Bouchy2025}. The NIRPS-DRS is the nominal pipeline for NIRPS data reduction for the ESO science archive through the VLT Data Flow System (DFS). APERO \citep{Cook2022} is the standard data reduction software for the SPIRou near-infrared spectrograph \citep{Donati2020}, and was adapted and made fully compatible with NIRPS. 
The RV extraction from the reduced data of HARPS and NIRPS was performed with both the cross-correlation method (CCF) and the LBL method of \citet{Artigau2022}, available as an open-source package (v0.65.003; \texttt{LBL}\footnote{\url{https://github.com/njcuk9999/lbl}}). For the CCF method, we used the CCFs from the HARPS and NIRPS DRS using an M2V and M1V mask respectively. The \texttt{LBL} package is compatible with both NIRPS and HARPS. The method is conceptually similar to template matching \citep[e.g.,][]{Anglada2012,Astudillo2017}, while being more resilient to outlying spectral features (e.g., telluric residuals, cosmic rays, detector defects) as the template fitting is performed line by line, which facilitates the identification and removal of outliers. For NIRPS, we used the template of a brighter star with a similar spectral type, GL~514 (M1V), from NIRPS-GTO observations. For HARPS, we instead employed the template of GL~699 (M4V) from public data obtained via the ESO archive \citep{Delmotte2006}. An additional telluric correction was performed for HARPS inside the LBL code by fitting a TAPAS atmospheric model \citep{Bertaux2014}.

Finally, for the analysis presented in Sect.~\ref{sect:RVs_analysis}, we used the HARPS data processed using the DRS pipeline in combination with LBL, and the NIRPS data reduced with APERO, which provides a slighly better telluric absorption correction, also in conjunction with LBL. We employed nightly binned data and applied a preprocessing step to exclude points with higher uncertainties than the majority, using a 95\% percentile error-based filtering on RVs  and the second-order derivative D2V indicator, defined in \citet{Artigau2022}. Variations in the second-order derivative can be associated with changes in the FWHM from the CCF method. 
All the data will be publicly available through the DACE platform\footnote{\url{https://dace.unige.ch/}} after publication.

\section{Stellar characterization}\label{Sect:stellar_charac}
\subsection{Spectroscopic parameters}\label{subSect:stellar_teff}

The derivation of spectroscopic stellar parameters was done by applying different techniques to the HARPS and NIRPS spectra. For the first technique, we combined all the individual HARPS spectra with the task {\tt scombine} within IRAF\footnote{IRAF is distributed by National Optical Astronomy Observatories, operated by the Association of Universities for Research in Astronomy, Inc., under contract with the National Science Foundation, USA.} to obtain a high S/N spectrum. We used the machine learning tool {\tt ODUSSEAS}\footnote{\url{https://github.com/AlexandrosAntoniadis/ODUSSEAS}} \citep{Antoniadis20,Antoniadis24} to derive the effective temperature (\teff) and metallicity (\feh). This tool measures the pseudo equivalent widths (EWs) of a set of $\sim$4000 lines in the optical spectra. Then, it applies a machine learning model trained with the same lines measured and calibrated in a reference sample of 65 M dwarfs observed with HARPS for which their \feh were obtained from photometric calibrations \citep{Neves12} and their \teff\ from interferometric calibrations \citep{Khata21}. With this method, we derived a \teff\,=\,3620$\pm$94K and \feh=\,0.14$\pm$0.11\,dex.

\begin{table}[t]
    \centering
    \caption{TOI-756 stellar abundances measured with NIRPS.}
    \label{tab:toi756_abundances}
    \begin{tabular}{lccc}
        \hline \hline
        \textbf{Element} & \textbf{[X/H]}$^*$  & \textbf{\# of lines} \\
        \hline
        Fe I  &  0.20 $\pm$ 0.03  &  15  \\
        Mg I  &  0.22 $\pm$ 0.03  &  5   \\
        Si I  &  0.39 $\pm$ 0.12  &  6   \\
        Ca I  & 0.12 $\pm$ 0.23   &  4   \\
        Ti I  &  0.30 $\pm$  0.10   &  14   \\
        Al I  &  0.10 $\pm$ 0.18   &  1   \\
        Na I  &  0.11 $\pm$ 0.18  &  2   \\
        C I  &  0.29 $\pm$ 0.18  &  2   \\
        K I   &  0.38 $\pm$ 0.18   &  4   \\
        OH  &  -0.43 $\pm$ 0.03  &  37  \\
        \hline
        \multicolumn{3}{l}{$^*$Relative-to-solar abundances}
    \end{tabular}
\end{table}

For the second technique, the combined telluric-corrected NIRPS spectrum obtained with APERO is used to determine the stellar parameters and abundances. Following the methodology of \citet{jahandarComprehensiveHighresolutionChemical2024, jahandarChemicalFingerprintsDwarfs2025}, initially developed for SPIRou spectra \citep{Donati2020},
we retrieve the effective temperature \teff, overall metallicity [M/H] and chemical abundances of TOI-756. We first determine \teff\,and [M/H] by fitting individual spectral lines to a grid of PHOENIX ACES stellar models \citep{Husser_2013} convolved to the resolution of NIRPS. The models are interpolated to fixed $\log g=4.75$ based on the value obtained in Sect.~\ref{subSect:stellar_mass}. We find \teff\,=\,$3710\pm33$\,K and [M/H]\,=\,$0.17\pm0.08$\,dex. While \teff\, is in agreement with the value derived from HARPS, we observe a significant discrepancy between the measurements of the different bands, obtaining $3575\pm23$\,K for $Y$ and $J$, and $3803\pm17$\,K for $H$. This discrepancy could be due to the lack of $K$-band coverage in NIRPS, as this spectral range was found to be crucial in the determination of \teff\ with SPIRou \citep{jahandarComprehensiveHighresolutionChemical2024}. To better reflect the bimodality of the distribution, we inflated the uncertainty on the temperature to the half-distance of the two chromatic measurements (\teff\,=\,$3710\pm113$\,K). The temperatures obtained for NIRPS and HARPS were then combined with a weighted average to give the adopted effective temperature, \teff\,=\,$3657\pm72$\,K. However, for the abundance analysis, we used the effective temperature obtained from HARPS to be more conservative.

The abundances of chemical species are determined by fitting the PHOENIX grid to individual spectral lines \citep{jahandarComprehensiveHighresolutionChemical2024, jahandarChemicalFingerprintsDwarfs2025} with fixed \teff\, of 3620\,K. The stellar abundances measured from the NIRPS spectrum are recorded in Table \ref{tab:toi756_abundances}, although it should be noted that the assumption of a fixed \teff\, likely results in an underestimation of the uncertainties.

\subsection{Mass, radius, and age}\label{subSect:stellar_mass}

\begin{figure}[t]
  \centering
    \includegraphics[width=0.48\textwidth]{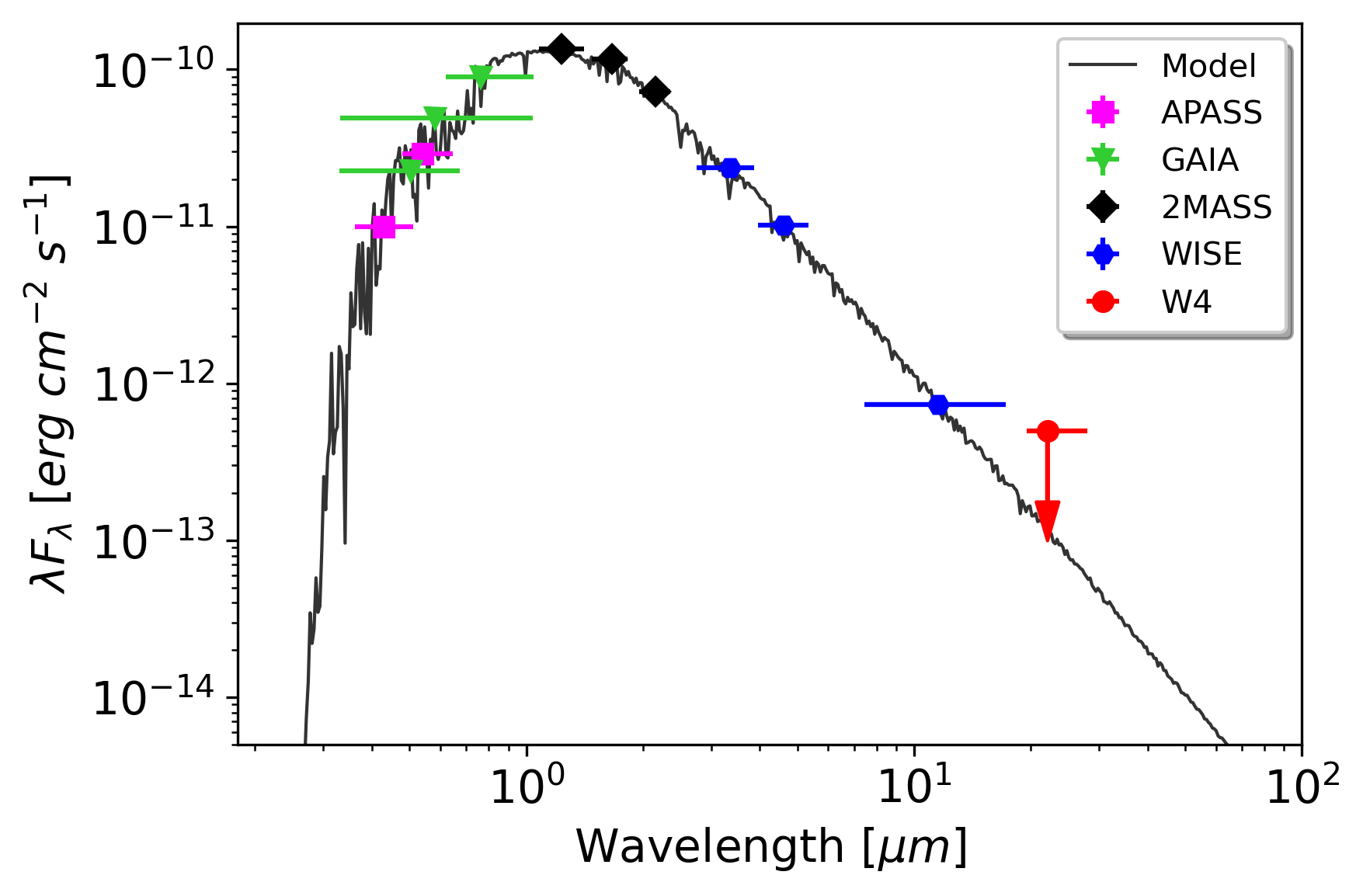}
  \caption{SED of TOI-756, constructed using broadband photometric data from APASS (magenta), Gaia (green), 2MASS (black), and WISE (blue). The upper limit for the WISE W4 band is indicated by a red dot. The horizontal error bars represent the passband widths of the respective filters. Below the SED, the residuals are shown, normalized to the photometric errors. The SED was modeled using the BT Settl atmospheric model \citep{Allard2012} with \teff\,=\,3600\,K, [M/H]\,=\,0\,dex, and $log\,g_\star$\,=\,4.5\,$\mathrm{cm}\ \mathrm{s}^{-2}$.} 
  \label{fig:SED}  
\end{figure}

To derive the mass and radius of the star, we constructed the spectral energy distribution (SED) using the flux densities from the photometric bands \textit{GBP}, \textit{G}, and \textit{GRP} from the GAIA mission \citep{Gaia2018}, \textit{B} and \textit{V} from APASS \citep{Henden2015}, \textit{J}, \textit{H}, and \textit{Ks} from the 2MASS project \citep{2MASS2006}, and \textit{W1}, \textit{W2}, \textit{W3}, and \textit{W4} from the WISE mission \citep{WISE2010}. For the SED modeling process, we employed the Virtual Observatory Spectral Analyzer (VOSA) tool \citep{Bayo2008}. Theoretical models such as BTSettl \citep{Allard2012}, Kurucz \citep{Kurucz1993}, and Castelli \& Kurucz \citep{Castelli2003} are used to construct the synthetic SEDs, where the most suited model was BT Settl with $T_\text{eff}$ = 3600~K, [M/H] = 0 dex, and $log(g)$ = 4.5 $\mathrm{cm}\ \mathrm{s}^{-2}$. VOSA uses a $\chi^2$ minimization technique to achieve the best fit between the theoretical curve and the observational data, taking into account the observed flux, the theoretical flux predicted by the model, the observational error related to the flux, the number of photometric points, input parameters, the object's radius, and the distance between the observer and the object. The resulting analysis is shown in Fig.\ref{fig:SED}.

We then integrated the observed SED to obtain the bolometric luminosity: $L_\star$\,=\,0.03805\,$\pm$\,0.00011\,$L_\odot$. We obtained the stellar radius $R_\star$\,=\,0.501\,$\pm
$\,0.014\,$R_\odot$ using the Stefan-Boltzmann law, $L_\star$\,=\,$4\pi R² \sigma \teff^4$. Finally, the stellar mass ($M_\star$\,=\,0.505\,$\pm$\,0.019\,$M_\odot$) was estimated using Equation 6 from \citet{Schweitzer2019}.

We independently derived the mass and radius of TOI-756 using the empirically established M-dwarf mass-luminosity and radius-luminosity relations from \citet{Mann2019} and \citet{Mann2015}, respectively. To do this, we utilized the Gaia stellar parallax and the $K_s$ magnitude from 2MASS to calculate the absolute $K_s$ magnitude ($M_K$). We employed a Monte Carlo method to propagate the uncertainties and incorporated the intrinsic errors of the relations, which are 2.89\% for the radius and 2.2\% for the mass, into our results.

The results of these two independent methods are compiled in Table~\ref{tab:stellar_params}. The two methods give very consistent radius and mass values. We combined the resulting stellar masses and radii, taking the larger uncertainties to be conservative, as final stellar parameters. Together with the results of Sect.~\ref{subSect:stellar_teff}, we derived the associated stellar luminosity ($L_*$), gravity ($log~g_*$), and density ($\rho_*$). These final parameters are listed in Table~\ref{tab:Stellar parameters}.

Additionally, to estimate the age of the star, we used the code \texttt{isoAR}\footnote{\url{https://github.com/rabrahm/isoAR}} from \citet{Brahm2018} using the T$_\text{eff}$ and [Fe/H] derived in Sect.~\ref{subSect:stellar_teff}, plus Gaia photometry and parsec isochrones. The resulting age of 3.2$^{+5.5}_{-2.3}$ Gyr is poorly constrained, which is expected for an M dwarf.

\begin{table}[t]
    \centering
    \caption{TOI-756 stellar parameters derived by different methods.}
    \label{tab:stellar_params}
    \small
    \renewcommand{\arraystretch}{1.2}
    \setlength{\tabcolsep}{3.5pt}
    \begin{tabular}{lcc}
        \hline\hline
         & \textbf{SED} & \textbf{Mann's relations}  \\
        \hline
        $R_*$ ($R_\odot$)  & $0.501\pm0.014$ & $0.508\pm0.015$ \\
        $M_*$ ($M_\odot$)  & $0.505\pm0.019$ & $0.505\pm0.012$ \\
        $L_*$ ($L_\odot$)  & $0.0381 \pm 0.0001$  & $0.042^{+0.0042}_{-0.0039}$  \\
        $\rho_*$ ($g\ cm^{-3}$) & $5.66^{+0.54}_{-0.49}$ & $5.45^{+0.53}_{-0.47}$  \\
        $\log g_*$ ($cm\ s^{-2}$)  & $4.742\pm0.029$ & $4.731^{+0.028}_{-0.027}$  \\
        \hline
    \end{tabular}
\end{table}

Finally, space velocities U$_{LSR}$, V$_{LSR}$, and W$_{LSR}$\footnote{$U_{LSR}$ is directed radially inwards towards the Galactic center, $V_{LSR}$ along the direction of Galactic rotation, and $W_{LSR}$ vertically upwards towards the Galactic North pole.}, were calculated using positions, parallaxes, proper motions, and RVs from Gaia DR3. To relate the space velocities to the local standard of rest (LSR), the Sun’s velocity components relative to the LSR (U$_\odot$, V$_\odot$, W$_\odot$) = (11.10, 12.24, 7.25)~km\,s$^{-1}$ from \citet{Schronrich2010} were added. We obtained : $U_{LSR}= -57.56 \pm 0.64$ km\,s$^{-1}$, $V_{LSR} = -44.82 \pm 0.97$ km\,s$^{-1}$ and $W_{LSR} = 22.18 \pm 0.36$ km\,s$^{-1}$. According to the probabilistic approach of \citet{Bensby2014}, the galactic kinematic indicates that TOI-756 is a thin disc population star. 
All the adopted astrometric and photometric stellar properties of TOI-756 are listed in Table~\ref{tab:Stellar parameters}. According to all these parameters and the Table 5 from \citet{Pecaut2013}, \hbox{TOI-756} corresponds well to a M1V star.

In terms of stellar activity, TOI-756 appears to be rather quiet, with no identifiable rotation period. Inspecting the TESS SAP light curves and ASAS \citep{Pojmanski1997} data with Lomb-Scargle periodograms reveal no significant peaks. Similarly, no notable signals are observed in the RV indicators from HARPS and NIRPS. Additionally, we used the HARPS DRS data to compute the $log_{R'hk}$ from the S-index measuring Ca H and K emission. We used the relations from \citet{Suarez2015,Suarez2016}, and found a median value of -5.16, in line with the absence of activity of the star.

\section{Photometric and radial velocity analysis}\label{sect:photo_rv_analysis}

We utilized the software package \texttt{juliet} \citep{Espinoza2019_juliet} to model both the photometric and RV data. This algorithm integrates several publicly available tools for modeling transits (\texttt{batman}; \citealt{Kreidberg2015_batman}), RVs (\texttt{radvel}; \citealt{Fulton2018_radvel}), and Gaussian processes (GPs; \texttt{george}; \citealt{Ambikasaran2015_george}; \texttt{celerite}, \citealt{Foreman-Mackey2017_celerite}). To compare different models, \texttt{juliet} efficiently calculates the Bayesian evidence (ln Z) using \texttt{dynesty} \citep{Speagle2020_dynesty}, a Python package that estimates Bayesian posteriors and evidence through nested sampling methods. Unlike traditional approaches that begin with an initial parameter vector centered around a likelihood maximum found via optimization, nested sampling algorithms draw samples directly from the priors. Throughout our analyses, we ensured that we had a sufficient number of live points $N_{live}$ relative to the number of free parameters $d$ ($N_{live} \geq 25 \times d$), preventing us from missing peaks in the parameter space. We conducted several analyses: starting with only the photometry, and using the resulting planet parameters as priors for a subsequent RV analysis and then a joint fit of the two.

\subsection{Photometry analysis}\label{sect:photometry_fit}

First, we used \texttt{juliet} to model the photometry. We used the TESS PDCSAP fluxes of the four sectors where our planet candidate was initially detected, the two transits from LCO-CTIO telescope in $\textit{i'}$ and  $\textit{g'}$ bands and the four transits from \hbox{ExTrA} telescopes. The transit model fits the stellar density $\rho_\star$ along with the planetary and jitter parameters. We adopted a few parametrization modifications when dealing with the transit photometry. Rather than fitting directly for the planet-to-star radius ratio ($p = R_p / R_\star$) and the impact parameter of the orbit ($b = a/R_\star \text{cos} i$), \texttt{juliet} uses the parametrization introduced in \citet{Espinoza2018} and fits for the parameters $r_1$ and $r_2$ to guarantee full exploration of physically plausible values in the ($p$,$b$) plane. Additionally, we implemented a "power-2" limb-darkening law in \texttt{juliet}, as shown to be the best for fitting cold star intensity profiles \citep{Morello2017}. We derived the "power-2" stellar limb-darkening coefficients and their uncertainties for each photometric filter used using the \texttt{LDCU}\footnote{\url{https://github.com/delinea/LDCU}} code \citep{Deline2022}. The \texttt{LDCU} code is a modified version of the Python routine implemented by \citet{Espinoza2015} that computes the limb-darkening coefficients and their corresponding uncertainties using a set of stellar intensity profiles accounting for the uncertainties on the stellar parameters. The stellar intensity profiles are generated based on two libraries of synthetic stellar spectra: ATLAS \citep{Kurucz1979} and PHOENIX \citep{Husser_2013}. We utilized the limb-darkening coefficients determined from \texttt{LDCU} as Gaussian priors for the fit. Since the TESS PDCSAP light curves are already corrected for contamination, we fixed the TESS dilution factor to one. We applied the same assumption to the ground-based photometry, as the apertures are free from contaminating sources. We added in quadrature jitter terms $\sigma_{i}$ to all the photometric uncertainties, which may be underestimated due to additional systematics. To account for the remaining photometric variability, we included a GP for Sectors 10 and 11 of TESS PDCSAP fluxes using a Matérn 3/2 kernel with hyper-parameters amplitude ($\sigma_{GP}$) and timescale ($\rho_{GP}$). Including GPs for Sectors 37 and 64 led to negligible amplitudes, so we did not apply the GP correction to these sectors. We detrended the LCO-CTIO transit in the $\textit{g'}$ band with airmass and the ExTrA data was detrended with a GP using a Matérn 3/2 kernel as suggested by \citet{Cointepas2021}. The resulting detrending can be seen in Appendix~\ref{appendix:tess_fit}. 

We first assumed a circular orbit so we fixed the eccentricity to zero and used normal priors around the ExoFOP values for the period and transit epoch. We used a normal prior for stellar density using the value derived in Sect.~\ref{Sect:stellar_charac}. In the first instance, we fit the TESS photometry alone to constrain these parameters and then used the resulting posteriors to jointly fit all the photometry (see Appendix~\ref{appendix:tess_fit}). We first used the classical ($p$,$b$) parametrization to let the planet-to-star ratio $p$ to be different among photometric filters to check for possible false positives. We found consistent transit depths among the different photometric bands: $p_{TESS} = 0.050\pm 0.001$, $p_{LCO-i'} = 0.052\pm 0.003$, $p_{LCO-g'} = 0.048\pm 0.004$ and $p_{ExTrA} = 0.051\pm 0.005$. We then fitted only one set of $r_1$ and $r_2$ parameters for the planet. For the joint photometry fit, we only took TESS data around the transits ($\pm$\,3 hours around the transit times calculated with the resulting period and transit epoch of the TESS-only fit) in order to reduce the fit time of the joint analysis with \texttt{juliet}.

\subsection{Radial velocity analysis}\label{sect:RVs_analysis}

\begin{figure*}[t]
\centering
    \includegraphics[width=1.\textwidth]{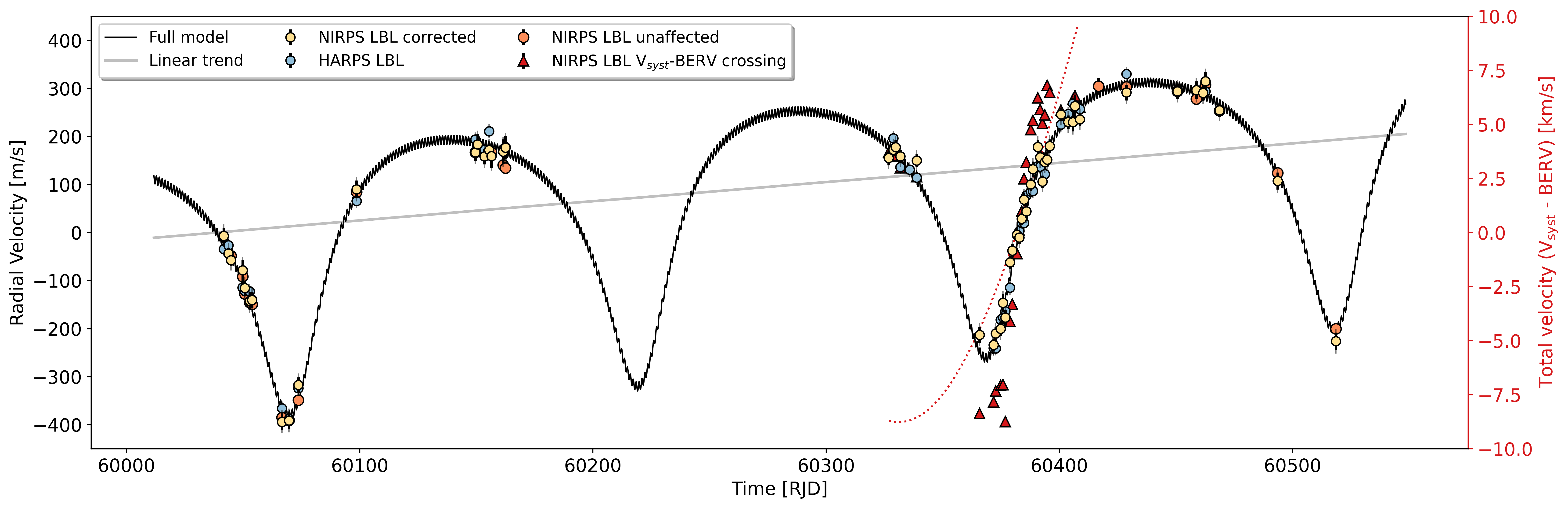}
     \caption{RV data from HARPS (blue dots) and NIRPS measurements. The NIRPS data is separated in unaffected datapoints (orange dots) and affected datapoints (red triangles) by the crossing of the V$_\text{syst}$ and BERV velocities during the observations. Yellow dots are NIRPS data with the correction explained in Sect.~\ref{sect:NIRPS_RVs_correction}. The dotted red line together with the right \textit{y} axis represent the total velocity of TOI-756, showing the crossing of the BERV with the V$_\mathrm{syst}$. The complete inferred model in Sect.~\ref{sect:joint_fit_juliet}, which comprises signals from the two planets along with the linear model for the acceleration (in gray), is represented by a black solid line.}
     \label{fig:RVs}
\end{figure*}

The reduction and preprocessing of the RVs are explained in Sect.~\ref{sect:RVs_observations} and the resulting datasets are shown in Fig.~\ref{fig:RVs}. A significant RV variation was detected for TOI-756, suggesting the presence of an additional companion to the TESS sub-Neptune. Several possible orbital periods were initially explored, motivating further observations to constrain this signal alongside that of the 1.24-day sub-Neptune. These efforts led to the confirmation of an eccentric companion on a $\sim$150-day orbit. In addition, the RVs reveal an acceleration, hinting at a third, more distant object.

\begin{figure}
  \centering
    \includegraphics[width=0.45\textwidth]{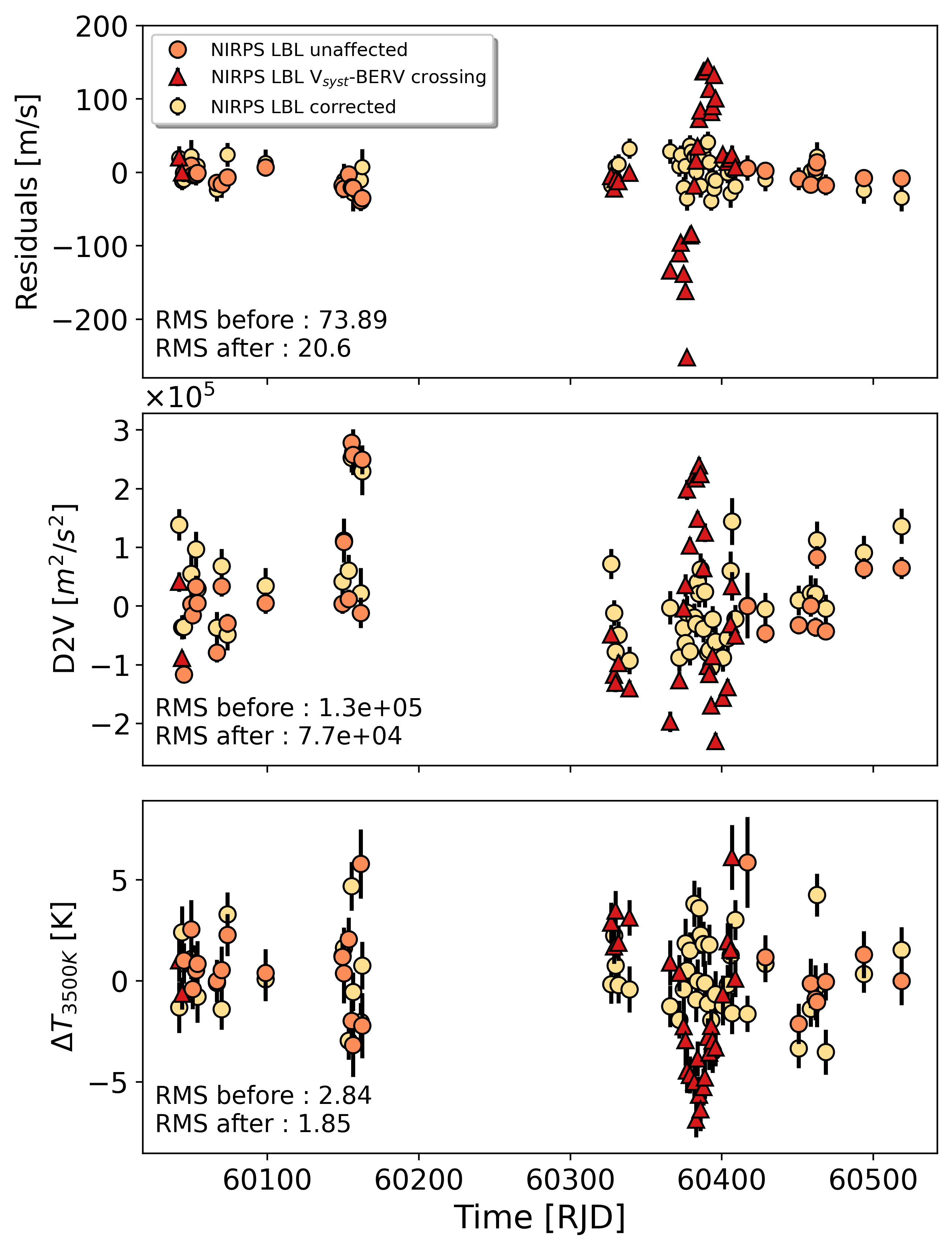}
  \caption{NIRPS data residuals (upper panel) and LBL indicators D2V (middle panel), along with dTemp ($\Delta T$) calculated at 3500~K close to the effective temperature of TOI-756 (bottom panel). Orange dots are the indicators from unaffected NIRPS data, red triangles are the affected spectra by V$_\text{syst}$-BERV crossing and yellow dots are the indicators with the correction explained in Sect.~\ref{sect:NIRPS_RVs_correction}.}
  \label{fig:RVs_indicators}
\end{figure}

\subsubsection{Telluric contamination of the NIR data during the V$_\text{syst}$-BERV crossing.}

One of the main challenges of NIRPS and near-infrared spectroscopy is the contamination of the stellar spectrum by molecular species in Earth's atmosphere. This issue is particularly pronounced when observing M dwarfs, whose spectra also contain absorption features from species such as H$_2$O and CH$_4$. The NIRPS-DRS includes a telluric absorption correction based on \citet{Allart2022}, but the observed spectra of faint M dwarfs are further dominated by strong emission lines from Earth's atmosphere, notably from OH. These emission lines are typically corrected during the reduction process (Srivastava et al. in prep), but their non-LTE nature makes them more challenging to correct, as they cannot be modeled using standard telluric line approaches.

However, this contamination becomes especially problematic when aiming to measure precise stellar radial velocities, particularly when telluric lines coincide with the absorption lines of the star. This situation arises when the barycentric Earth radial velocity (BERV) crosses the systemic velocity (V$_\text{syst}$) of the observed star. In such cases, the correction is challenging, as blending between the stellar and telluric lines distorts the stellar line profiles, leading to erroneous RV estimates. Since the telluric emission features are more numerous and stronger in $H$ compared to the $J$, there will be a chromatic offset induced in the calculated RVs between the "red" wavelengths, and "blue" wavelengths.

We observed this phenomenon between RJD = 60365 and 60416, when the Keplerian fit of the outer companion of TOI-756 showed clear outliers of $\sim$100-200\,m\,s$^{-1}$ in the NIRPS RVs (see top panel of Fig.~\ref{fig:RVs_indicators}). This effect is illustrated in Fig.~\ref{fig:RVs}, the NIRPS data is represented as red triangles when the absolute total velocity |V$_{\text{tot}}| = |\text{V}_\text{syst} - BERV | \leq 10~$km\,s$^{-1}$, corresponding to approximately twice the typical spectra line FWHM for slow rotating M dwarfs (5~km\,s$^{-1}$). The HARPS data are shown using blue dots and the non-affected NIRPS data (V$_{\text{tot}}>10$~km\,s$^{-1}$) using orange dots. In this Figure~\ref{fig:RVs}, we plotted the relative velocity to the systemic velocity of TOI-756, which was found to be $\text{V}_\text{syst}\sim15.2$~km\,s$^{-1}$. This phenomenon also induces distortions in the stellar line profiles, which are evident in the systemic variations of different LBL indicators, such as D2V and $\Delta T$ \citep{Artigau2024}, as shown in Fig.~\ref{fig:RVs_indicators}. 

\begin{figure}[t]
  \centering
    \includegraphics[width=0.45\textwidth]{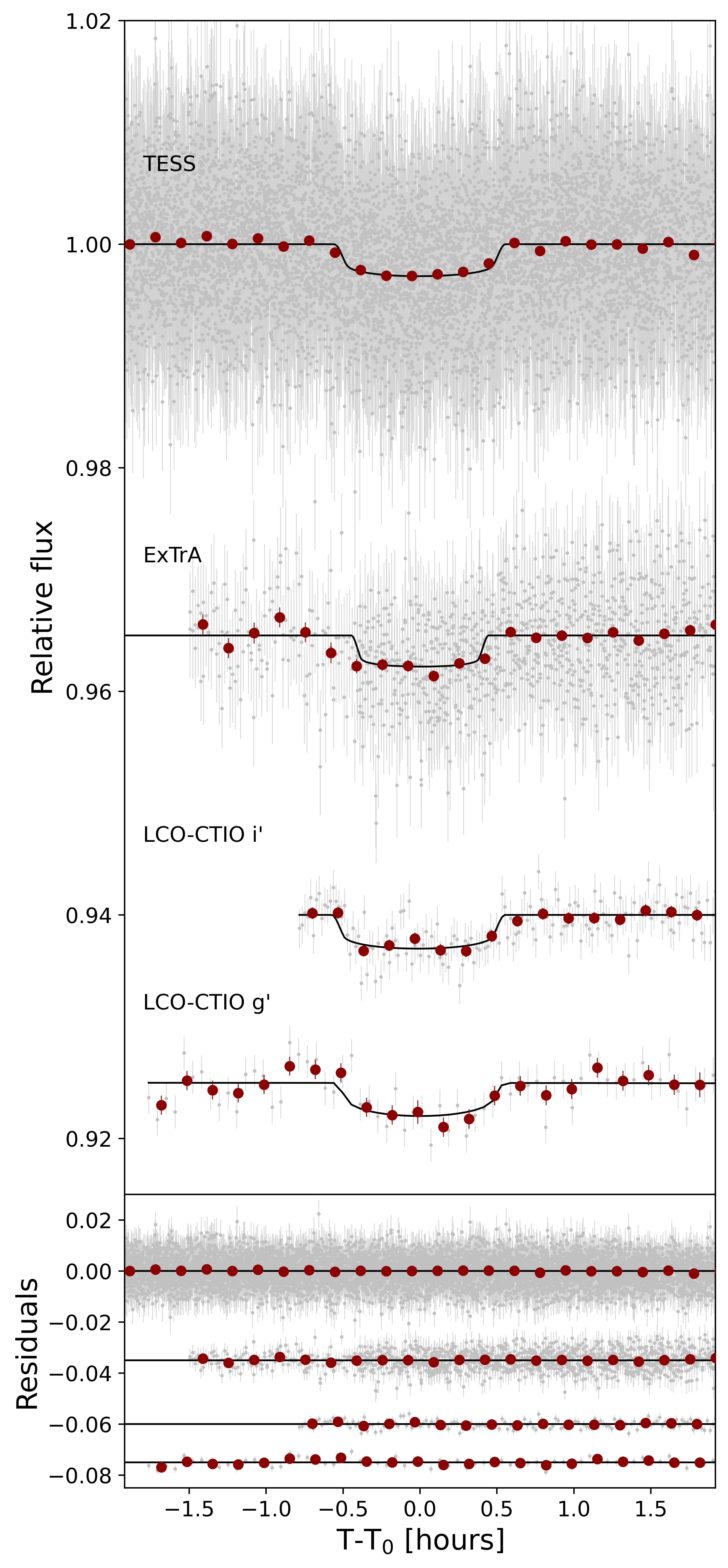}
  \caption{Top panel: Phase-folded TESS, ExTrA, and LCO-CTIO light curves of TOI-756~b (gray points). Dark red circles are data binned to 10 min. The black lines represent the median model of each instrument from the joint fit. Bottom panel: Residuals of the data compared to the model. An arbitrary offset has been added to the ground-based photometry for clarity.} 
  \label{fig:Photometry}
\end{figure}

\subsubsection{Correction with removing affected lines in the LBL}\label{sect:NIRPS_RVs_correction}

The advantage of using the line-by-line (LBL) technique to derive radial velocities and other spectral indicators lies in its ability to provide individual measurements for each spectral line across the entire spectrum. The observed discrepancy, where NIRPS RVs appear underestimated and then overestimated relative to the RV fit, is caused by the crossing of stellar absorption lines with atmospheric OH emission lines, primarily arising from excitation of rotational-vibrational modes of the OH molecule.

To mitigate this effect, we utilized the HITRAN\footnote{\url{https://hitran.org/}} database \citep{Gordon2022} to identify OH lines present within the NIRPS spectral range. We selected the 25\% most intense OH lines and removed the LBL-derived measurements in a $\pm$20~km\,s$^{-1}$ window around these lines, accounting for the approximate width of both OH and stellar lines ($\sim$10~km\,s$^{-1}$ each). This process reduced the number of spectral lines used in the LBL analysis from 26,301 to 17,253, inevitably increasing the RV uncertainties. We then recomputed the final radial velocity and indicator values for each epoch using the same LBL method, which robustly averages the per-line values while down-weighting outliers \citep[Appendix B of][]{Artigau2022}. We applied this correction for all the NIRPS data for consistency. The corrected data are displayed as yellow dots in Figures \ref{fig:RVs} and \ref{fig:RVs_indicators}.

This correction successfully brought NIRPS RVs into agreement with HARPS and effectively removed the systematic distortions previously visible in the residuals of the keplerian fit and in the spectral indicators (Fig.~\ref{fig:RVs_indicators}). The corrected indicators now show consistent values across epochs, with no residual systematics during the V$_\text{sys}$–BERV crossing. Additionally, the root mean square (RMS) of the residuals and the indicators, displayed in the same figure, demonstrates a significant decrease after correction.

We applied the same technique to the mask used to derived the NIRPS DRS cross-correlation function (CCF) data, using the OH line list from the DRS telluric correction module. While this also mitigated the effect, the CCF method relies on significantly fewer spectral lines than the LBL approach, leading to much larger error bars after correction. Consequently, we opted to retain the LBL-derived values for our analysis.

\subsubsection{Radial velocity analysis and joint modeling with \texttt{juliet}}\label{sect:joint_fit_juliet}

\texttt{juliet} was also used to model these RV datasets. We used a two-planet plus a linear trend model. At first, we fixed the eccentricity of the TESS planet TOI-756~b to zero. We accounted for the evident eccentricity of the outer companion by fitting for the parameters $\sqrt{e}cos(\omega)$, $\sqrt{e}sin(\omega)$, as implemented in \texttt{juliet}. This parametrization has been shown to improve the exploration of the eccentricity–argument of periastron parameter space \citep{Lucy1971}. We used the period and transit epoch results of TOI-756~b of the photometry fit (Sect.~\ref{sect:photometry_fit}) as priors for the RV fits.

We compared several analyses: (1) NIRPS data corrected using the method described in Sect.~\ref{sect:NIRPS_RVs_correction} combined with HARPS data; (2) uncorrected NIRPS data with the affected points removed, also combined with HARPS data; (3) HARPS data only; and (4) corrected NIRPS data only. We present the posteriors of the main changing planetary parameters of the different fits in Table~\ref{tab:rvs_fit}. We did not put the period and transit epoch of planet b in this table because they are very similar for these analyses since they are highly constrained by the photometry fit priors. The resulting parameters exhibit good consistency within 1$\sigma$ for all our different analyses.

Since the NIRPS corrected data combined with the HARPS data are fully consistent with the other fits and included all RV points, we choose this dataset as the final one. We do a joint fit of the RVs and photometry with \texttt{juliet}: NIRPS corrected RVs, HARPS RVs, TESS, LCO-CTIO and ExTrA. 
In order to prevent any potential Lucy-Sweeney bias in the eccentricity measurement \citep{Lucy1971,Hara2019}, we fixed the orbital eccentricity of the planet b to zero. However, to explore the possibility of non-circular orbit, we ran a separate analysis without any constraints on the eccentricity. The logarithmic evidence for the eccentric model is lower than for the circular one (55,557 vs. 55,580), and the fitted jitter values for both HARPS and NIRPS RVs are higher in the eccentric case, further supporting the preference for the circular model. The fitted eccentricity for TOI-756~b is $e_b = 0.096^{+0.092}_{-0.067}$. Therefore, the condition $e > 2.45~\sigma_e$ \citep{Lucy1971} is not satisfied, which suggests that the RV data are compatible with a circular orbit. For now, the current data do not provide sufficient precision to draw a firm conclusion regarding the orbital eccentricity. Additionally, in Table~\ref{tab:planetaryparams}, we show that the 3-$\sigma$ upper limit on the eccentricity for TOI-756~b is 0.51. The fitted and derived parameters for TOI-756~b and TOI-756~c are presented in Table~\ref{tab:planetaryparams}.

\begin{table*}[t]
    \centering
    \caption{Posterior planetary parameters of the different RV fits.}
    \label{tab:rvs_fit}
    \small
    \renewcommand{\arraystretch}{1.2}
    \begin{tabular}{lcccc}
        \hline\hline
         & \textbf{NIRPS corrected + HARPS} & \textbf{NIRPS unaffected + HARPS} & \textbf{HARPS-only} & \textbf{NIRPS corrected-only} \\
        \hline
        $K_b$ $\mathrm{m\,s}^{-1}$  & $9.4^{+2.3}_{-2.5}$ & $8.4 \pm 2.2$ & $10.0^{+3.1}_{-3.2}$ & $9.2^{+3.9}_{-3.8}$ \\
        $P_c$ (days)  & $149.66^{+0.28}_{-0.26}$ & $149.61\pm0.20$ & $149.36^{+0.38}_{-0.37}$ & $149.92^{+0.36}_{-0.39}$\\
        $T_{0,c}$ (RJD)  & $60350.27^{+0.69}_{-0.68}$  & $60350.29^{+0.52}_{-0.53}$ & $60199.80^{+0.70}_{-0.68}$ & $60201.59^{+0.78}_{-0.81}$ \\
        $K_c$ (m/s) & $272.6^{+3.6}_{-3.4}$ & $273.3^{+2.5}_{-2.6}$ & $275.7^{+5.0}_{-4.9}$ & $268.9 \pm 5.1$ \\
        $e_c$   & $0.46\pm0.01$ & $0.46\pm0.01$ & $0.44\pm0.01$ & $0.48^{+0.02}_{-0.01}$\\
        $\omega_c$ ($^\circ$)& $-167.9 \pm 1.3$ & $-166.6^{+1.2}_{-1.3}$ & $-166.7 \pm 1.7$ & $-169.24 \pm 1.9$ \\
        Acceleration $\mathrm{m\,s}^{-1}\,\mathrm{yr}^{-1}$ & $144.6\pm 5.8$ & $148.3^{+3.7}_{-4.0}$ & $145.7^{+8.0}_{-8.4}$ & $145.4^{+8.0}_{-8.8}$ \\
        \hline
    \end{tabular}
\end{table*}

\begin{figure*}[t]
\centering
    \includegraphics[width=0.48\textwidth]{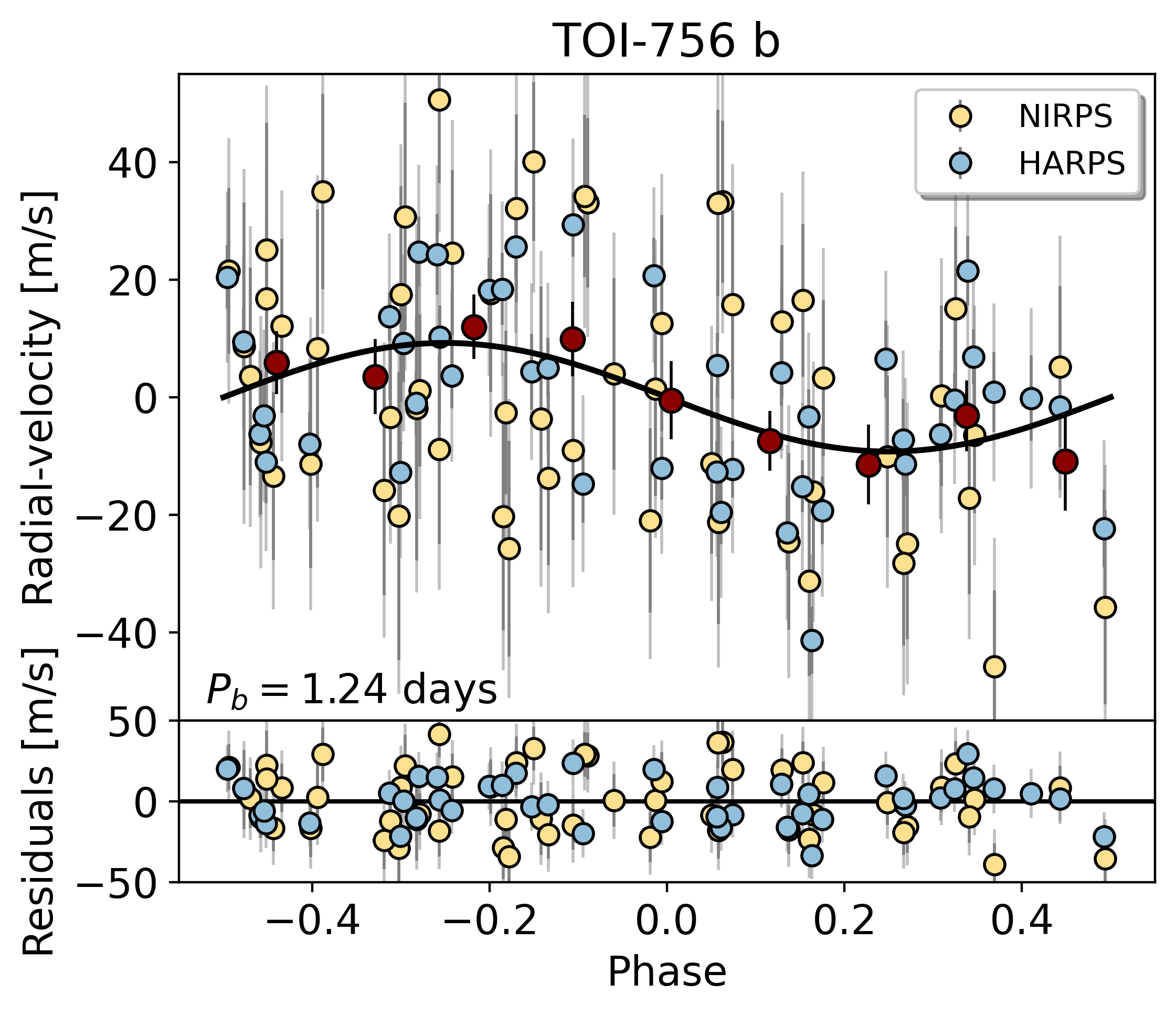}
   \includegraphics[width=0.48\textwidth]{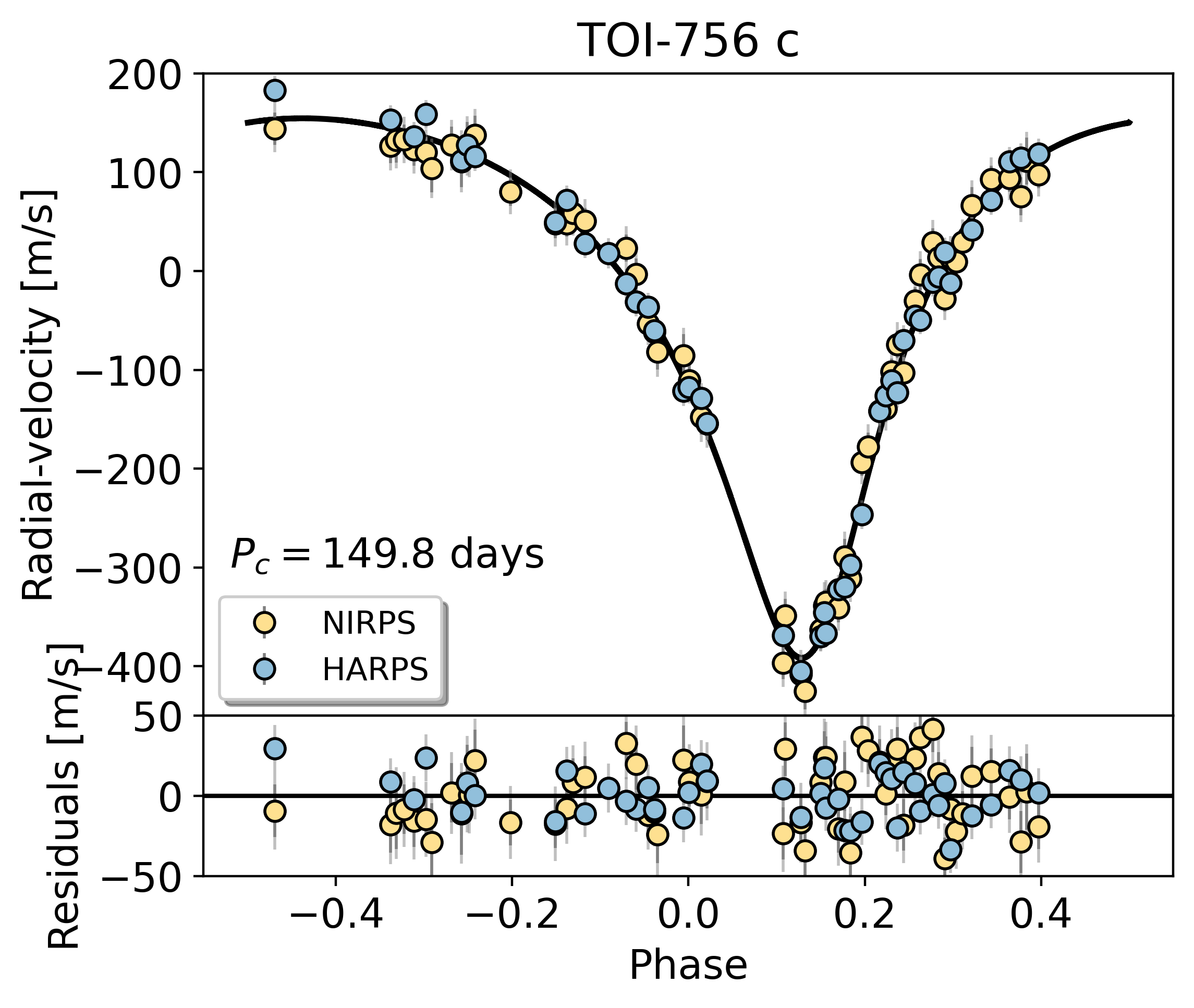}
     \caption{Phase-folded RVs with the resulting model and its residuals for TOI-756~b (left) and TOI-756~c (right). In red dots, binned data combining HARPS (blue dots) and NIRPS (yellow dots). The error-bars in light gray account for the fitted jitters.}
     \label{fig:phase_folded_Rvs}
\end{figure*}

The priors and posteriors of the joint fit can be found in Table \ref{tab:posteriors_jointfit}. 
Fig.\ref{fig:Photometry} shows the phase-folded light-curves of the photometry fit. Fig.\ref{fig:RVs} shows RV data together with the resulting model from the joint fit and Fig.~\ref{fig:phase_folded_Rvs} the phase-folded RV curves for the two planets.

We searched for a possible transit of planet c in the TESS data by phase-folding the light curve using its orbital period and time of conjunction. Although TESS observations cover the expected transit window, no transit signal is visible in the data, allowing us to exclude a transiting configuration. Furthermore, assuming coplanar orbits aligned with the inclination of TOI~756\,b (85.5$^\circ$), we estimated the expected impact parameter of the outer planet based on its semi-major axis and the stellar radius. We obtained $b_c = 14.3 \pm 0.7$, a value well above 1.

\section{Discussion}\label{sect:discussion}

We present the discovery and characterization of the transiting sub-Neptune TOI-756\,b and the non-transiting eccentric cold giant TOI-756\,c, both orbiting the M1V star TOI-756. TOI-756\,b has an orbital period of 1.24 days, a radius of $2.81\pm0.10$ \re and a mass of $9.8\pm1.7$ \me. The outer companion, TOI-756\,c, follows an eccentric (0.45) 149-day orbit and has a minimum mass of $4.05\pm0.11$ \mjup. Using the stellar parameters (Table~\ref{tab:Stellar parameters}), we determine the semi-major axes of TOI-756\,b and TOI-756\,c to be $0.0180\pm0.0002$ au and $0.439\pm0.005$ au, respectively. Assuming zero albedo and full heat redistribution, the equilibrium temperature of TOI-756\,b is $934\pm24$~K,  with a stellar insolation of $127\pm13$~\se. For TOI-756\,c, we estimate an equilibrium temperature of $194\pm5$~K and a stellar insolation of $0.24\pm0.02$~\se averaged along the eccentric orbit. In addition, the RVs present an acceleration of $145.6\pm5.2~\mathrm{m\,s}^{-1}\,\mathrm{yr}^{-1}$ hinting at an additional component in the system.

\subsection{NIRPS + HARPS performances}
The characterization of this system was made possible by TESS and ground-based facilities for the photometric analysis of the inner transiting planet, as well as by the combination of HARPS and NIRPS for RV follow-up. The synergy between these two spectrographs enabled us to precisely characterize an early-M with a peculiar planetary-system configuration. The benefit of this combination is evident in Table~\ref{tab:rvs_fit}: using HARPS (NIRPS) alone, the semi-amplitude of TOI-756~b is determined with a precision of 31\% (42\%), whereas combining HARPS and NIRPS improves this to 17\% in the joint RV and photometry fit. All other fitted parameters also benefit from improved precision thanks to this joint analysis. This study highlights the added value of NIRPS (to HARPS) in characterizing a low-mass planet around a faint M dwarf (\textit{V} = 14.6, \textit{J} = 11.1), compared to typical radial velocity targets. Independently, the two instruments have similar median photon noises. The median photon noise is 5.5\,m\,s$^{-1}$ for HARPS and 15.4\,m\,s$^{-1}$ for NIRPS, but the latter having increased from 8.4\,m\,s$^{-1}$ due to the removal of affected lines during the LBL computation (see Sect. \ref{sect:NIRPS_RVs_correction}). The fitted jitter values for both instruments are similar, around 15 m\,s$^{-1}$, which matches the photon noise for NIRPS but is elevated compared to HARPS photon noise. This suggests the presence of atmospheric residuals or enhanced stellar activity in the optical range. Given the low S/N regime in which HARPS is operating, increased background sky contamination and possible interference from the Moon are to be expected. At such low S/N, the LBL method is also likely to underestimate the uncertainties associated with the derived RVs.

\begin{table}[h]
\tiny
\caption{Fitted and derived parameters for TOI-756 b and TOI-756 c.}
\centering
\renewcommand{\arraystretch}{1.5}
\setlength{\tabcolsep}{1pt}
\begin{center}
\begin{tabular}{lcc}
\hline\hline
\textbf{Parameter} &\textbf{TOI-756 b} & \textbf{TOI-756 c} \\
 \hline
    Orbital period, $\it{P_{\mathrm{orb}}}$ (days)\dotfill  & $ 1.2392495\pm0.0000007$&$ 149.40 \pm 0.16 $ \\
    Time of conjunction, $T_{0}$ (RJD)\dotfill & $58570.65234^{+0.00035}_{-0.00037}$ & $60498.882^{+0.57}_{-0.52}$ \\
    Planet radius, $\it{R_{\mathrm{p}}}$ (\re)\dotfill  & $2.81 \pm 0.10$& - \\
    Planet mass, $\it{M_{\mathrm{p}}}$ (\me)\dotfill  & $ 9.83^{+1.8}_{-1.6}$  & -  \\
    Planet min. mass, $\it{M_{\mathrm{p}}\sin(i)}$ (\mjup) \dotfill  &  -  &$ 4.05 \pm 0.11$  \\
    Planet density, $\rho_\mathrm{p}$ (g~cm$^{-3}$)\dotfill  & $ 2.42^{+0.53}_{-0.45}$ & -\\
    RV semi-amplitude, ($\mathrm{m}\,\mathrm{s}^{-1}$) \dotfill & $9.2^{+1.7}_{-1.5}$ & $273.3 \pm 2.6$ \\
    Orbital inclination, $i$ ($^\circ$)\dotfill & $85.53^{+0.19} _{-0.18}$  & -  \\
    Scaled planetary radius, $R_{P}$/$R_{*}$ \dotfill &  $0.05113^{+0.0008}_{-0.0009}$& - \\
    Impact parameter, $\textit{b}$\dotfill  &$ 0.589^{+0.018} _{-0.021}$ & - \\
    Semi-major axis, $\it{a}$ (au)\dotfill  & $0.0180 \pm 0.0002$&$ 0.439\pm0.005$\\
    Eccentricity \dotfill  & $ 0~(< 0.51, 3\sigma)$ & $0.445\pm0.008$\\
    Argument of periastron, $\omega$ (deg) \dotfill  & $90$ (fixed) & $-167.77^{+0.99}_{-0.93}$\\
    Insolation$^{(a)}$, $\it{S_{\mathrm{p}}}$ (\se)\dotfill  &  $ 127\pm13$& $ 0.24\pm0.02$\\
    Equilibrium temperature$^{(a)}$, $T_{\mathrm{eq}}$ (K)\dotfill  & $ 934^{+23} _{-24}$ &$ 194\pm5$\\
    TSM $^{(b)}$\dotfill  & $ 63^{+13} _{-10}$ & -\\
    Transit duration, $T_{14}$ (h)\dotfill  & $ 1.10\pm0.01$ & -\\

      \hline
\end{tabular}
\begin{tablenotes}
\item
\textbf{Notes:} $^{(a)}$ Insolation and equilibrium temperature are calculated as in \citet{Parc2024}, assuming global circulation and a Bond albedo of A$_\mathrm{B}$ = 0. $^{(b)}$ Transmission spectroscopy metric (TSM) calculated following \citet{Kempton2018}.
\end{tablenotes}
\label{tab:planetaryparams}
\end{center}
\end{table}

\subsection{TOI-756~b: Internal structure, composition, and population context}
\subsubsection{A volatile-rich sub-Neptune around an M dwarf}

\begin{figure}[t]
  \centering
    \includegraphics[width=0.48\textwidth]{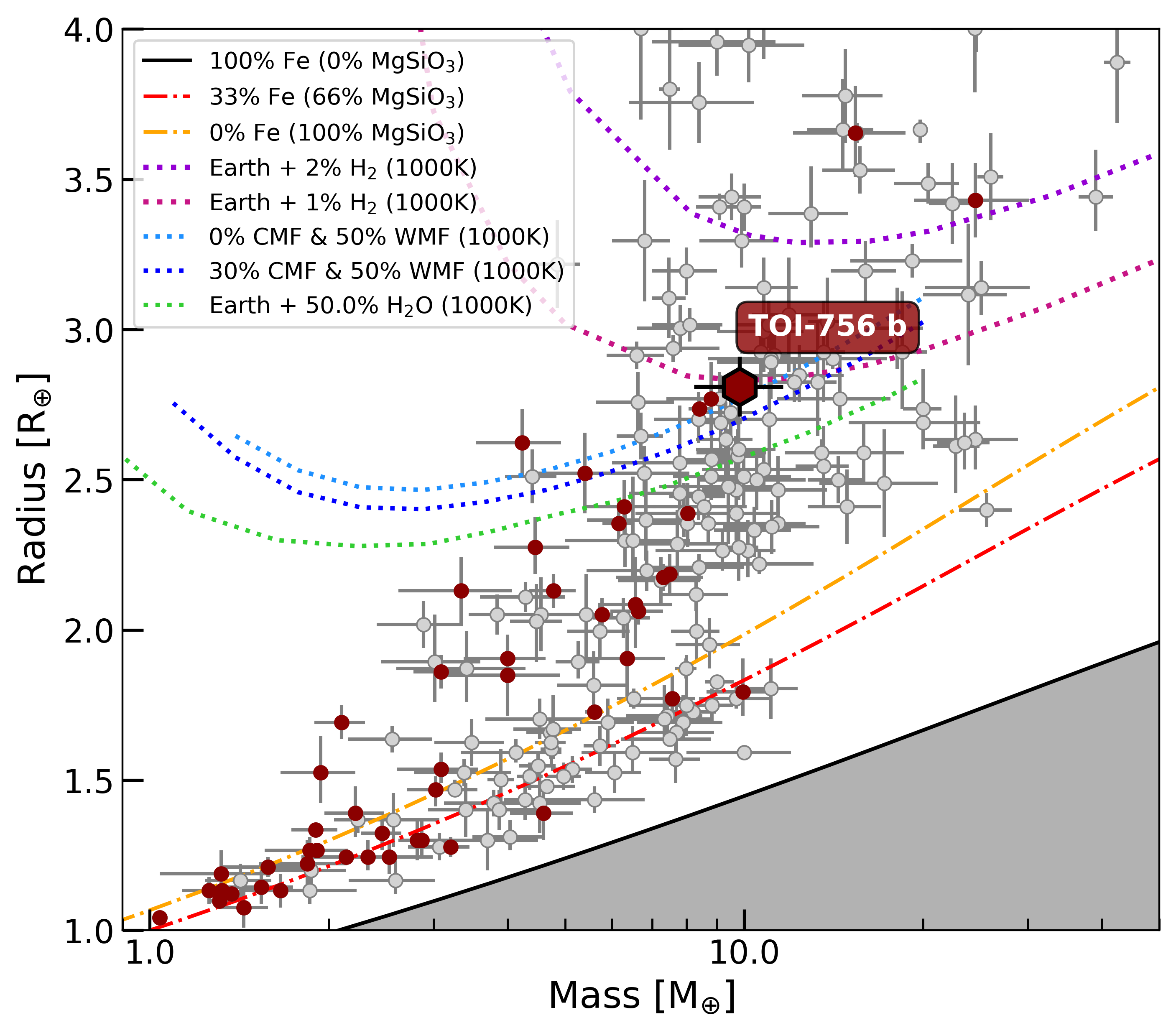}
  \caption{Mass-radius diagram of small exoplanets (with radii ranging from 1–4~\re) with precise densities from the PlanetS catalog. The red (gray) dots correspond to exoplanets orbiting M dwarfs (FGK dwarfs). The composition lines of pure silicates (yellow dashed), Earth-like planets (red dashed), and pure iron (solid black) from \citet{Zeng2016,Zeng2019} are displayed. The red hexagon represent TOI-756~b. Two compositions line that incorporate both water and terrestrial elements from \citet{Aguichine2021} models, matching the equilibrium temperature of the planet, are plotted as light and dark blue dotted lines. Two compositions of Earth with an hydrogen-rich atmospheres from \citet{Zeng2019} are represented in pink and purple dotted lines. This plot has been generated with \texttt{mr-plotter} (\url{https://github.com/castro-gzlz/mr-plotter/}).}
  \label{fig:MR-diagram}
\end{figure}

The characterization of TOI-756~b adds to the currently small population of known transiting sub-Neptunes ($2~\re < R_p < 4~\re$) around M dwarfs, as shown in the Mass-Radius (M-R) diagram (Fig.~\ref{fig:MR-diagram}) where the red (gray) dots represent planets from the PlanetS catalog \citep{Parc2024,Otegi2020} orbiting M dwarfs (FGK dwarfs). Inside this population, \citet{Parc2024} identified statistical evidence for small sub-Neptunes ($1.8~\re < R_p < 2.8~\re$) being less dense around M dwarfs than around FGK dwarfs with a p-value of 0.013, rejecting the null hypothesis. This means that the densities of small sub-Neptunes orbiting M and FGK dwarfs belong to different distributions. We updated this analysis with the up-to-date PlanetS catalog and by including TOI-756~b, which had a density of 2.42 $\mathrm{g}\,\mathrm{cm}^{-3}$. We choose to increase the upper radius limit of this sample to 2.9~\re (2.8~\re having been chosen to capture all small sub-Neptunes around M dwarfs at that time). We find, with the same Mann–Whitney U test \citep{Wilcoxon1945,Mann1947}, an improved p-value of 0.006 for this trend. However, these two analysis are not taking into account the uncertainties on the density measurements. We did a Mann–Whitney U test on 10000 samples using a bootstrapping method to draw density values in the density distributions of each planet and obtained a median p-value of 0.015, a still significant value. However, the sample remains small and the increase of the well-characterized planets in this parameter space is one of the objectives of the NIRPS GTO SP2 subprogram "sub-Neptunes" described in Sect.~\ref{sect:wp2subprog}. 

We plot the mass and radius of TOI-756~b, alongside with the small planets of the PlanetS catalog in Fig.~\ref{fig:MR-diagram}. With its density of 2.42 $\mathrm{g}\,\mathrm{cm}^{-3}$ (and looking at the composition lines shown in the same figure), TOI-756~b lies above the 50\% water plus Earth-composition line at 1000~K (at 2$\sigma$) from \citet{Aguichine2021} (dark blue dotted line). A pure silicate interior with a 50\% steam atmosphere can explain within 1$\sigma$ the radius and mass of TOI-756~b for this model (light blue dotted line). Furthermore, it corresponds well within 1$\sigma$ to the Earth-composition with a H$_2$/He dominated-atmosphere of 1\% the mass from \citet{Zeng2019} (pink dotted line). As models from \citet{Aguichine2021} include pure steam atmosphere with no solubility between the atmosphere and the mantle+core compared to \citet{Luo2024} models (e.g., green dotted line), and are static in time \citep[compared to ][]{Aguichine2024}, they can be considered to over-estimate the radii of the planets. They can thus be interpreted as an upper limit of M-R composition lines for water-rich models. In addition, due to its high equilibrium temperature of approximately 934~K, any water present in the atmosphere of TOI-756~b is expected to be in a supercritical state. In conclusion, it is more likely that TOI-756~b needs an amount of hydrogen/helium in its atmosphere to explain its density, in the form of pure H/He envelope or mixed supercritical H$_2$O and H/He. This places the planet within the "miscible-envelope sub-Neptunes" category defined by \citet{Benneke2024}. Atmospheric characterization could confirm this classification by revealing a mean molecular weight significantly higher than that of Jupiter (2.2) or Neptune (2.53–2.69). We investigate this in greater detail in the following section.

Interestingly, \citet{Schlecker2021} found a difference in the bulk composition of inner small planets with and without cold Jupiters. High-density small planets point to the existence of outer giant planets in the same system. Conversely, a present cold Jupiter gives rise to rocky, volatile-depleted inner super-Earths, by obstructing inward migration of icy planets that form on distant orbits. However, TOI-756~c lies beyond the system's ice line, and its formation may have contributed to the inward delivery of water-rich material, as proposed for the Solar System by \citet{Raymond2017}. This process could account for the potentially ice-rich composition of TOI-756~b. As shown by \citet{Bitsch2021}, the water content of an inner sub-Neptune can provide valuable insights into the formation location and timescale of an outer giant planet relative to the water ice line, offering constraints on planet formation theories.

\subsubsection{Detailed interior modeling}\label{sect:interior}

We perform a detailed interior characterization of TOI-756~b using a Bayesian inference approach, adopting the \texttt{emcee} affine-invariant ensemble sampler \citep{Foreman-Mackey_2013} coupled to a three-layer interior structure model. The planetary interior is assumed to be composed of an Fe-Ni metallic core and a silicate mantle (\texttt{SUPEREARTH} \citealt{Valencia_2007}), while the outermost layer consists of either a hydrogen-helium envelope or a water vapor atmosphere modeled using the CEPAM code \citep{Guillot_1995}, with equations of state from \citet{Saumon_1995} for H/He and \citet{French_2009} for H$_2$O. In all cases, we assume that the rocky interior contains no volatiles and follow the numerical set-up given in \citet{Plotnykov_2020}.

To explore the range of possible atmospheric mass fraction (AMF) values, we consider two sub-Neptune composition scenarios: (1) the planet has a H/He envelope (75$\%$ of H$_2$ to 25$\%$ He) and (2) the planet has pure H$_2$O envelope. For these scenarios, we impose stellar-informed priors on the rocky interior based on the host star’s refractory abundances taken from Table~\ref{tab:toi756_abundances}, namely,
 \[
 \mathrm{Fe/Mg_{planet}} \sim \mathcal{N}(\mathrm{Fe/Mg_{star}},\sigma_{\rm star}^2);\
 \mathrm{Fe/Si_{planet}} \sim \mathcal{N}(\mathrm{Fe/Si_{star}},\sigma_{\rm star}^2),
 \]
where all ratios are by weight. Additionally, the mantle mineralogy is allowed to vary in terms of the Bridgmanite to Wustite ratio (MgSiO$_3$ vs MgO, xWu). These assumptions effectively constrain the rocky core-mass fraction ($\mathrm{CMF}=\frac{\mathrm{rcmf}}{\mathrm{rcmf}+1}$, where rcmf is the core to mantle mass ratio, $M_{core}/M_{mantle}$) of the planet and mitigate problem of compositional degeneracy.

Considering case (1) where TOI-756~b has retained its primordial H/He envelope, we recover a well-constrained AMF~$=0.023 \pm 0.003$ (3 wt$\%$), with a corresponding CMF~$=0.27 \pm 0.03$. Note that this strong constraint may partly result from underestimated abundance uncertainties, as the values were derived assuming a fixed T$_\mathrm{eff}$ (see Sect.~\ref{subSect:stellar_mass}). However, if we impose no prior on the rocky composition, the envelope has an AMF~$=0.03 \pm 0.01$ (3 wt$\%$), while the interior has almost a uniform distribution of CMF~$=0.5 \pm 0.3$. These results suggesting strong evidence that this planet has a volatile envelope based on mass–radius data alone. For the case where the envelope is composed of pure water vapor (2), we find that AMF~$=0.79 \pm 0.10$ and CMF~$=0.27 \pm 0.03$. This very high value of AMF of pure water vapor is highly unlikely when linked to formation theories. Our analysis suggests that the presence of H/He in the atmosphere is more plausible, although a combination of both scenarios (1) and (2) remains a possibility. Regardless, this confirms that TOI-756~b requires a volatile envelope to account for its density, and that the abundances derived from NIRPS spectra have allowed us to better constrain both the CMF and AMF in the case of a pure H/He envelope. The resulting corner plots of this analysis are shown Fig.~\ref{fig:corner_interiors}.

\subsubsection{A planet at the radius cliff and in the Neptune desert}

TOI-756~b is a very interesting target in this sample of small sub-Neptunes around M dwarfs. Indeed, it is a unique object close to the "radius cliff", a steep drop in planet occurrence between 2.5 and 4.0~\re \citep{Borucki2011,Howard2012,Fulton2017}. This still poorly explored demographic feature seems to vary in location with the host star's spectral type as also seen with the radius valley \citep{Ho2024,Parc2024,Burn2024}. \citet{Kite2019} proposed that atmospheric sequestration into magma could explain this phenomenon, as larger atmospheres reach the critical base pressure needed for H$_2$ to dissolve into the core. Ongoing studies seek to understand variations in atmospheric observables across these features to better comprehend their underlying physics. 

\begin{figure}[t]
  \centering
    \includegraphics[width=0.48\textwidth]{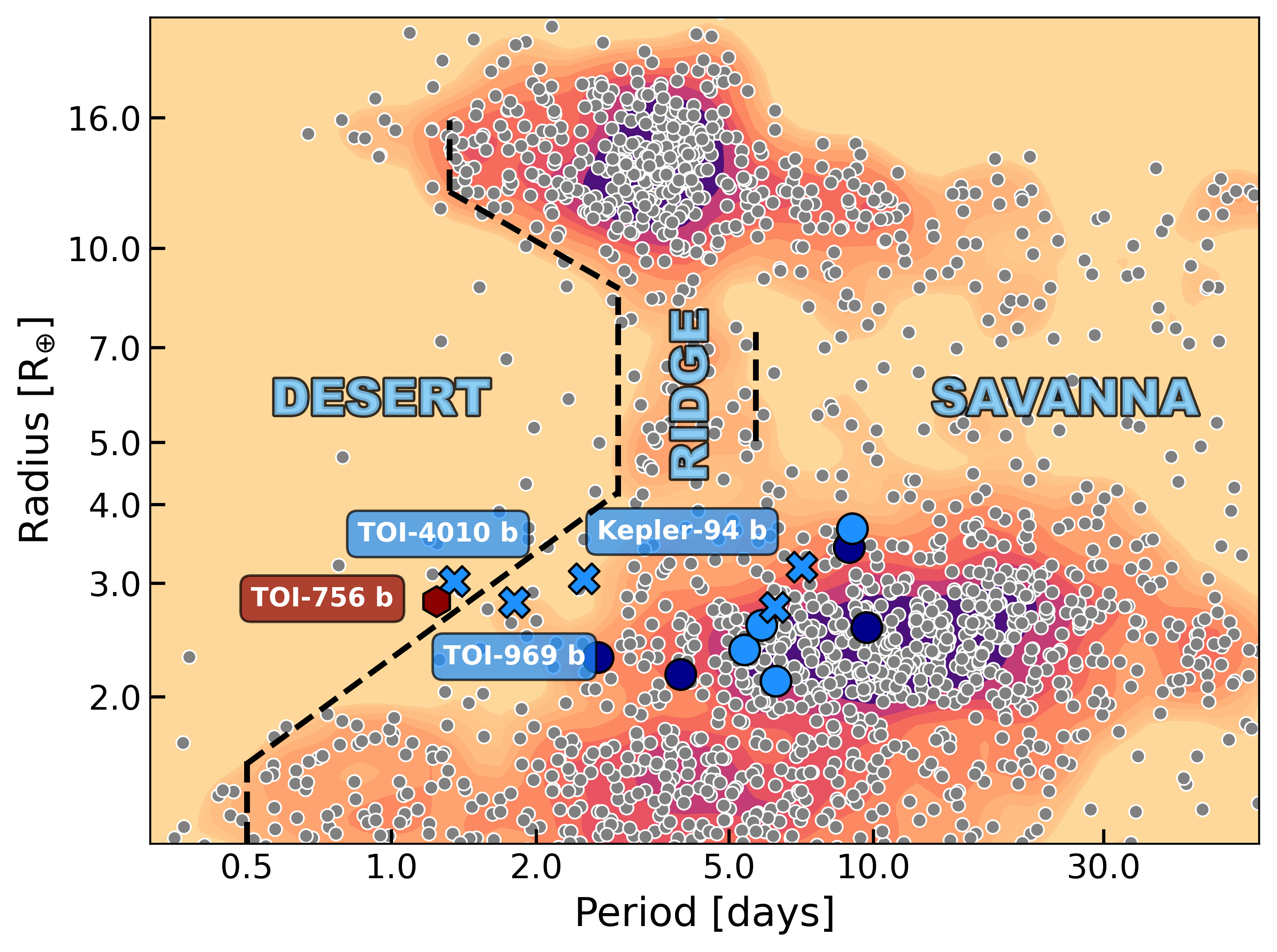}
  \caption{Planet radius as a function of orbital period for known exoplanets from NASA Exoplanet Archive with a radius precision below 8\%. We highlighted the Neptunian desert, ridge, and savanna regions from \citet{Castro2024}. The colorcode represents the observed density of planets. TOI-756~b is depicted as a dark red hexagon. Light blue symbols represent sub-Neptunes in systems hosting eccentric giant companions: crosses indicate systems with only giant planets, while circles correspond to systems containing both small and giant planets. In contrast, dark blue circles represent sub-Neptunes in systems with non-eccentric giant planets. The description of the selection is made in Sect.~\ref{sect:discussion_system}. This plot has been generated with \texttt{nep-des} (\url{https://github.com/castro-gzlz/nep-des/}).}
  \label{fig:nep_desert}
\end{figure}

Moreover, TOI-756~b lies at the very lower edge of the Neptune desert. We plotted its location together with the boundaries of the desert as defined in \citet{Castro2024} in Fig.~\ref{fig:nep_desert}. The main hy pothesis for the shaping of the lower edge of the Neptune desert is through hydrodynamical atmospheric escape, driven by intense stellar X-ray and extreme ultraviolet (XUV) irradiation \citep[e.g.,][]{Yelle2004,Tian2005,Owen2012,McDonald2019}. This process can strip the gaseous envelopes of close-in Neptune-sized planets, leaving behind smaller, denser remnants such as sub-Neptunes or bare rocky cores \citep{Lopez2013}. Therefore, TOI-756~b may have lost at least part of its gaseous envelope as a result of prolonged exposure to XUV irradiation from its host star. Notably, M dwarfs remain active for significantly longer periods than Sun-like stars \citep{Ribas2005}, extending the timescale over which atmospheric escape operate. Additionally, the possible non-zero eccentricity of the sub-Neptune, along with the eccentricity of TOI-756~c and the presence of a third body, may suggest dynamical activity, potentially involving high-eccentricity tidal migration (HEM). HEM, which includes mechanisms such as planet–planet scattering \citep[e.g.,][]{Gratia2017}, Kozai-Lidov migration \citep[e.g.,][]{Wu2003}, and secular chaos \citep[e.g.,][]{Wu2011}, can occur at any stage after disk dispersal, from early evolutionary phases to several billion years later. HEM typically leads to strongly misaligned orbits, erasing any memory of the system’s primordial configuration. In this scenario, a distant massive companion excites the eccentricity of the inner planet via gravitational perturbations, which is then followed by tidal circularization and inward migration due to energy dissipation induced by stellar tides \citep[e.g.,][]{Rasio1996}. This process can be investigated by measuring the spin–orbit alignment of the transiting planet using Rossiter–McLaughlin (RM) observations \citep{Rossiter1924,McLaughlin1924}. These dynamical processes are considered to be key factors in shaping the Neptune desert. Additionally, the boundaries of the Neptune desert may vary with the spectral type of the host star, and to date, there has been no comprehensive study of the Neptune desert around M dwarfs. Indeed, for a given orbital period, a planet orbiting an M dwarf receives intuitively less stellar flux than planets around other types of stars, which could affect the atmospheric escape kick off. Interestingly, TOI-756~b does not show the high density commonly found in planets within the Neptune desert, such as TOI-849~b \citep{Armstrong2020}. Its ability to retain an atmosphere despite strong irradiation could be explained by its orbit around a metal-rich star, since metal-rich atmospheres are thought to be more resistant to photoevaporative mass loss \citep{Owen2018, Wilson2022}.

Atmospheric characterization of planets located within the radius cliff and Neptune desert could help test theories regarding the origins of these demographic features. With a Transmission Spectroscopic Metric (TSM) of 63 \citep{Kempton2018}, TOI-756~b stands out as a promising target for future transmission spectroscopy studies, for instance with JWST \citep{JWST}.

\subsection{TOI-756 system: A unique system in exoplanet zoology}
\subsubsection{Population of transiting sub-Neptunes with an outer companion}\label{sect:discussion_system}

The TOI-756 system with its transiting sub-Neptune, its cold giant non-transiting companion, and an additional component, all orbiting an M dwarf in a wide binary system, is a very unique system in exoplanet zoology. We searched the NASA Exoplanet Archive\footnote{\url{https://exoplanetarchive.ipac.caltech.edu/}} for the multi-planetary systems with a transiting sub-Neptune ($2~\re < R_p < 4~\re$) orbiting with a period of less than 10 days and with a giant outer companion orbiting at more than 100 days detected by transit or radial velocity (or both) with $R_p > 4~\re$ or $M_p\sin(i) > 20~\me$. We plotted in Fig.~\ref{fig:nep_desert} the radii and orbital periods of the sub-Neptunes of these systems. We found 13 systems but none are orbiting an M dwarf.  TOI-756 is currently the only confirmed system with a transiting sub-Neptune and a cold giant orbiting an M dwarf. This remains true even if we remove all constraints on the periods of the inner and outer planets. An additional but unconfirmed system with this peculiar architecture has been identified: the K2-43 system. K2-43~c is a sub-Neptune \citep[R$_\mathrm{p}$ = 2.4~\re, P = 2.2 d ;][]{Hedges2019}, and more recently a single transit event with a depth corresponding to a Jupiter-sized planet has been detected in the TESS data (TOI-5523.01). 

This system adds up to the small sample of the recent study of \citet{Bryan2025}, investigating the stellar mass and metallicity trends for small planets with a gas giant companion. They found a higher gas giant frequency around metal-rich M dwarfs for both samples (with gas giant (GG) or with gas giant plus small planet (GG|SE)), but they find no significant difference in gas giant occurrence rate between P(GG) and P(GG|SE). While they find no significant correlation between small planets and outer gas giants around M dwarfs, previous work has found a significant positive correlation between these planet populations around more massive stars that are metal-rich: \citet{Bryan2024} and \citet{Chachan2023} hypothesized that this positive correlation should persist and may even strengthen for lower-mass stars. This follows the well known metallicity-giant planet correlation seen for FGK stars \citep[e.g.,][]{Sousa2021} and M dwarfs \citep[e.g.,][]{Neves2013}. We are offering an additional system around a metal-rich M dwarf to address a largely underexplored parameter space, aiding studies that investigate the correlations and occurrence rates of specific populations in relation to stellar parameters.

In addition to being a unique multi-planet system, TOI-756 is an M dwarf hosting a rare giant component. Planet of and above Jupiter's mass are remarkably rare around M dwarfs. Core-accretion theory predicts that giant planets should be less common around M dwarfs than around FGK-type stars, primarily due to the lower surface density of solids and longer formation timescales in protoplanetary disks around low-mass stars \citep[e.g.,][]{Laughlin2004,Ida2005}. This trend is supported by recent population synthesis models, which not only confirm the low occurrence rate of giant planets in such environments but also suggest it may drop to nearly zero for host stars with masses between 0.1 and 0.3~\msol \citep{Burn2021}. They generally form in the outer region of the disk beyond the ice line \citep{Alexander2012,Bitsch2015}, where there is more material for them to form, but we are biased against detecting them with the transit method as transit probability decreases at long orbital periods. This probability is even lower around M dwarfs since they are small stars. However, RV campaigns will certainly provide more of these outer companions to small transiting planets, but also confirm giant TESS candidates. The RV follow-up of TESS giant planet candidates is another one of the subprograms of SP2 of the NIRPS-GTO, thanks to the unique sensitivity of NIRPS in the infrared, which allows us to characterize such planets around host stars with $J<12$. The discovery of TOI-756~c together with the other discoveries of giants around M dwarfs with NIRPS (Frensch et al. in prep) will help to test the hypothesis of their formation and evolution.

Coming back to the identified systems similar to TOI-756, an interesting thing to note is that 11 cold giants (in 9 of these 13 systems, including TOI-756) have a detected non-zero eccentricity ($e > 0.1$). We highlighted these systems in light blue in Fig.~\ref{fig:nep_desert}. Systems represented by crosses consist solely of a sub-Neptune accompanied by one or more giant planets, whereas systems shown as circles include both small and giant planets in addition to the sub-Neptune. In addition, we emphasize three systems that share strong similarities with the TOI-756 system: TOI-4010 \citep{Kunimoto2023}, TOI-969 \citep{LilloBox2023}, and Kepler-94 \citep{Weiss2024}. All three systems consist of a sub-Neptune located near the lower boundary of the Neptune desert, accompanied by a giant planet with an orbital period exceeding 100 days and a non-zero eccentricity.
Regarding this class of systems, \citet{Bitsch2023} used N-body simulations that combine pebble and gas accretion with planetary migration. They found that systems hosting outer giant planets tend to produce more systems with predominantly a single inner planet and exhibit higher eccentricities for all planets, compared to simulations without outer giants. In addition, unstable systems (with high eccentricities) mostly host only one inner sub-Neptune (and for most systems, this inner planet is transiting). Additional observations of TOI-756 to precisely constrain the eccentricity of TOI-756~b could be a good test case of these results considering the large eccentricity of the planet c. Here again, \citet{Bitsch2023} predicted that systems with truly single close-in planets are more likely to host outer gas giants. Conversely, \citet{Schlecker2021} predicted that planetary systems around stars with high metallicity frequently contain warm and dynamically active giant planets that can disrupt inner planetary systems and then are less likely to harbor inner small planets. The RV follow up of transiting close-in planets by the NIRPS-GTO SP2 will help to test these predictions by planetary formation models.

\begin{figure}[t]
  \centering
    \includegraphics[width=0.48\textwidth]{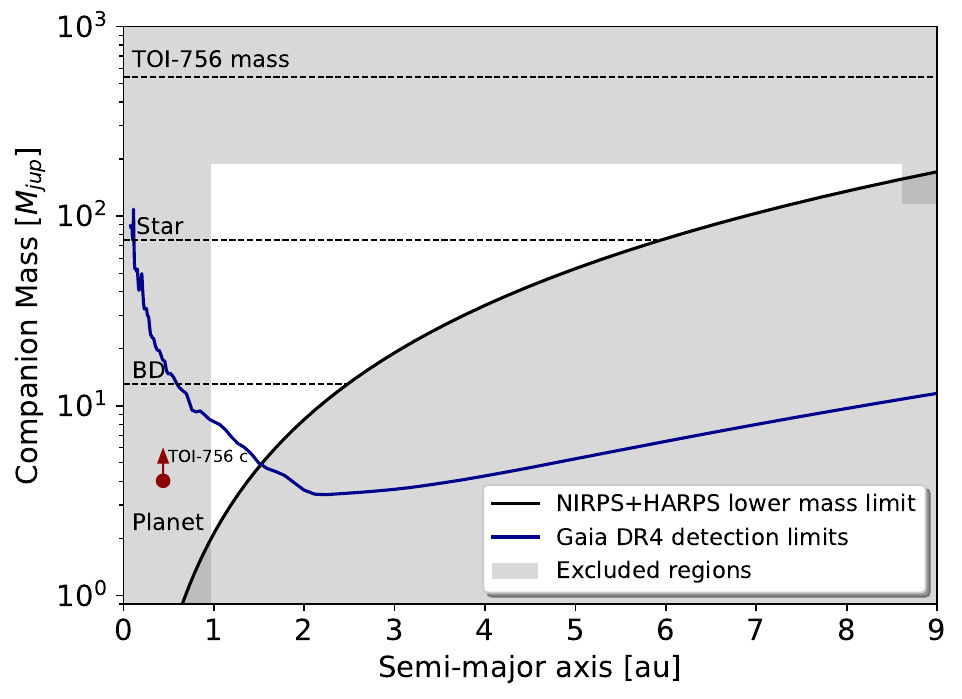}
  \caption{Limits on the companion mass of TOI-756 system as a function of semi-major axis. The lower limit, indicated by the solid black line, is calculated based on the RV linear trend. The excluded gray areas include this constraint of the RV linear trend, the timespan of observations of the RV (left rectangle), the limit from high-contrast imaging (upper right rectangle), and the absence of a double peak in the CCF (upper rectangle). We plotted the Gaia DR4 astrometric detection limits as a solid dark blue line for a star at 86 pc with a RUWE of 1.25 \citep{Wallace2025}. The dotted black lines are the different mass limit categories separating planetary, brown dwarf (BD), and stellar natures.}
  \label{fig:companion_mass_limits}
\end{figure}

\subsubsection{Constraints on the RV acceleration}

In addition to the sub-Neptune and the eccentric giant planet, NIRPS and HARPS have revealed an acceleration in the RV of TOI-756. To determine if the wide binary \hbox{WT~352} could, plausibly, be responsible for the acceleration, we used the following equation \citep{Torres1999}:

\begin{equation}\label{eq:mass_comp}
    M_{\text{comp}} = 5.34\times10^{-6} M_{\odot} \left( \frac{d}{\text{pc}} \frac{\rho}{\text{arcsec}} \right)^2 \left| \frac{\dot{v}}{\text{m s}^{-1} \text{ yr}^{-1}} \right| F(i, e, \omega, \phi)
\end{equation}

where $d$ is the distance to the system, $\rho$ is the projected separation of the companion on the sky, and $\dot{v}$  is the best-fit RV trend. $F(i, e, \omega, \phi)$ is an equation that depends on the unknown orbital parameters of the companion and has a minimum value of $\sqrt{27}/2$, which we use in our calculations here. We convert the projected separation on the sky to a minimum semi-major axis using the Gaia DR3 distance of TOI-756. With an acceleration of $145.6~\mathrm{m\,s}^{-1}\,\mathrm{yr}^{-1}$ and the separation of 11.09\,arcsec, we found a mass of 1857~\msol. We therefore conclude that the co-moving companion (a $\sim$ M3/4V star with $M_\star \sim 0.3~\msol$) cannot be responsible for the trend in this system. We again used Eq.~\ref{eq:mass_comp} to draw the black curve in Fig.~\ref{fig:companion_mass_limits} representing the lower mass limit permitted by the RV trend as a function of the semi-major axis. We see that we cannot exclude a planetary nature for this additional companion. In addition, we can exclude the left rectangle of the figure corresponding to our time-span of observations of 480 days. An additional constraint on the mass of this companion comes from high-resolution imaging of TOI-756, which rules out companions more than 5 magnitudes fainter (down to approximately an M5.5/6V star, around 0.11~\msol ) at a separation of 0.1 arcseconds ($\sim$ 8.6 au) (see Sect.~\ref{sect:high_res}). Moreover, the RV measurements do not show any evidence of a blended companion of a mass similar to the primary with no increased contrast, FWHM, or bisector deviations of the CCFs compared to a similar single star (e.g., GL514, the star used for the LBL template). We expect to detect a second peak in the CCF with a contrast up to 10 times smaller ($\sim$3--5\%) than that of the primary peak, corresponding to a companion with a magnitude difference of about 2.5. Given that TOI-756 is an M1 star, this sensitivity would allow us to detect a companion as late as M4.5--M5 ($\sim$0.16--0.18~\msol). As a result, we can exclude the presence of early M-type stellar companions as the source of the observed RV trend (see Fig.~\ref{fig:companion_mass_limits}).

Moreover, in Fig.~\ref{fig:companion_mass_limits}, we plotted the expected Gaia DR4 detection limits at the distance of TOI-756 and for a RUWE of 1.25 (RUWE$_\text{TOI-756}$ = 1.24) from \citet{Wallace2025}. We can see that Gaia will be capable of resolving the orbit and parameters of almost any object within the parameter space for this companion. In addition, we are still monitoring TOI-756 system, with NIRPS and HARPS to further constrain this additional component. The combination of radial velocity and astrometric data will be crucial for a precise characterization of the full system. In particular, it may enable us to resolve the orbits of both TOI-756~c and the additional companion, allowing us to measure their mutual inclination and gain insight into the formation, evolution, and dynamical history of this rare system orbiting a low-mass star.

\subsubsection{The binarity effect on formation and evolution}

The presence of the widely separated stellar companion at $\sim$11" raises important questions about its influence on the formation and dynamical evolution of planets within the TOI-756 system.

Binarity is known to truncate the circumstellar disk, shorten its lifetime, and reduce planet occurrence rates \citep{Cieza2009, Harris2012, Moe&Krater2021}. While the wide physical separation between the two stars suggests that the stellar companion had a limited direct impact on the protoplanetary disk of TOI-756, studies have shown that binary companions can still influence planet formation and evolution even at separations up to 1000 au, potentially hindering the formation of massive planetary cores \citep{Sullivan2023}.

On the other hand, recent studies \citep{Sullivan2024} show that the radius gap in wide binaries (separation $>$ 300 au) appears to be shifted toward smaller radii. This suggests that the presence of a stellar companion can influence disk conditions and, consequently, the formation and evolution of planets. 

However, despite expectations that a stellar companion at this distance could have a noticeable gravitational effect, \hbox{WT~352} is significantly less massive (M$_\star \sim 0.3$~\msol) than the primary and is located at a separation that could exceed 1000 au. In this case, the companion did not inhibit the formation of giant planets or sub-stellar objects around the primary star. This suggests that the gravitational effect of a companion depends not only on its separation, but also on its mass relative to the primary star.
Furthermore, while upcoming Gaia DR4 data will refine astrometric measurements, it is unlikely to provide significant new constraints on the orbital parameters of the binary, given that its expected orbital period is on the order of $\sim$40,000 years — too long for measurable motion within Gaia’s observational timeline \citep{ElBadry2024}. 

Recently, \citet{Behmard2022,Chirstian2022} investigated the potential alignment between the orbital planes of planetary systems and their visual binary companions. The TOI-756 system, along with its wide binary companion WT~352, is included in their sample. \citet{Chirstian2022} reported a significant misalignment in this system, with a mutual inclination of $i = 118^{+34}_{-17}$ degrees (5th to 95th percentile range). While their analysis reveals an excess of aligned systems among binaries with separations less than 700 au, they find that the distribution of mutual inclinations becomes consistent with uniformity for wider binaries ($a > 700$ au). This could account for the observed misalignment in TOI-756, given its projected binary separation of approximately 955 au.

\section{Conclusions}\label{sect:ccl}

We present the "Sub-Neptunes" subprogram of the NIRPS-GTO SP2 program, which aims to improve our understanding of the diversity in composition and internal structure of small planets around M dwarfs. By enabling the study of targets hosting TESS sub-Neptune candidates with $V<15.4$, NIRPS (the red arm of HARPS) expands the reach of RV follow-up beyond the traditional limits of optical spectrographs installed on 4-meter-class telescopes.

We report the first results of the RV follow-up program of the NIRPS-GTO, presenting the characterization of a two-planet system orbiting the M-dwarf TOI-756, which is the primary component of a wide binary system with the star WT~352. TOI-756 b was initially identified by TESS and subsequently confirmed with the ground-based photometry facilities LCO-CTIO and ExTrA, as well with RV measurements obtained with NIRPS and HARPS, which enabled the determination of its mass. Additionally, NIRPS and HARPS allowed the identification of a second non-transiting planet in the system TOI-756 c, as well as an supplementary RV acceleration hinting at an extra third component in the system that could be planetary as well.
TOI-756 b is a sub-Neptune with a radius of 2.81~\re and a mass of 9.8~\me orbiting with a period of 1.24 days around its star. TOI-756 c is a cold eccentric giant planet orbiting at 149 days with a minimum mass of 4.05~\mjup and an eccentricity of 0.45. 

TOI-756 b, with a density of 2.42 g.cm$^{-3}$, is in line with the recently identified trend of low-density sub-Neptunes around M dwarfs compared to FGK dwarfs. In addition, this peculiar target lies in the radius cliff of M-dwarf planets and in the Neptune desert. TOI-756 b most likely requires a certain amount of hydrogen and helium in its atmosphere to account for its observed density, either as a pure H/He envelope or as a mixture of supercritical H$_2$O and H/He. 

The TOI-756 system is particularly unique, as it is the only known confirmed system hosting both a transiting sub-Neptune and an outer giant planet around an M dwarf. This makes it a valuable case for comparison with planet formation and evolution models, as well as for studying correlations between planetary populations and stellar parameters such as stellar mass and metallicity. In addition, TOI-756~c enhances the small population of giant planets around M dwarfs, a population whose formation mechanisms are still not fully understood. Identifying more of these planets is vital for constraining our models of their formation and evolutionary processes, providing deeper insights into the pathways that shape such systems. The astrometric measurements from the Gaia DR4 release will be key to combine with the RVs to further characterize this unique system. 

TOI-756~b is also a promising candidate for future atmospheric characterization through transmission spectroscopy with JWST, which could help confirm or rule out the presence of a primordial H/He-dominated atmosphere. It also offers an opportunity to test hypotheses regarding the radius cliff and the Neptune desert population and to constrain the formation and evolution models of small planets orbiting alongside an eccentric outer companion.

In this study, we demonstrate the capabilities of the unique NIRPS and HARPS combination to obtain precise RVs of M dwarfs, enabling the confirmation and characterization of candidates detected by current photometric surveys such as TESS, as well as upcoming missions like PLATO \citep{Rauer2014}.

\begin{acknowledgements}
We thank the anonymous referee for their valuable comments, which helped improve the manuscript.
This work has been carried out within the framework of the NCCR PlanetS supported by the Swiss National Science Foundation under grants 51NF40\_182901 and 51NF40\_205606.\\
NJC, ÉA, AL, RD, FBa, BB, LMa, RA, LB, AB, CC, AD-B, LD, PLam, OL, LMo, JS-A, PV, TV \& JPW acknowledge the financial support of the FRQ-NT through the Centre de recherche en astrophysique du Québec as well as the support from the Trottier Family Foundation and the Trottier Institute for Research on Exoplanets.\\
ÉA, RD, FBa, LMa, TA, J-SM, MO, JS-A \& PV acknowledges support from Canada Foundation for Innovation (CFI) program, the Université de Montréal and Université Laval, the Canada Economic Development (CED) program and the Ministere of Economy, Innovation and Energy (MEIE).\\
AL acknowledges support from the Fonds de recherche du Québec (FRQ) - Secteur Nature et technologies under file \#349961.\\
The Board of Observational and Instrumental Astronomy (NAOS) at the Federal University of Rio Grande do Norte's research activities are supported by continuous grants from the Brazilian funding agency CNPq. This study was partially funded by the Coordena\c{c}ão de Aperfei\c{c}oamento de Pessoal de N\'ivel Superior—Brasil (CAPES) — Finance Code 001 and the CAPES-Print program.\\
0\\
SCB, ED-M, NCS, EC, ARCS \& JGd acknowledge the support from FCT - Funda\c{c}ão para a Ci\^encia e a Tecnologia through national funds by these grants: UIDB/04434/2020, UIDP/04434/2020. Co-funded by the European Union (ERC, FIERCE, 101052347). Views and opinions expressed are however those of the author(s) only and do not necessarily reflect those of the European Union or the European Research Council. Neither the European Union nor the granting authority can be held responsible for them.\\
SCB acknowledges the support from Funda\c{c}ão para a Ci\^encia e Tecnologia (FCT) in the form of a work contract through the Scientific Employment Incentive program with reference 2023.06687.CEECIND and DOI \href{https://doi.org/10.54499/2023.06687.CEECIND/CP2839/CT0002}{10.54499/2023.06687.CEECIND/CP2839/CT0002.}\\
XB, XDe, ACar, TF \& VY acknowledge funding from the French ANR under contract number ANR\-18\-CE31\-0019 (SPlaSH), and the French National Research Agency in the framework of the Investissements d'Avenir program (ANR-15-IDEX-02), through the funding of the ``Origin of Life" project of the Grenoble-Alpes University.\\
BLCM \& AMM acknowledge CAPES postdoctoral fellowships.\\
BLCM acknowledges CNPq research fellowships (Grant No. 305804/2022-7).\\
NBC acknowledges support from an NSERC Discovery Grant, a Canada Research Chair, and an Arthur B. McDonald Fellowship, and thanks the Trottier Space Institute for its financial support and dynamic intellectual environment.\\
DBF acknowledges financial support from the Brazilian agency CNPq-PQ (Grant No. 305566/2021-0). Continuous grants from the Brazilian agency CNPq support the STELLAR TEAM of the Federal University of Ceara's research activities.\\
JRM acknowledges CNPq research fellowships (Grant No. 308928/2019-9).\\
ED-M  further acknowledges the support from FCT through Stimulus FCT contract 2021.01294.CEECIND. ED-M  acknowledges the support by the Ram\'on y Cajal contract RyC2022-035854-I funded by MICIU/AEI/10.13039/501100011033 and by ESF+.\\
XDu acknowledges the support from the European Research Council (ERC) under the European Union’s Horizon 2020 research and innovation programme (grant agreement SCORE No 851555) and from the Swiss National Science Foundation under the grant SPECTRE (No 200021\_215200).\\
DE acknowledge support from the Swiss National Science Foundation for project 200021\_200726. The authors acknowledge the financial support of the SNSF.\\
JIGH, RR, ASM, FGT, NN, VMP, JLR \& AKS  acknowledge financial support from the Spanish Ministry of Science, Innovation and Universities (MICIU) projects PID2020-117493GB-I00 and PID2023-149982NB-I00.\\
ICL acknowledges CNPq research fellowships (Grant No. 313103/2022-4).\\
CMo acknowledges the funding from the Swiss National Science Foundation under grant 200021\_204847 “PlanetsInTime”.\\
KAM acknowledges support from the Swiss National Science Foundation (SNSF) under the Postdoc Mobility grant P500PT\_230225.\\
RA acknowledges the Swiss National Science Foundation (SNSF) support under the Post-Doc Mobility grant P500PT\_222212 and the support of the Institut Trottier de Recherche sur les Exoplan\`etes (IREx).\\
We acknowledge funding from the European Research Council under the ERC Grant Agreement n. 337591-ExTrA.\\
LB acknowledges the support of the Natural Sciences and Engineering Research Council of Canada (NSERC).\\
This project has received funding from the European Research Council (ERC) under the European Union's Horizon 2020 research and innovation programme (project {\sc Spice Dune}, grant agreement No 947634). This material reflects only the authors' views and the Commission is not liable for any use that may be made of the information contained therein.\\
ARCS acknowledges the support from Funda\c{c}ao para a Ci\^encia e a Tecnologia (FCT) through the fellowship 2021.07856.BD.\\
LD acknowledges the support of the Natural Sciences and Engineering Research Council of Canada (NSERC) and from the Fonds de recherche du Québec (FRQ) - Secteur Nature et technologies.\\
FG acknowledges support from the Fonds de recherche du Québec (FRQ) - Secteur Nature et technologies under file \#350366.\\
H.J.H. acknowledges funding from eSSENCE (grant number eSSENCE@LU 9:3), the Swedish National Research Council (project number 2023-05307), The Crafoord foundation and the Royal Physiographic Society of Lund, through The Fund of the Walter Gyllenberg Foundation.\\
LMo  acknowledges the support of the Natural Sciences and Engineering Research Council of Canada (NSERC), [funding reference number 589653].\\
NN acknowledges financial support by Light Bridges S.L, Las Palmas de Gran Canaria.\\
NN acknowledges funding from Light Bridges for the Doctoral Thesis "Habitable Earth-like planets with ESPRESSO and NIRPS", in cooperation with the Instituto de Astrof\'isica de Canarias, and the use of Indefeasible Computer Rights (ICR) being commissioned at the ASTRO POC project in the Island of Tenerife, Canary Islands (Spain). The ICR-ASTRONOMY used for his research was provided by Light Bridges in cooperation with Hewlett Packard Enterprise (HPE).\\
CPi acknowledges support from the NSERC Vanier scholarship, and the Trottier Family Foundation. CPi  also acknowledges support from the E. Margaret Burbidge Prize Postdoctoral Fellowship from the Brinson Foundation.\\
AKS acknowledges financial support from La Caixa Foundation (ID 100010434) under the grant LCF/BQ/DI23/11990071.\\
TV acknowledges support from the Fonds de recherche du Québec (FRQ) - Secteur Nature et technologies under file \#320056.
KaC acknowledges support from the TESS mission via subaward s3449 from MIT. \\
Funding for the TESS mission is provided by NASA's Science Mission Directorate. We acknowledge the use of public TESS data from pipelines at the TESS Science Office and at the TESS Science Processing Operations Center. Resources supporting this work were provided by the NASA High-End Computing (HEC) Program through the NASA Advanced Supercomputing (NAS) Division at Ames Research Center for the production of the SPOC data products. This research has made use of the Exoplanet Follow-up Observation Program (ExoFOP; DOI: 10.26134/ExoFOP5) website, which is operated by the California Institute of Technology, under contract with the National Aeronautics and Space Administration under the Exoplanet Exploration Program. This paper includes data collected by the TESS mission that are publicly available from the Mikulski Archive for Space Telescopes (MAST). This work makes use of observations from the LCOGT network. Part of the LCOGT telescope time was granted by NOIRLab through the Mid-Scale Innovations Program (MSIP). MSIP is funded by NSF.
This research has made use of the NASA Exoplanet Archive, which is operated by the California Institute of Technology, under contract with the National Aeronautics and Space Administration under the Exoplanet Exploration Program.
\end{acknowledgements}

\bibliographystyle{aa} 

\bibliography{bib}

\begin{thebibliography}{180}
\expandafter\ifx\csname natexlab\endcsname\relax\def\natexlab#1{#1}\fi

\bibitem[{{Aguichine} {et~al.}(2025){Aguichine}, {Batalha}, {Fortney}, {Nettelmann}, {Owen}, \& {Kempton}}]{Aguichine2024}
{Aguichine}, A., {Batalha}, N., {Fortney}, J.~J., {et~al.} 2025, \apj, 988, 186

\bibitem[{{Aguichine} {et~al.}(2021){Aguichine}, {Mousis}, {Deleuil}, \& {Marcq}}]{Aguichine2021}
{Aguichine}, A., {Mousis}, O., {Deleuil}, M., \& {Marcq}, E. 2021, \apj, 914, 84

\bibitem[{{Alexander} \& {Pascucci}(2012)}]{Alexander2012}
{Alexander}, R.~D. \& {Pascucci}, I. 2012, \mnras, 422, L82

\bibitem[{{Alibert} \& {Benz}(2017)}]{Alibert2017}
{Alibert}, Y. \& {Benz}, W. 2017, \aap, 598, L5

\bibitem[{{Allard} {et~al.}(2012){Allard}, {Homeier}, \& {Freytag}}]{Allard2012}
{Allard}, F., {Homeier}, D., \& {Freytag}, B. 2012, Philosophical Transactions of the Royal Society of London Series A, 370, 2765

\bibitem[{{Allart} {et~al.}(2022){Allart}, {Lovis}, {Faria}, {Dumusque}, {Sosnowska}, {Figueira}, {Silva}, {Mehner}, {Pepe}, {Cristiani}, {Rebolo}, {Santos}, {Adibekyan}, {Cupani}, {Di Marcantonio}, {D'Odorico}, {Gonz{\'a}lez Hern{\'a}ndez}, {Martins}, {Milakovi{\'c}}, {Nunes}, {Sozzetti}, {Su{\'a}rez Mascare{\~n}o}, {Tabernero}, \& {Zapatero Osorio}}]{Allart2022}
{Allart}, R., {Lovis}, C., {Faria}, J., {et~al.} 2022, \aap, 666, A196

\bibitem[{{Aller} {et~al.}(2020){Aller}, {Lillo-Box}, {Jones}, {Miranda}, \& {Barcel{\'o} Forteza}}]{Aller2020}
{Aller}, A., {Lillo-Box}, J., {Jones}, D., {Miranda}, L.~F., \& {Barcel{\'o} Forteza}, S. 2020, \aap, 635, A128

\bibitem[{{Ambikasaran} {et~al.}(2015){Ambikasaran}, {Foreman-Mackey}, {Greengard}, {Hogg}, \& {O'Neil}}]{Ambikasaran2015_george}
{Ambikasaran}, S., {Foreman-Mackey}, D., {Greengard}, L., {Hogg}, D.~W., \& {O'Neil}, M. 2015, IEEE Transactions on Pattern Analysis and Machine Intelligence, 38, 252

\bibitem[{{Anglada-Escud{\'e}} \& {Butler}(2012)}]{Anglada2012}
{Anglada-Escud{\'e}}, G. \& {Butler}, R.~P. 2012, \apjs, 200, 15

\bibitem[{{Antoniadis-Karnavas} {et~al.}(2024){Antoniadis-Karnavas}, {Sousa}, {Delgado-Mena}, {Santos}, \& {Andreasen}}]{Antoniadis24}
{Antoniadis-Karnavas}, A., {Sousa}, S.~G., {Delgado-Mena}, E., {Santos}, N.~C., \& {Andreasen}, D.~T. 2024, \aap, 690, A58

\bibitem[{{Antoniadis-Karnavas} {et~al.}(2020){Antoniadis-Karnavas}, {Sousa}, {Delgado-Mena}, {Santos}, {Teixeira}, \& {Neves}}]{Antoniadis20}
{Antoniadis-Karnavas}, A., {Sousa}, S.~G., {Delgado-Mena}, E., {et~al.} 2020, \aap, 636, A9

\bibitem[{{Armstrong} {et~al.}(2020){Armstrong}, {Lopez}, {Adibekyan}, {Booth}, {Bryant}, {Collins}, {Deleuil}, {Emsenhuber}, {Huang}, {King}, {Lillo-Box}, {Lissauer}, {Matthews}, {Mousis}, {Nielsen}, {Osborn}, {Otegi}, {Santos}, {Sousa}, {Stassun}, {Veras}, {Ziegler}, {Acton}, {Almenara}, {Anderson}, {Barrado}, {Barros}, {Bayliss}, {Belardi}, {Bouchy}, {Brice{\~n}o}, {Brogi}, {Brown}, {Burleigh}, {Casewell}, {Chaushev}, {Ciardi}, {Collins}, {Col{\'o}n}, {Cooke}, {Crossfield}, {D{\'\i}az}, {Delgado Mena}, {Demangeon}, {Dorn}, {Dumusque}, {Eigm{\"u}ller}, {Fausnaugh}, {Figueira}, {Gan}, {Gandhi}, {Gill}, {Gonzales}, {Goad}, {G{\"u}nther}, {Helled}, {Hojjatpanah}, {Howell}, {Jackman}, {Jenkins}, {Jenkins}, {Jensen}, {Kennedy}, {Latham}, {Law}, {Lendl}, {Lozovsky}, {Mann}, {Moyano}, {McCormac}, {Meru}, {Mordasini}, {Osborn}, {Pollacco}, {Queloz}, {Raynard}, {Ricker}, {Rowden}, {Santerne}, {Schlieder}, {Seager}, {Sha}, {Tan}, {Tilbrook}, {Ting}, {Udry}, {Vanderspek}, {Watson}, {West}, {Wilson}, {Winn},
  {Wheatley}, {Villasenor}, {Vines}, \& {Zhan}}]{Armstrong2020}
{Armstrong}, D.~J., {Lopez}, T.~A., {Adibekyan}, V., {et~al.} 2020, \nat, 583, 39

\bibitem[{{Artigau} {et~al.}(2024){Artigau}, {Cadieux}, {Cook}, {Doyon}, {Dauplaise}, {Arnold}, {Cadieux}, {Donati}, {Cristofari}, {Delfosse}, {Fouqu{\'e}}, {Moutou}, {Larue}, \& {Allart}}]{Artigau2024}
{Artigau}, {\'E}., {Cadieux}, C., {Cook}, N.~J., {et~al.} 2024, \aj, 168, 252

\bibitem[{{Artigau} {et~al.}(2022){Artigau}, {Cadieux}, {Cook}, {Doyon}, {Vandal}, {Donati}, {Moutou}, {Delfosse}, {Fouqu{\'e}}, {Martioli}, {Bouchy}, {Parsons}, {Carmona}, {Dumusque}, {Astudillo-Defru}, {Bonfils}, \& {Mignon}}]{Artigau2022}
{Artigau}, {\'E}., {Cadieux}, C., {Cook}, N.~J., {et~al.} 2022, \aj, 164, 84

\bibitem[{{Astudillo-Defru} {et~al.}(2017){Astudillo-Defru}, {D{\'\i}az}, {Bonfils}, {Almenara}, {Delisle}, {Bouchy}, {Delfosse}, {Forveille}, {Lovis}, {Mayor}, {Murgas}, {Pepe}, {Santos}, {S{\'e}gransan}, {Udry}, \& {W{\"u}nsche}}]{Astudillo2017}
{Astudillo-Defru}, N., {D{\'\i}az}, R.~F., {Bonfils}, X., {et~al.} 2017, \aap, 605, L11

\bibitem[{{Baraffe} {et~al.}(1998){Baraffe}, {Chabrier}, {Allard}, \& {Hauschildt}}]{Baraffe1998}
{Baraffe}, I., {Chabrier}, G., {Allard}, F., \& {Hauschildt}, P.~H. 1998, \aap, 337, 403

\bibitem[{{Baraffe} {et~al.}(2015){Baraffe}, {Homeier}, {Allard}, \& {Chabrier}}]{Baraffe2015}
{Baraffe}, I., {Homeier}, D., {Allard}, F., \& {Chabrier}, G. 2015, \aap, 577, A42

\bibitem[{{Batalha} {et~al.}(2019){Batalha}, {Lewis}, {Fortney}, {Batalha}, {Kempton}, {Lewis}, \& {Line}}]{Batalha2019}
{Batalha}, N.~E., {Lewis}, T., {Fortney}, J.~J., {et~al.} 2019, \apjl, 885, L25

\bibitem[{{Batalha} {et~al.}(2013){Batalha}, {Rowe}, {Bryson}, {Barclay}, {Burke}, {Caldwell}, {Christiansen}, {Mullally}, {Thompson}, {Brown}, {Dupree}, {Fabrycky}, {Ford}, {Fortney}, {Gilliland}, {Isaacson}, {Latham}, {Marcy}, {Quinn}, {Ragozzine}, {Shporer}, {Borucki}, {Ciardi}, {Gautier}, {Haas}, {Jenkins}, {Koch}, {Lissauer}, {Rapin}, {Basri}, {Boss}, {Buchhave}, {Carter}, {Charbonneau}, {Christensen-Dalsgaard}, {Clarke}, {Cochran}, {Demory}, {Desert}, {Devore}, {Doyle}, {Esquerdo}, {Everett}, {Fressin}, {Geary}, {Girouard}, {Gould}, {Hall}, {Holman}, {Howard}, {Howell}, {Ibrahim}, {Kinemuchi}, {Kjeldsen}, {Klaus}, {Li}, {Lucas}, {Meibom}, {Morris}, {Pr{\v{s}}a}, {Quintana}, {Sanderfer}, {Sasselov}, {Seader}, {Smith}, {Steffen}, {Still}, {Stumpe}, {Tarter}, {Tenenbaum}, {Torres}, {Twicken}, {Uddin}, {Van Cleve}, {Walkowicz}, \& {Welsh}}]{Batalha2013}
{Batalha}, N.~M., {Rowe}, J.~F., {Bryson}, S.~T., {et~al.} 2013, \apjs, 204, 24

\bibitem[{{Bayo} {et~al.}(2008){Bayo}, {Rodrigo}, {Barrado Y Navascu{\'e}s}, {Solano}, {Guti{\'e}rrez}, {Morales-Calder{\'o}n}, \& {Allard}}]{Bayo2008}
{Bayo}, A., {Rodrigo}, C., {Barrado Y Navascu{\'e}s}, D., {et~al.} 2008, \aap, 492, 277

\bibitem[{{Behmard} {et~al.}(2022){Behmard}, {Dai}, \& {Howard}}]{Behmard2022}
{Behmard}, A., {Dai}, F., \& {Howard}, A.~W. 2022, \aj, 163, 160

\bibitem[{{Benneke} {et~al.}(2024){Benneke}, {Roy}, {Coulombe}, {Radica}, {Piaulet}, {Ahrer}, {Pierrehumbert}, {Krissansen-Totton}, {Schlichting}, {Hu}, {Yang}, {Christie}, {Thorngren}, {Young}, {Pelletier}, {Knutson}, {Miguel}, {Evans-Soma}, {Dorn}, {Gagnebin}, {Fortney}, {Komacek}, {MacDonald}, {Raul}, {Cloutier}, {Acuna}, {Lafreni{\`e}re}, {Cadieux}, {Doyon}, {Welbanks}, \& {Allart}}]{Benneke2024}
{Benneke}, B., {Roy}, P.-A., {Coulombe}, L.-P., {et~al.} 2024, arXiv e-prints, arXiv:2403.03325

\bibitem[{{Bensby} {et~al.}(2014){Bensby}, {Feltzing}, \& {Oey}}]{Bensby2014}
{Bensby}, T., {Feltzing}, S., \& {Oey}, M.~S. 2014, \aap, 562, A71

\bibitem[{{Bertaux} {et~al.}(2014){Bertaux}, {Lallement}, {Ferron}, {Boonne}, \& {Bodichon}}]{Bertaux2014}
{Bertaux}, J.~L., {Lallement}, R., {Ferron}, S., {Boonne}, C., \& {Bodichon}, R. 2014, \aap, 564, A46

\bibitem[{{Bitsch} \& {Izidoro}(2023)}]{Bitsch2023}
{Bitsch}, B. \& {Izidoro}, A. 2023, \aap, 674, A178

\bibitem[{{Bitsch} {et~al.}(2015){Bitsch}, {Lambrechts}, \& {Johansen}}]{Bitsch2015}
{Bitsch}, B., {Lambrechts}, M., \& {Johansen}, A. 2015, \aap, 582, A112

\bibitem[{{Bitsch} {et~al.}(2021){Bitsch}, {Raymond}, {Buchhave}, {Bello-Arufe}, {Rathcke}, \& {Schneider}}]{Bitsch2021}
{Bitsch}, B., {Raymond}, S.~N., {Buchhave}, L.~A., {et~al.} 2021, \aap, 649, L5

\bibitem[{{Bonfils} {et~al.}(2015){Bonfils}, {Almenara}, {Jocou}, {Wunsche}, {Kern}, {Delboulb{\'e}}, {Delfosse}, {Feautrier}, {Forveille}, {Gluck}, {Lafrasse}, {Magnard}, {Maurel}, {Moulin}, {Murgas}, {Rabou}, {Rochat}, {Roux}, \& {Stadler}}]{Bonfils2015}
{Bonfils}, X., {Almenara}, J.~M., {Jocou}, L., {et~al.} 2015, in Society of Photo-Optical Instrumentation Engineers (SPIE) Conference Series, Vol. 9605, Techniques and Instrumentation for Detection of Exoplanets VII, ed. S.~{Shaklan}, 96051L

\bibitem[{{Bonfils} {et~al.}(2013){Bonfils}, {Delfosse}, {Udry}, {Forveille}, {Mayor}, {Perrier}, {Bouchy}, {Gillon}, {Lovis}, {Pepe}, {Queloz}, {Santos}, {S{\'e}gransan}, \& {Bertaux}}]{Bonfils2013}
{Bonfils}, X., {Delfosse}, X., {Udry}, S., {et~al.} 2013, \aap, 549, A109

\bibitem[{{Borucki} {et~al.}(2010){Borucki}, {Koch}, {Basri}, {Batalha}, {Brown}, {Caldwell}, {Caldwell}, {Christensen-Dalsgaard}, {Cochran}, {DeVore}, {Dunham}, {Dupree}, {Gautier}, {Geary}, {Gilliland}, {Gould}, {Howell}, {Jenkins}, {Kondo}, {Latham}, {Marcy}, {Meibom}, {Kjeldsen}, {Lissauer}, {Monet}, {Morrison}, {Sasselov}, {Tarter}, {Boss}, {Brownlee}, {Owen}, {Buzasi}, {Charbonneau}, {Doyle}, {Fortney}, {Ford}, {Holman}, {Seager}, {Steffen}, {Welsh}, {Rowe}, {Anderson}, {Buchhave}, {Ciardi}, {Walkowicz}, {Sherry}, {Horch}, {Isaacson}, {Everett}, {Fischer}, {Torres}, {Johnson}, {Endl}, {MacQueen}, {Bryson}, {Dotson}, {Haas}, {Kolodziejczak}, {Van Cleve}, {Chandrasekaran}, {Twicken}, {Quintana}, {Clarke}, {Allen}, {Li}, {Wu}, {Tenenbaum}, {Verner}, {Bruhweiler}, {Barnes}, \& {Prsa}}]{Borucki2010}
{Borucki}, W.~J., {Koch}, D., {Basri}, G., {et~al.} 2010, Science, 327, 977

\bibitem[{{Borucki} {et~al.}(2011){Borucki}, {Koch}, {Basri}, {Batalha}, {Brown}, {Bryson}, {Caldwell}, {Christensen-Dalsgaard}, {Cochran}, {DeVore}, {Dunham}, {Gautier}, {Geary}, {Gilliland}, {Gould}, {Howell}, {Jenkins}, {Latham}, {Lissauer}, {Marcy}, {Rowe}, {Sasselov}, {Boss}, {Charbonneau}, {Ciardi}, {Doyle}, {Dupree}, {Ford}, {Fortney}, {Holman}, {Seager}, {Steffen}, {Tarter}, {Welsh}, {Allen}, {Buchhave}, {Christiansen}, {Clarke}, {Das}, {D{\'e}sert}, {Endl}, {Fabrycky}, {Fressin}, {Haas}, {Horch}, {Howard}, {Isaacson}, {Kjeldsen}, {Kolodziejczak}, {Kulesa}, {Li}, {Lucas}, {Machalek}, {McCarthy}, {MacQueen}, {Meibom}, {Miquel}, {Prsa}, {Quinn}, {Quintana}, {Ragozzine}, {Sherry}, {Shporer}, {Tenenbaum}, {Torres}, {Twicken}, {Van Cleve}, {Walkowicz}, {Witteborn}, \& {Still}}]{Borucki2011}
{Borucki}, W.~J., {Koch}, D.~G., {Basri}, G., {et~al.} 2011, \apj, 736, 19

\bibitem[{{Bouchy} {et~al.}(2017){Bouchy}, {Doyon}, {Artigau}, {Melo}, {Hernandez}, {Wildi}, {Delfosse}, {Lovis}, {Figueira}, {Canto Martins}, {Gonz{\'a}lez Hern{\'a}ndez}, {Thibault}, {Reshetov}, {Pepe}, {Santos}, {de Medeiros}, {Rebolo}, {Abreu}, {Adibekyan}, {Bandy}, {Benz}, {Blind}, {Bohlender}, {Boisse}, {Bovay}, {Broeg}, {Brousseau}, {Cabral}, {Chazelas}, {Cloutier}, {Coelho}, {Conod}, {Cumming}, {Delabre}, {Genolet}, {Hagelberg}, {Jayawardhana}, {K{\"a}ufl}, {Lafreni{\`e}re}, {de Castro Le{\~a}o}, {Malo}, {de Medeiros Martins}, {Matthews}, {Metchev}, {Oshagh}, {Ouellet}, {Parro}, {Rasilla Pi{\~n}eiro}, {Santos}, {Sarajlic}, {Segovia}, {Sordet}, {Udry}, {Valencia}, {Vall{\'e}e}, {Venn}, {Wade}, \& {Saddlemyer}}]{Bouchy2017}
{Bouchy}, F., {Doyon}, R., {Artigau}, {\'E}., {et~al.} 2017, The Messenger, 169, 21

\bibitem[{{Bouchy} {et~al.}(2025){Bouchy}, {Doyon}, {Pepe}, {Melo}, \& {Artigau}}]{Bouchy2025}
{Bouchy}, F., {Doyon}, R., {Pepe}, F., {Melo}, C., \& {Artigau}, {\'E}. 2025, \aap

\bibitem[{{Brahm} {et~al.}(2018){Brahm}, {Espinoza}, {Jord{\'a}n}, {Rojas}, {Sarkis}, {D{\'\i}az}, {Rabus}, {Drass}, {Lachaume}, {Soto}, {Jenkins}, {Jones}, {Henning}, {Pantoja}, \& {Vu{\v{c}}kovi{\'c}}}]{Brahm2018}
{Brahm}, R., {Espinoza}, N., {Jord{\'a}n}, A., {et~al.} 2018, \mnras, 477, 2572

\bibitem[{{Brown} {et~al.}(2013){Brown}, {Baliber}, {Bianco}, {Bowman}, {Burleson}, {Conway}, {Crellin}, {Depagne}, {De Vera}, {Dilday}, {Dragomir}, {Dubberley}, {Eastman}, {Elphick}, {Falarski}, {Foale}, {Ford}, {Fulton}, {Garza}, {Gomez}, {Graham}, {Greene}, {Haldeman}, {Hawkins}, {Haworth}, {Haynes}, {Hidas}, {Hjelstrom}, {Howell}, {Hygelund}, {Lister}, {Lobdill}, {Martinez}, {Mullins}, {Norbury}, {Parrent}, {Paulson}, {Petry}, {Pickles}, {Posner}, {Rosing}, {Ross}, {Sand}, {Saunders}, {Shobbrook}, {Shporer}, {Street}, {Thomas}, {Tsapras}, {Tufts}, {Valenti}, {Vander Horst}, {Walker}, {White}, \& {Willis}}]{Brown2013}
{Brown}, T.~M., {Baliber}, N., {Bianco}, F.~B., {et~al.} 2013, \pasp, 125, 1031

\bibitem[{{Brugger} {et~al.}(2017){Brugger}, {Mousis}, {Deleuil}, \& {Deschamps}}]{Brugger2017}
{Brugger}, B., {Mousis}, O., {Deleuil}, M., \& {Deschamps}, F. 2017, \apj, 850, 93

\bibitem[{{Bryan} \& {Lee}(2024)}]{Bryan2024}
{Bryan}, M.~L. \& {Lee}, E.~J. 2024, \apjl, 968, L25

\bibitem[{{Bryan} \& {Lee}(2025)}]{Bryan2025}
{Bryan}, M.~L. \& {Lee}, E.~J. 2025, \apjl, 982, L7

\bibitem[{{Bryant} {et~al.}(2023){Bryant}, {Bayliss}, \& {Van Eylen}}]{Bryant2023}
{Bryant}, E.~M., {Bayliss}, D., \& {Van Eylen}, V. 2023, \mnras, 521, 3663

\bibitem[{{Burn} {et~al.}(2024){Burn}, {Mordasini}, {Mishra}, {Haldemann}, {Venturini}, {Emsenhuber}, \& {Henning}}]{Burn2024}
{Burn}, R., {Mordasini}, C., {Mishra}, L., {et~al.} 2024, Nature Astronomy, 8, 463

\bibitem[{{Burn} {et~al.}(2021){Burn}, {Schlecker}, {Mordasini}, {Emsenhuber}, {Alibert}, {Henning}, {Klahr}, \& {Benz}}]{Burn2021}
{Burn}, R., {Schlecker}, M., {Mordasini}, C., {et~al.} 2021, \aap, 656, A72

\bibitem[{{Castelli} \& {Kurucz}(2003)}]{Castelli2003}
{Castelli}, F. \& {Kurucz}, R.~L. 2003, in IAU Symposium, Vol. 210, Modelling of Stellar Atmospheres, ed. N.~{Piskunov}, W.~W. {Weiss}, \& D.~F. {Gray}, A20

\bibitem[{{Castro-Gonz{\'a}lez} {et~al.}(2024){Castro-Gonz{\'a}lez}, {Bourrier}, {Lillo-Box}, {Delisle}, {Armstrong}, {Barrado}, \& {Correia}}]{Castro2024}
{Castro-Gonz{\'a}lez}, A., {Bourrier}, V., {Lillo-Box}, J., {et~al.} 2024, \aap, 689, A250

\bibitem[{{Chachan} \& {Lee}(2023)}]{Chachan2023}
{Chachan}, Y. \& {Lee}, E.~J. 2023, \apjl, 952, L20

\bibitem[{{Christian} {et~al.}(2022){Christian}, {Vanderburg}, {Becker}, {Yahalomi}, {Pearce}, {Zhou}, {Collins}, {Kraus}, {Stassun}, {de Beurs}, {Ricker}, {Vanderspek}, {Latham}, {Winn}, {Seager}, {Jenkins}, {Abe}, {Agabi}, {Amado}, {Baker}, {Barkaoui}, {Benkhaldoun}, {Benni}, {Berberian}, {Berlind}, {Bieryla}, {Esparza-Borges}, {Bowen}, {Brown}, {Buchhave}, {Burke}, {Buttu}, {Cadieux}, {Caldwell}, {Charbonneau}, {Chazov}, {Chimaladinne}, {Collins}, {Combs}, {Conti}, {Crouzet}, {de Leon}, {Deljookorani}, {Diamond}, {Doyon}, {Dragomir}, {Dransfield}, {Essack}, {Evans}, {Fukui}, {Gan}, {Esquerdo}, {Gillon}, {Girardin}, {Guerra}, {Guillot}, {K. Habich}, {Henriksen}, {Hoch}, {Isogai}, {Jehin}, {Jensen}, {Johnson}, {Livingston}, {Kielkopf}, {Kim}, {Kawauchi}, {Krushinsky}, {Kunzle}, {Laloum}, {Leger}, {Lewin}, {Mallia}, {Massey}, {Mori}, {McLeod}, {M{\'e}karnia}, {Mireles}, {Mishevskiy}, {Tamura}, {Murgas}, {Narita}, {Naves}, {Nelson}, {Osborn}, {Palle}, {Parviainen}, {Plavchan}, {Pozuelos}, {Rabus}, {Relles},
  {Rodr{\'\i}guez L{\'o}pez}, {Quinn}, {Schmider}, {Schlieder}, {Schwarz}, {Shporer}, {Sibbald}, {Srdoc}, {Stibbards}, {Stickler}, {Suarez}, {Stockdale}, {Tan}, {Terada}, {Triaud}, {Tronsgaard}, {Waalkes}, {Wang}, {Watanabe}, {Wenceslas}, {Wingham}, {Wittrock}, \& {Ziegler}}]{Chirstian2022}
{Christian}, S., {Vanderburg}, A., {Becker}, J., {et~al.} 2022, \aj, 163, 207

\bibitem[{{Ciardi} {et~al.}(2015){Ciardi}, {Beichman}, {Horch}, \& {Howell}}]{Ciardi2015}
{Ciardi}, D.~R., {Beichman}, C.~A., {Horch}, E.~P., \& {Howell}, S.~B. 2015, \apj, 805, 16

\bibitem[{{Cieza} {et~al.}(2009){Cieza}, {Padgett}, {Allen}, {McCabe}, {Brooke}, {Carey}, {Chapman}, {Fukagawa}, {Huard}, {Noriga-Crespo}, {Peterson}, \& {Rebull}}]{Cieza2009}
{Cieza}, L.~A., {Padgett}, D.~L., {Allen}, L.~E., {et~al.} 2009, \apjl, 696, L84

\bibitem[{{Cloutier} \& {Menou}(2020)}]{Cloutier2020}
{Cloutier}, R. \& {Menou}, K. 2020, \aj, 159, 211

\bibitem[{{Cointepas} {et~al.}(2021){Cointepas}, {Almenara}, {Bonfils}, {Bouchy}, {Astudillo-Defru}, {Murgas}, {Otegi}, {Wyttenbach}, {Anderson}, {Artigau}, {Canto Martins}, {Charbonneau}, {Collins}, {Collins}, {Correia}, {Curaba}, {Delboulb{\'e}}, {Delfosse}, {D{\'\i}az}, {Dorn}, {Doyon}, {Feautrier}, {Figueira}, {Forveille}, {Gaisne}, {Gan}, {Gluck}, {Helled}, {Hellier}, {Jocou}, {Kern}, {Lafrasse}, {Law}, {Le{\~a}o}, {Lovis}, {Magnard}, {Mann}, {Maurel}, {de Medeiros}, {Melo}, {Moulin}, {Pepe}, {Rabou}, {Rochat}, {Rodriguez}, {Roux}, {Santos}, {S{\'e}gransan}, {Stadler}, {Ting}, {Twicken}, {Udry}, {Waalkes}, {West}, {W{\"u}nsche}, {Ziegler}, {Ricker}, {Vanderspek}, {Latham}, {Seager}, {Winn}, \& {Jenkins}}]{Cointepas2021}
{Cointepas}, M., {Almenara}, J.~M., {Bonfils}, X., {et~al.} 2021, \aap, 650, A145

\bibitem[{{Collins}(2019)}]{Collins2019}
{Collins}, K. 2019, in American Astronomical Society Meeting Abstracts, Vol. 233, American Astronomical Society Meeting Abstracts \#233, 140.05

\bibitem[{{Collins} {et~al.}(2017){Collins}, {Kielkopf}, {Stassun}, \& {Hessman}}]{Collins2017}
{Collins}, K.~A., {Kielkopf}, J.~F., {Stassun}, K.~G., \& {Hessman}, F.~V. 2017, \aj, 153, 77

\bibitem[{{Cook} {et~al.}(2022){Cook}, {Artigau}, {Doyon}, {Hobson}, {Martioli}, {Bouchy}, {Moutou}, {Carmona}, {Usher}, {Fouqu{\'e}}, {Arnold}, {Delfosse}, {Boisse}, {Cadieux}, {Vandal}, {Donati}, \& {Desli{\`e}res}}]{Cook2022}
{Cook}, N.~J., {Artigau}, {\'E}., {Doyon}, R., {et~al.} 2022, \pasp, 134, 114509

\bibitem[{{Deline} {et~al.}(2022){Deline}, {Hooton}, {Lendl}, {Morris}, {Salmon}, {Olofsson}, {Broeg}, {Ehrenreich}, {Beck}, {Brandeker}, {Hoyer}, {Sulis}, {Van Grootel}, {Bourrier}, {Demangeon}, {Demory}, {Heng}, {Parviainen}, {Serrano}, {Singh}, {Bonfanti}, {Fossati}, {Kitzmann}, {Sousa}, {Wilson}, {Alibert}, {Alonso}, {Anglada}, {B{\'a}rczy}, {Barrado Navascues}, {Barros}, {Baumjohann}, {Beck}, {Bekkelien}, {Benz}, {Billot}, {Bonfils}, {Cabrera}, {Charnoz}, {Collier Cameron}, {Corral van Damme}, {Csizmadia}, {Davies}, {Deleuil}, {Delrez}, {de Roche}, {Erikson}, {Fortier}, {Fridlund}, {Futyan}, {Gandolfi}, {Gillon}, {G{\"u}del}, {Gutermann}, {Hasiba}, {Isaak}, {Kiss}, {Laskar}, {Lecavelier des Etangs}, {Lovis}, {Magrin}, {Maxted}, {Munari}, {Nascimbeni}, {Ottensamer}, {Pagano}, {Pall{\'e}}, {Peter}, {Piotto}, {Pollacco}, {Queloz}, {Ragazzoni}, {Rando}, {Rauer}, {Ribas}, {Santos}, {Scandariato}, {S{\'e}gransan}, {Simon}, {Smith}, {Steller}, {Szab{\'o}}, {Thomas}, {Udry}, {Walter}, \& {Walton}}]{Deline2022}
{Deline}, A., {Hooton}, M.~J., {Lendl}, M., {et~al.} 2022, \aap, 659, A74

\bibitem[{{Delmotte} {et~al.}(2006){Delmotte}, {Dolensky}, {Padovani}, {Retzlaff}, {Rit{\'e}}, {Rosati}, {Slijkhuis}, {Wicenec}, {Fernique}, \& {Micol}}]{Delmotte2006}
{Delmotte}, N., {Dolensky}, M., {Padovani}, P., {et~al.} 2006, in Astronomical Society of the Pacific Conference Series, Vol. 351, Astronomical Data Analysis Software and Systems XV, ed. C.~{Gabriel}, C.~{Arviset}, D.~{Ponz}, \& S.~{Enrique}, 690

\bibitem[{{Donati} {et~al.}(2020){Donati}, {Kouach}, {Moutou}, {Doyon}, {Delfosse}, {Artigau}, {Baratchart}, {Lacombe}, {Barrick}, {H{\'e}brard}, {Bouchy}, {Saddlemyer}, {Par{\`e}s}, {Rabou}, {Micheau}, {Dolon}, {Reshetov}, {Challita}, {Carmona}, {Striebig}, {Thibault}, {Martioli}, {Cook}, {Fouqu{\'e}}, {Vermeulen}, {Wang}, {Arnold}, {Pepe}, {Boisse}, {Figueira}, {Bouvier}, {Ray}, {Feugeade}, {Morin}, {Alencar}, {Hobson}, {Castilho}, {Udry}, {Santos}, {Hernandez}, {Benedict}, {Vall{\'e}e}, {Gallou}, {Dupieux}, {Larrieu}, {Perruchot}, {Sottile}, {Moreau}, {Usher}, {Baril}, {Wildi}, {Chazelas}, {Malo}, {Bonfils}, {Loop}, {Kerley}, {Wevers}, {Dunn}, {Pazder}, {Macdonald}, {Dubois}, {Carri{\'e}}, {Valentin}, {Henault}, {Yan}, \& {Steinmetz}}]{Donati2020}
{Donati}, J.~F., {Kouach}, D., {Moutou}, C., {et~al.} 2020, \mnras, 498, 5684

\bibitem[{{Dorn} {et~al.}(2015){Dorn}, {Khan}, {Heng}, {Connolly}, {Alibert}, {Benz}, \& {Tackley}}]{Dorn2015}
{Dorn}, C., {Khan}, A., {Heng}, K., {et~al.} 2015, \aap, 577, A83

\bibitem[{{Dressing} \& {Charbonneau}(2013)}]{Dressing2013}
{Dressing}, C.~D. \& {Charbonneau}, D. 2013, \apj, 767, 95

\bibitem[{{Dressing} \& {Charbonneau}(2015)}]{Dressing2015}
{Dressing}, C.~D. \& {Charbonneau}, D. 2015, \apj, 807, 45

\bibitem[{{El-Badry} {et~al.}(2024){El-Badry}, {Lam}, {Holl}, {Halbwachs}, {Rix}, {Mazeh}, \& {Shahaf}}]{ElBadry2024}
{El-Badry}, K., {Lam}, C., {Holl}, B., {et~al.} 2024, The Open Journal of Astrophysics, 7, 100

\bibitem[{{El-Badry} {et~al.}(2021){El-Badry}, {Rix}, \& {Heintz}}]{ElBadry2021}
{El-Badry}, K., {Rix}, H.-W., \& {Heintz}, T.~M. 2021, \mnras, 506, 2269

\bibitem[{{Espinoza}(2018)}]{Espinoza2018}
{Espinoza}, N. 2018, Research Notes of the American Astronomical Society, 2, 209

\bibitem[{{Espinoza} \& {Jord{\'a}n}(2015)}]{Espinoza2015}
{Espinoza}, N. \& {Jord{\'a}n}, A. 2015, \mnras, 450, 1879

\bibitem[{{Espinoza} {et~al.}(2019){Espinoza}, {Kossakowski}, \& {Brahm}}]{Espinoza2019_juliet}
{Espinoza}, N., {Kossakowski}, D., \& {Brahm}, R. 2019, \mnras, 490, 2262

\bibitem[{{Foreman-Mackey} {et~al.}(2017){Foreman-Mackey}, {Agol}, {Ambikasaran}, \& {Angus}}]{Foreman-Mackey2017_celerite}
{Foreman-Mackey}, D., {Agol}, E., {Ambikasaran}, S., \& {Angus}, R. 2017, \aj, 154, 220

\bibitem[{{Foreman-Mackey} {et~al.}(2013){Foreman-Mackey}, {Hogg}, {Lang}, \& {Goodman}}]{Foreman-Mackey_2013}
{Foreman-Mackey}, D., {Hogg}, D.~W., {Lang}, D., \& {Goodman}, J. 2013, \pasp, 125, 306

\bibitem[{{French} {et~al.}(2009){French}, {Mattsson}, {Nettelmann}, \& {Redmer}}]{French_2009}
{French}, M., {Mattsson}, T.~R., {Nettelmann}, N., \& {Redmer}, R. 2009, \prb, 79, 054107

\bibitem[{{Fulton} {et~al.}(2018){Fulton}, {Petigura}, {Blunt}, \& {Sinukoff}}]{Fulton2018_radvel}
{Fulton}, B.~J., {Petigura}, E.~A., {Blunt}, S., \& {Sinukoff}, E. 2018, \pasp, 130, 044504

\bibitem[{{Fulton} {et~al.}(2017){Fulton}, {Petigura}, {Howard}, {Isaacson}, {Marcy}, {Cargile}, {Hebb}, {Weiss}, {Johnson}, {Morton}, {Sinukoff}, {Crossfield}, \& {Hirsch}}]{Fulton2017}
{Fulton}, B.~J., {Petigura}, E.~A., {Howard}, A.~W., {et~al.} 2017, \aj, 154, 109

\bibitem[{{Gaia Collaboration} {et~al.}(2018){Gaia Collaboration}, {Brown}, {Vallenari}, {Prusti}, {de Bruijne}, {Babusiaux}, {Bailer-Jones}, {Biermann}, {Evans}, {Eyer}, {Jansen}, {Jordi}, {Klioner}, {Lammers}, {Lindegren}, {Luri}, {Mignard}, {Panem}, {Pourbaix}, {Randich}, {Sartoretti}, {Siddiqui}, {Soubiran}, {van Leeuwen}, {Walton}, {Arenou}, {Bastian}, {Cropper}, {Drimmel}, {Katz}, {Lattanzi}, {Bakker}, {Cacciari}, {Casta{\~n}eda}, {Chaoul}, {Cheek}, {De Angeli}, {Fabricius}, {Guerra}, {Holl}, {Masana}, {Messineo}, {Mowlavi}, {Nienartowicz}, {Panuzzo}, {Portell}, {Riello}, {Seabroke}, {Tanga}, {Th{\'e}venin}, {Gracia-Abril}, {Comoretto}, {Garcia-Reinaldos}, {Teyssier}, {Altmann}, {Andrae}, {Audard}, {Bellas-Velidis}, {Benson}, {Berthier}, {Blomme}, {Burgess}, {Busso}, {Carry}, {Cellino}, {Clementini}, {Clotet}, {Creevey}, {Davidson}, {De Ridder}, {Delchambre}, {Dell'Oro}, {Ducourant}, {Fern{\'a}ndez-Hern{\'a}ndez}, {Fouesneau}, {Fr{\'e}mat}, {Galluccio}, {Garc{\'\i}a-Torres},
  {Gonz{\'a}lez-N{\'u}{\~n}ez}, {Gonz{\'a}lez-Vidal}, {Gosset}, {Guy}, {Halbwachs}, {Hambly}, {Harrison}, {Hern{\'a}ndez}, {Hestroffer}, {Hodgkin}, {Hutton}, {Jasniewicz}, {Jean-Antoine-Piccolo}, {Jordan}, {Korn}, {Krone-Martins}, {Lanzafame}, {Lebzelter}, {L{\"o}ffler}, {Manteiga}, {Marrese}, {Mart{\'\i}n-Fleitas}, {Moitinho}, {Mora}, {Muinonen}, {Osinde}, {Pancino}, {Pauwels}, {Petit}, {Recio-Blanco}, {Richards}, {Rimoldini}, {Robin}, {Sarro}, {Siopis}, {Smith}, {Sozzetti}, {S{\"u}veges}, {Torra}, {van Reeven}, {Abbas}, {Abreu Aramburu}, {Accart}, {Aerts}, {Altavilla}, {{\'A}lvarez}, {Alvarez}, {Alves}, {Anderson}, {Andrei}, {Anglada Varela}, {Antiche}, {Antoja}, {Arcay}, {Astraatmadja}, {Bach}, {Baker}, {Balaguer-N{\'u}{\~n}ez}, {Balm}, {Barache}, {Barata}, {Barbato}, {Barblan}, {Barklem}, {Barrado}, {Barros}, {Barstow}, {Bartholom{\'e} Mu{\~n}oz}, {Bassilana}, {Becciani}, {Bellazzini}, {Berihuete}, {Bertone}, {Bianchi}, {Bienaym{\'e}}, {Blanco-Cuaresma}, {Boch}, {Boeche}, {Bombrun}, {Borrachero},
  {Bossini}, {Bouquillon}, {Bourda}, {Bragaglia}, {Bramante}, {Breddels}, {Bressan}, {Brouillet}, {Br{\"u}semeister}, {Brugaletta}, {Bucciarelli}, {Burlacu}, {Busonero}, {Butkevich}, {Buzzi}, {Caffau}, {Cancelliere}, {Cannizzaro}, {Cantat-Gaudin}, {Carballo}, {Carlucci}, {Carrasco}, {Casamiquela}, {Castellani}, {Castro-Ginard}, {Charlot}, {Chemin}, {Chiavassa}, {Cocozza}, {Costigan}, {Cowell}, {Crifo}, {Crosta}, {Crowley}, {Cuypers}, {Dafonte}, {Damerdji}, {Dapergolas}, {David}, {David}, {de Laverny}, {De Luise}, {De March}, {de Martino}, {de Souza}, {de Torres}, {Debosscher}, {del Pozo}, {Delbo}, {Delgado}, {Delgado}, {Di Matteo}, {Diakite}, {Diener}, {Distefano}, {Dolding}, {Drazinos}, {Dur{\'a}n}, {Edvardsson}, {Enke}, {Eriksson}, {Esquej}, {Eynard Bontemps}, {Fabre}, {Fabrizio}, {Faigler}, {Falc{\~a}o}, {Farr{\`a}s Casas}, {Federici}, {Fedorets}, {Fernique}, {Figueras}, {Filippi}, {Findeisen}, {Fonti}, {Fraile}, {Fraser}, {Fr{\'e}zouls}, {Gai}, {Galleti}, {Garabato}, {Garc{\'\i}a-Sedano}, {Garofalo},
  {Garralda}, {Gavel}, {Gavras}, {Gerssen}, {Geyer}, {Giacobbe}, {Gilmore}, {Girona}, {Giuffrida}, {Glass}, {Gomes}, {Granvik}, {Gueguen}, {Guerrier}, {Guiraud}, {Guti{\'e}rrez-S{\'a}nchez}, {Haigron}, {Hatzidimitriou}, {Hauser}, {Haywood}, {Heiter}, {Helmi}, {Heu}, {Hilger}, {Hobbs}, {Hofmann}, {Holland}, {Huckle}, {Hypki}, {Icardi}, {Jan{\ss}en}, {Jevardat de Fombelle}, {Jonker}, {Juh{\'a}sz}, {Julbe}, {Karampelas}, {Kewley}, {Klar}, {Kochoska}, {Kohley}, {Kolenberg}, {Kontizas}, {Kontizas}, {Koposov}, {Kordopatis}, {Kostrzewa-Rutkowska}, {Koubsky}, {Lambert}, {Lanza}, {Lasne}, {Lavigne}, {Le Fustec}, {Le Poncin-Lafitte}, {Lebreton}, {Leccia}, {Leclerc}, {Lecoeur-Taibi}, {Lenhardt}, {Leroux}, {Liao}, {Licata}, {Lindstr{\o}m}, {Lister}, {Livanou}, {Lobel}, {L{\'o}pez}, {Managau}, {Mann}, {Mantelet}, {Marchal}, {Marchant}, {Marconi}, {Marinoni}, {Marschalk{\'o}}, {Marshall}, {Martino}, {Marton}, {Mary}, {Massari}, {Matijevi{\v{c}}}, {Mazeh}, {McMillan}, {Messina}, {Michalik}, {Millar}, {Molina}, {Molinaro},
  {Moln{\'a}r}, {Montegriffo}, {Mor}, {Morbidelli}, {Morel}, {Morris}, {Mulone}, {Muraveva}, {Musella}, {Nelemans}, {Nicastro}, {Noval}, {O'Mullane}, {Ord{\'e}novic}, {Ord{\'o}{\~n}ez-Blanco}, {Osborne}, {Pagani}, {Pagano}, {Pailler}, {Palacin}, {Palaversa}, {Panahi}, {Pawlak}, {Piersimoni}, {Pineau}, {Plachy}, {Plum}, {Poggio}, {Poujoulet}, {Pr{\v{s}}a}, {Pulone}, {Racero}, {Ragaini}, {Rambaux}, {Ramos-Lerate}, {Regibo}, {Reyl{\'e}}, {Riclet}, {Ripepi}, {Riva}, {Rivard}, {Rixon}, {Roegiers}, {Roelens}, {Romero-G{\'o}mez}, {Rowell}, {Royer}, {Ruiz-Dern}, {Sadowski}, {Sagrist{\`a} Sell{\'e}s}, {Sahlmann}, {Salgado}, {Salguero}, {Sanna}, {Santana-Ros}, {Sarasso}, {Savietto}, {Schultheis}, {Sciacca}, {Segol}, {Segovia}, {S{\'e}gransan}, {Shih}, {Siltala}, {Silva}, {Smart}, {Smith}, {Solano}, {Solitro}, {Sordo}, {Soria Nieto}, {Souchay}, {Spagna}, {Spoto}, {Stampa}, {Steele}, {Steidelm{\"u}ller}, {Stephenson}, {Stoev}, {Suess}, {Surdej}, {Szabados}, {Szegedi-Elek}, {Tapiador}, {Taris}, {Tauran}, {Taylor},
  {Teixeira}, {Terrett}, {Teyssandier}, {Thuillot}, {Titarenko}, {Torra Clotet}, {Turon}, {Ulla}, {Utrilla}, {Uzzi}, {Vaillant}, {Valentini}, {Valette}, {van Elteren}, {Van Hemelryck}, {van Leeuwen}, {Vaschetto}, {Vecchiato}, {Veljanoski}, {Viala}, {Vicente}, {Vogt}, {von Essen}, {Voss}, {Votruba}, {Voutsinas}, {Walmsley}, {Weiler}, {Wertz}, {Wevers}, {Wyrzykowski}, {Yoldas}, {{\v{Z}}erjal}, {Ziaeepour}, {Zorec}, {Zschocke}, {Zucker}, {Zurbach}, \& {Zwitter}}]{Gaia2018}
{Gaia Collaboration}, {Brown}, A.~G.~A., {Vallenari}, A., {et~al.} 2018, \aap, 616, A1

\bibitem[{{Gaidos} {et~al.}(2016){Gaidos}, {Mann}, {Kraus}, \& {Ireland}}]{Gaidos2016}
{Gaidos}, E., {Mann}, A.~W., {Kraus}, A.~L., \& {Ireland}, M. 2016, \mnras, 457, 2877

\bibitem[{{Gardner} {et~al.}(2006){Gardner}, {Mather}, {Clampin}, {Doyon}, {Greenhouse}, {Hammel}, {Hutchings}, {Jakobsen}, {Lilly}, {Long}, {Lunine}, {McCaughrean}, {Mountain}, {Nella}, {Rieke}, {Rieke}, {Rix}, {Smith}, {Sonneborn}, {Stiavelli}, {Stockman}, {Windhorst}, \& {Wright}}]{JWST}
{Gardner}, J.~P., {Mather}, J.~C., {Clampin}, M., {et~al.} 2006, \ssr, 123, 485

\bibitem[{{Gordon} {et~al.}(2022){Gordon}, {Rothman}, {Hargreaves}, {Hashemi}, {Karlovets}, {Skinner}, {Conway}, {Hill}, {Kochanov}, {Tan}, {Wcis{\l}o}, {Finenko}, {Nelson}, {Bernath}, {Birk}, {Boudon}, {Campargue}, {Chance}, {Coustenis}, {Drouin}, {Flaud}, {Gamache}, {Hodges}, {Jacquemart}, {Mlawer}, {Nikitin}, {Perevalov}, {Rotger}, {Tennyson}, {Toon}, {Tran}, {Tyuterev}, {Adkins}, {Baker}, {Barbe}, {Can{\`e}}, {Cs{\'a}sz{\'a}r}, {Dudaryonok}, {Egorov}, {Fleisher}, {Fleurbaey}, {Foltynowicz}, {Furtenbacher}, {Harrison}, {Hartmann}, {Horneman}, {Huang}, {Karman}, {Karns}, {Kassi}, {Kleiner}, {Kofman}, {Kwabia-Tchana}, {Lavrentieva}, {Lee}, {Long}, {Lukashevskaya}, {Lyulin}, {Makhnev}, {Matt}, {Massie}, {Melosso}, {Mikhailenko}, {Mondelain}, {M{\"u}ller}, {Naumenko}, {Perrin}, {Polyansky}, {Raddaoui}, {Raston}, {Reed}, {Rey}, {Richard}, {T{\'o}bi{\'a}s}, {Sadiek}, {Schwenke}, {Starikova}, {Sung}, {Tamassia}, {Tashkun}, {Vander Auwera}, {Vasilenko}, {Vigasin}, {Villanueva}, {Vispoel}, {Wagner}, {Yachmenev}, \&
  {Yurchenko}}]{Gordon2022}
{Gordon}, I.~E., {Rothman}, L.~S., {Hargreaves}, R.~J., {et~al.} 2022, \jqsrt, 277, 107949

\bibitem[{{Gratia} \& {Fabrycky}(2017)}]{Gratia2017}
{Gratia}, P. \& {Fabrycky}, D. 2017, \mnras, 464, 1709

\bibitem[{{Guerrero} {et~al.}(2021){Guerrero}, {Seager}, {Huang}, {Vanderburg}, {Garcia Soto}, {Mireles}, {Hesse}, {Fong}, {Glidden}, {Shporer}, {Latham}, {Collins}, {Quinn}, {Burt}, {Dragomir}, {Crossfield}, {Vanderspek}, {Fausnaugh}, {Burke}, {Ricker}, {Daylan}, {Essack}, {G{\"u}nther}, {Osborn}, {Pepper}, {Rowden}, {Sha}, {Villanueva}, {Yahalomi}, {Yu}, {Ballard}, {Batalha}, {Berardo}, {Chontos}, {Dittmann}, {Esquerdo}, {Mikal-Evans}, {Jayaraman}, {Krishnamurthy}, {Louie}, {Mehrle}, {Niraula}, {Rackham}, {Rodriguez}, {Rowden}, {Sousa-Silva}, {Watanabe}, {Wong}, {Zhan}, {Zivanovic}, {Christiansen}, {Ciardi}, {Swain}, {Lund}, {Mullally}, {Fleming}, {Rodriguez}, {Boyd}, {Quintana}, {Barclay}, {Col{\'o}n}, {Rinehart}, {Schlieder}, {Clampin}, {Jenkins}, {Twicken}, {Caldwell}, {Coughlin}, {Henze}, {Lissauer}, {Morris}, {Rose}, {Smith}, {Tenenbaum}, {Ting}, {Wohler}, {Bakos}, {Bean}, {Berta-Thompson}, {Bieryla}, {Bouma}, {Buchhave}, {Butler}, {Charbonneau}, {Doty}, {Ge}, {Holman}, {Howard}, {Kaltenegger}, {Kane},
  {Kjeldsen}, {Kreidberg}, {Lin}, {Minsky}, {Narita}, {Paegert}, {P{\'a}l}, {Palle}, {Sasselov}, {Spencer}, {Sozzetti}, {Stassun}, {Torres}, {Udry}, \& {Winn}}]{Guerrero2021}
{Guerrero}, N.~M., {Seager}, S., {Huang}, C.~X., {et~al.} 2021, \apjs, 254, 39

\bibitem[{{Guillot} \& {Morel}(1995)}]{Guillot_1995}
{Guillot}, T. \& {Morel}, P. 1995, \aaps, 109, 109

\bibitem[{{Hara} {et~al.}(2019){Hara}, {Bou{\'e}}, {Laskar}, {Delisle}, \& {Unger}}]{Hara2019}
{Hara}, N.~C., {Bou{\'e}}, G., {Laskar}, J., {Delisle}, J.~B., \& {Unger}, N. 2019, \mnras, 489, 738

\bibitem[{{Harris} {et~al.}(2012){Harris}, {Andrews}, {Wilner}, \& {Kraus}}]{Harris2012}
{Harris}, R.~J., {Andrews}, S.~M., {Wilner}, D.~J., \& {Kraus}, A.~L. 2012, \apj, 751, 115

\bibitem[{{Hedges} {et~al.}(2019){Hedges}, {Saunders}, {Barentsen}, {Coughlin}, {Cardoso}, {Kostov}, {Dotson}, \& {Cody}}]{Hedges2019}
{Hedges}, C., {Saunders}, N., {Barentsen}, G., {et~al.} 2019, \apjl, 880, L5

\bibitem[{{Henden} {et~al.}(2015){Henden}, {Levine}, {Terrell}, \& {Welch}}]{Henden2015}
{Henden}, A.~A., {Levine}, S., {Terrell}, D., \& {Welch}, D.~L. 2015, in American Astronomical Society Meeting Abstracts, Vol. 225, American Astronomical Society Meeting Abstracts \#225, 336.16

\bibitem[{{Henry} {et~al.}(2006){Henry}, {Jao}, {Subasavage}, {Beaulieu}, {Ianna}, {Costa}, \& {M{\'e}ndez}}]{Henry2006}
{Henry}, T.~J., {Jao}, W.-C., {Subasavage}, J.~P., {et~al.} 2006, \aj, 132, 2360

\bibitem[{{Ho} {et~al.}(2024){Ho}, {Rogers}, {Van Eylen}, {Owen}, \& {Schlichting}}]{Ho2024}
{Ho}, C. S.~K., {Rogers}, J.~G., {Van Eylen}, V., {Owen}, J.~E., \& {Schlichting}, H.~E. 2024, \mnras, 531, 3698

\bibitem[{{Howard} {et~al.}(2012){Howard}, {Marcy}, {Bryson}, {Jenkins}, {Rowe}, {Batalha}, {Borucki}, {Koch}, {Dunham}, {Gautier}, {Van Cleve}, {Cochran}, {Latham}, {Lissauer}, {Torres}, {Brown}, {Gilliland}, {Buchhave}, {Caldwell}, {Christensen-Dalsgaard}, {Ciardi}, {Fressin}, {Haas}, {Howell}, {Kjeldsen}, {Seager}, {Rogers}, {Sasselov}, {Steffen}, {Basri}, {Charbonneau}, {Christiansen}, {Clarke}, {Dupree}, {Fabrycky}, {Fischer}, {Ford}, {Fortney}, {Tarter}, {Girouard}, {Holman}, {Johnson}, {Klaus}, {Machalek}, {Moorhead}, {Morehead}, {Ragozzine}, {Tenenbaum}, {Twicken}, {Quinn}, {Isaacson}, {Shporer}, {Lucas}, {Walkowicz}, {Welsh}, {Boss}, {Devore}, {Gould}, {Smith}, {Morris}, {Prsa}, {Morton}, {Still}, {Thompson}, {Mullally}, {Endl}, \& {MacQueen}}]{Howard2012}
{Howard}, A.~W., {Marcy}, G.~W., {Bryson}, S.~T., {et~al.} 2012, \apjs, 201, 15

\bibitem[{{Howell} {et~al.}(2011){Howell}, {Everett}, {Sherry}, {Horch}, \& {Ciardi}}]{Howell2011}
{Howell}, S.~B., {Everett}, M.~E., {Sherry}, W., {Horch}, E., \& {Ciardi}, D.~R. 2011, \aj, 142, 19

\bibitem[{Husser {et~al.}(2013)Husser, Wende-von Berg, Dreizler, Homeier, Reiners, Barman, \& Hauschildt}]{Husser_2013}
Husser, T.-O., Wende-von Berg, S., Dreizler, S., {et~al.} 2013, \aap, 553, A6

\bibitem[{{Ida} \& {Lin}(2005)}]{Ida2005}
{Ida}, S. \& {Lin}, D.~N.~C. 2005, \apj, 626, 1045

\bibitem[{Jahandar {et~al.}(2025)Jahandar, Doyon, Artigau, Cook, Cadieux, Donati, Cowan, Cloutier, Pelletier, {Alves-Brito}, Martins, Shang, \& Carmona}]{jahandarChemicalFingerprintsDwarfs2025}
Jahandar, F., Doyon, R., Artigau, {\'E}., {et~al.} 2025, The Astrophysical Journal, 978, 154

\bibitem[{Jahandar {et~al.}(2024)Jahandar, Doyon, Artigau, Cook, Cadieux, Lafreni{\`e}re, Forveille, Donati, Fouqu{\'e}, Carmona, Cloutier, Cristofari, Gaidos, {Gomes da Silva}, Malo, Martioli, {do Nascimento}, Pelletier, Vandal, \& Venn}]{jahandarComprehensiveHighresolutionChemical2024}
Jahandar, F., Doyon, R., Artigau, {\'E}., {et~al.} 2024, The Astrophysical Journal, 966, 56

\bibitem[{{Jenkins} {et~al.}(2016){Jenkins}, {Twicken}, {McCauliff}, {Campbell}, {Sanderfer}, {Lung}, {Mansouri-Samani}, {Girouard}, {Tenenbaum}, {Klaus}, {Smith}, {Caldwell}, {Chacon}, {Henze}, {Heiges}, {Latham}, {Morgan}, {Swade}, {Rinehart}, \& {Vanderspek}}]{Jenkins2016}
{Jenkins}, J.~M., {Twicken}, J.~D., {McCauliff}, S., {et~al.} 2016, in Society of Photo-Optical Instrumentation Engineers (SPIE) Conference Series, Vol. 9913, Software and Cyberinfrastructure for Astronomy IV, ed. G.~{Chiozzi} \& J.~C. {Guzman}, 99133E

\bibitem[{{Kempton} {et~al.}(2018){Kempton}, {Bean}, {Louie}, {Deming}, {Koll}, {Mansfield}, {Christiansen}, {L{\'o}pez-Morales}, {Swain}, {Zellem}, {Ballard}, {Barclay}, {Barstow}, {Batalha}, {Beatty}, {Berta-Thompson}, {Birkby}, {Buchhave}, {Charbonneau}, {Cowan}, {Crossfield}, {de Val-Borro}, {Doyon}, {Dragomir}, {Gaidos}, {Heng}, {Hu}, {Kane}, {Kreidberg}, {Mallonn}, {Morley}, {Narita}, {Nascimbeni}, {Pall{\'e}}, {Quintana}, {Rauscher}, {Seager}, {Shkolnik}, {Sing}, {Sozzetti}, {Stassun}, {Valenti}, \& {von Essen}}]{Kempton2018}
{Kempton}, E. M.~R., {Bean}, J.~L., {Louie}, D.~R., {et~al.} 2018, \pasp, 130, 114401

\bibitem[{{Khata} {et~al.}(2021){Khata}, {Mondal}, {Das}, \& {Baug}}]{Khata21}
{Khata}, D., {Mondal}, S., {Das}, R., \& {Baug}, T. 2021, \mnras, 507, 1869

\bibitem[{{Kite} {et~al.}(2019){Kite}, {Fegley}, {Schaefer}, \& {Ford}}]{Kite2019}
{Kite}, E.~S., {Fegley}, Jr., B., {Schaefer}, L., \& {Ford}, E.~B. 2019, \apjl, 887, L33

\bibitem[{{Kreidberg}(2015)}]{Kreidberg2015_batman}
{Kreidberg}, L. 2015, \pasp, 127, 1161

\bibitem[{{Kubyshkina} \& {Vidotto}(2021)}]{Kubyshkina2021}
{Kubyshkina}, D. \& {Vidotto}, A.~A. 2021, \mnras, 504, 2034

\bibitem[{{Kunimoto} {et~al.}(2023){Kunimoto}, {Vanderburg}, {Huang}, {Davis}, {Affer}, {Cameron}, {Charbonneau}, {Cosentino}, {Damasso}, {Dumusque}, {Fiorenzano}, {Ghedina}, {Haywood}, {Lienhard}, {L{\'o}pez-Morales}, {Mayor}, {Pepe}, {Pinamonti}, {Poretti}, {Maldonado}, {Rice}, {Sozzetti}, {Wilson}, {Udry}, {Baptista}, {Barkaoui}, {Becker}, {Benni}, {Bieryla}, {Bosch-Cabot}, {Ciardi}, {Collins}, {Collins}, {Evans}, {Dupuy}, {Goliguzova}, {Guerra}, {Kraus}, {Lissauer}, {Huber}, {Murgas}, {Palle}, {Quinn}, {Safonov}, {Schwarz}, {Shporer}, {Stassun}, {Jenkins}, {Latham}, {Ricker}, {Seager}, {Vanderspek}, {Winn}, {Essack}, {Lewis}, \& {Rose}}]{Kunimoto2023}
{Kunimoto}, M., {Vanderburg}, A., {Huang}, C.~X., {et~al.} 2023, \aj, 166, 7

\bibitem[{{Kurucz}(1979)}]{Kurucz1979}
{Kurucz}, R.~L. 1979, \apjs, 40, 1

\bibitem[{{Kurucz}(1993)}]{Kurucz1993}
{Kurucz}, R.~L. 1993, {SYNTHE spectrum synthesis programs and line data}

\bibitem[{{Laughlin} {et~al.}(2004){Laughlin}, {Bodenheimer}, \& {Adams}}]{Laughlin2004}
{Laughlin}, G., {Bodenheimer}, P., \& {Adams}, F.~C. 2004, \apjl, 612, L73

\bibitem[{{Lenzen} {et~al.}(2003){Lenzen}, {Hartung}, {Brandner}, {Finger}, {Hubin}, {Lacombe}, {Lagrange}, {Lehnert}, {Moorwood}, \& {Mouillet}}]{Lenzen2003}
{Lenzen}, R., {Hartung}, M., {Brandner}, W., {et~al.} 2003, in Society of Photo-Optical Instrumentation Engineers (SPIE) Conference Series, Vol. 4841, Instrument Design and Performance for Optical/Infrared Ground-based Telescopes, ed. M.~{Iye} \& A.~F.~M. {Moorwood}, 944--952

\bibitem[{{Li} {et~al.}(2019){Li}, {Tenenbaum}, {Twicken}, {Burke}, {Jenkins}, {Quintana}, {Rowe}, \& {Seader}}]{Li2019}
{Li}, J., {Tenenbaum}, P., {Twicken}, J.~D., {et~al.} 2019, \pasp, 131, 024506

\bibitem[{{Lillo-Box} {et~al.}(2023){Lillo-Box}, {Gandolfi}, {Armstrong}, {Collins}, {Nielsen}, {Luque}, {Korth}, {Sousa}, {Quinn}, {Acu{\~n}a}, {Howell}, {Morello}, {Hellier}, {Giacalone}, {Hoyer}, {Stassun}, {Palle}, {Aguichine}, {Mousis}, {Adibekyan}, {Azevedo Silva}, {Barrado}, {Deleuil}, {Eastman}, {Fukui}, {Hawthorn}, {Irwin}, {Jenkins}, {Latham}, {Muresan}, {Narita}, {Persson}, {Santerne}, {Santos}, {Savel}, {Osborn}, {Teske}, {Wheatley}, {Winn}, {Barros}, {Butler}, {Caldwell}, {Charbonneau}, {Cloutier}, {Crane}, {Demangeon}, {D{\'\i}az}, {Dumusque}, {Esposito}, {Falk}, {Gill}, {Hojjatpanah}, {Kreidberg}, {Mireles}, {Osborn}, {Ricker}, {Rodriguez}, {Schwarz}, {Seager}, {Serrano Bell}, {Shectman}, {Shporer}, {Vezie}, {Wang}, \& {Zhou}}]{LilloBox2023}
{Lillo-Box}, J., {Gandolfi}, D., {Armstrong}, D.~J., {et~al.} 2023, \aap, 669, A109

\bibitem[{{Lopez} \& {Fortney}(2013)}]{Lopez2013}
{Lopez}, E.~D. \& {Fortney}, J.~J. 2013, \apj, 776, 2

\bibitem[{{Lopez} \& {Rice}(2018)}]{Lopez2018}
{Lopez}, E.~D. \& {Rice}, K. 2018, \mnras, 479, 5303

\bibitem[{{Lovis} \& {Pepe}(2007)}]{Lovis2007}
{Lovis}, C. \& {Pepe}, F. 2007, \aap, 468, 1115

\bibitem[{{Lucy} \& {Sweeney}(1971)}]{Lucy1971}
{Lucy}, L.~B. \& {Sweeney}, M.~A. 1971, \aj, 76, 544

\bibitem[{{Luo} {et~al.}(2024){Luo}, {Dorn}, \& {Deng}}]{Luo2024}
{Luo}, H., {Dorn}, C., \& {Deng}, J. 2024, Nature Astronomy, 8, 1399

\bibitem[{{Mann} {et~al.}(2019){Mann}, {Dupuy}, {Kraus}, {Gaidos}, {Ansdell}, {Ireland}, {Rizzuto}, {Hung}, {Dittmann}, {Factor}, {Feiden}, {Martinez}, {Ru{\'\i}z-Rodr{\'\i}guez}, \& {Thao}}]{Mann2019}
{Mann}, A.~W., {Dupuy}, T., {Kraus}, A.~L., {et~al.} 2019, \apj, 871, 63

\bibitem[{{Mann} {et~al.}(2015){Mann}, {Feiden}, {Gaidos}, {Boyajian}, \& {von Braun}}]{Mann2015}
{Mann}, A.~W., {Feiden}, G.~A., {Gaidos}, E., {Boyajian}, T., \& {von Braun}, K. 2015, \apj, 804, 64

\bibitem[{Mann \& Whitney(1947)}]{Mann1947}
Mann, H.~B. \& Whitney, D.~R. 1947, The Annals of Mathematical Statistics, 18, 50

\bibitem[{{Marcy} {et~al.}(2014){Marcy}, {Weiss}, {Petigura}, {Isaacson}, {Howard}, \& {Buchhave}}]{Marcy2014}
{Marcy}, G.~W., {Weiss}, L.~M., {Petigura}, E.~A., {et~al.} 2014, Proceedings of the National Academy of Science, 111, 12655

\bibitem[{{Mayor} {et~al.}(2003){Mayor}, {Pepe}, {Queloz}, {Bouchy}, {Rupprecht}, {Lo Curto}, {Avila}, {Benz}, {Bertaux}, {Bonfils}, {Dall}, {Dekker}, {Delabre}, {Eckert}, {Fleury}, {Gilliotte}, {Gojak}, {Guzman}, {Kohler}, {Lizon}, {Longinotti}, {Lovis}, {Megevand}, {Pasquini}, {Reyes}, {Sivan}, {Sosnowska}, {Soto}, {Udry}, {van Kesteren}, {Weber}, \& {Weilenmann}}]{Mayor2003}
{Mayor}, M., {Pepe}, F., {Queloz}, D., {et~al.} 2003, The Messenger, 114, 20

\bibitem[{{Mayor} \& {Queloz}(1995)}]{Mayor1995}
{Mayor}, M. \& {Queloz}, D. 1995, \nat, 378, 355

\bibitem[{{McCully} {et~al.}(2018){McCully}, {Volgenau}, {Harbeck}, {Lister}, {Saunders}, {Turner}, {Siiverd}, \& {Bowman}}]{McCully2018}
{McCully}, C., {Volgenau}, N.~H., {Harbeck}, D.-R., {et~al.} 2018, in Society of Photo-Optical Instrumentation Engineers (SPIE) Conference Series, Vol. 10707, Software and Cyberinfrastructure for Astronomy V, ed. J.~C. {Guzman} \& J.~{Ibsen}, 107070K

\bibitem[{{McDonald} {et~al.}(2019){McDonald}, {Kreidberg}, \& {Lopez}}]{McDonald2019}
{McDonald}, G.~D., {Kreidberg}, L., \& {Lopez}, E. 2019, \apj, 876, 22

\bibitem[{{McLaughlin}(1924)}]{McLaughlin1924}
{McLaughlin}, D.~B. 1924, \apj, 60, 22

\bibitem[{{Mignon} {et~al.}(2025){Mignon}, {Delfosse}, {Meunier}, {Chaverot}, {Burn}, {Bonfils}, {Bouchy}, {Astudillo-Defru}, {Lo Curto}, {Gaisne}, {Udry}, {Forveille}, {Segransan}, {Lovis}, {Santos}, \& {Mayor}}]{Mignon2025}
{Mignon}, L., {Delfosse}, X., {Meunier}, N., {et~al.} 2025, \aap, in press

\bibitem[{{Moe} \& {Kratter}(2021)}]{Moe&Krater2021}
{Moe}, M. \& {Kratter}, K.~M. 2021, \mnras, 507, 3593

\bibitem[{{Morello} {et~al.}(2017){Morello}, {Tsiaras}, {Howarth}, \& {Homeier}}]{Morello2017}
{Morello}, G., {Tsiaras}, A., {Howarth}, I.~D., \& {Homeier}, D. 2017, \aj, 154, 111

\bibitem[{{Morris} {et~al.}(2020){Morris}, {Twicken}, {Smith}, {Clarke}, {Jenkins}, {Bryson}, {Girouard}, \& {Klaus}}]{Morris2020}
{Morris}, R.~L., {Twicken}, J.~D., {Smith}, J.~C., {et~al.} 2020, {Kepler Data Processing Handbook: Photometric Analysis}, Kepler Science Document KSCI-19081-003, id. 6. Edited by Jon M. Jenkins.

\bibitem[{{Mugrauer} \& {Michel}(2020)}]{Mugrauer2020}
{Mugrauer}, M. \& {Michel}, K.-U. 2020, Astronomische Nachrichten, 341, 996

\bibitem[{{Mulders} {et~al.}(2015){Mulders}, {Pascucci}, \& {Apai}}]{Mulders2015}
{Mulders}, G.~D., {Pascucci}, I., \& {Apai}, D. 2015, \apj, 814, 130

\bibitem[{{Neves} {et~al.}(2012){Neves}, {Bonfils}, {Santos}, {Delfosse}, {Forveille}, {Allard}, {Nat{\'a}rio}, {Fernandes}, \& {Udry}}]{Neves12}
{Neves}, V., {Bonfils}, X., {Santos}, N.~C., {et~al.} 2012, \aap, 538, A25

\bibitem[{{Neves} {et~al.}(2013){Neves}, {Bonfils}, {Santos}, {Delfosse}, {Forveille}, {Allard}, \& {Udry}}]{Neves2013}
{Neves}, V., {Bonfils}, X., {Santos}, N.~C., {et~al.} 2013, \aap, 551, A36

\bibitem[{{Otegi} {et~al.}(2020){Otegi}, {Bouchy}, \& {Helled}}]{Otegi2020}
{Otegi}, J.~F., {Bouchy}, F., \& {Helled}, R. 2020, \aap, 634, A43

\bibitem[{{Owen} \& {Jackson}(2012)}]{Owen2012}
{Owen}, J.~E. \& {Jackson}, A.~P. 2012, \mnras, 425, 2931

\bibitem[{{Owen} \& {Murray-Clay}(2018)}]{Owen2018}
{Owen}, J.~E. \& {Murray-Clay}, R. 2018, \mnras, 480, 2206

\bibitem[{{Owen} \& {Wu}(2017)}]{Owen2017}
{Owen}, J.~E. \& {Wu}, Y. 2017, \apj, 847, 29

\bibitem[{{Parc} {et~al.}(2024){Parc}, {Bouchy}, {Venturini}, {Dorn}, \& {Helled}}]{Parc2024}
{Parc}, L., {Bouchy}, F., {Venturini}, J., {Dorn}, C., \& {Helled}, R. 2024, \aap, 688, A59

\bibitem[{{Pascucci} {et~al.}(2016){Pascucci}, {Testi}, {Herczeg}, {Long}, {Manara}, {Hendler}, {Mulders}, {Krijt}, {Ciesla}, {Henning}, {Mohanty}, {Drabek-Maunder}, {Apai}, {Sz{\H{u}}cs}, {Sacco}, \& {Olofsson}}]{Pascucci2016}
{Pascucci}, I., {Testi}, L., {Herczeg}, G.~J., {et~al.} 2016, \apj, 831, 125

\bibitem[{{Pass} {et~al.}(2023){Pass}, {Winters}, {Charbonneau}, {Irwin}, {Latham}, {Berlind}, {Calkins}, {Esquerdo}, \& {Mink}}]{Pass2023}
{Pass}, E.~K., {Winters}, J.~G., {Charbonneau}, D., {et~al.} 2023, \aj, 166, 11

\bibitem[{{Pecaut} \& {Mamajek}(2013)}]{Pecaut2013}
{Pecaut}, M.~J. \& {Mamajek}, E.~E. 2013, \apjs, 208, 9

\bibitem[{{Pepe} {et~al.}(2021){Pepe}, {Cristiani}, {Rebolo}, {Santos}, {Dekker}, {Cabral}, {Di Marcantonio}, {Figueira}, {Lo Curto}, {Lovis}, {Mayor}, {M{\'e}gevand}, {Molaro}, {Riva}, {Zapatero Osorio}, {Amate}, {Manescau}, {Pasquini}, {Zerbi}, {Adibekyan}, {Abreu}, {Affolter}, {Alibert}, {Aliverti}, {Allart}, {Allende Prieto}, {{\'A}lvarez}, {Alves}, {Avila}, {Baldini}, {Bandy}, {Barros}, {Benz}, {Bianco}, {Borsa}, {Bourrier}, {Bouchy}, {Broeg}, {Calderone}, {Cirami}, {Coelho}, {Conconi}, {Coretti}, {Cumani}, {Cupani}, {D'Odorico}, {Damasso}, {Deiries}, {Delabre}, {Demangeon}, {Dumusque}, {Ehrenreich}, {Faria}, {Fragoso}, {Genolet}, {Genoni}, {G{\'e}nova Santos}, {Gonz{\'a}lez Hern{\'a}ndez}, {Hughes}, {Iwert}, {Kerber}, {Knudstrup}, {Landoni}, {Lavie}, {Lillo-Box}, {Lizon}, {Maire}, {Martins}, {Mehner}, {Micela}, {Modigliani}, {Monteiro}, {Monteiro}, {Moschetti}, {Murphy}, {Nunes}, {Oggioni}, {Oliveira}, {Oshagh}, {Pall{\'e}}, {Pariani}, {Poretti}, {Rasilla}, {Rebord{\~a}o}, {Redaelli}, {Santana Tschudi},
  {Santin}, {Santos}, {S{\'e}gransan}, {Schmidt}, {Segovia}, {Sosnowska}, {Sozzetti}, {Sousa}, {Span{\`o}}, {Su{\'a}rez Mascare{\~n}o}, {Tabernero}, {Tenegi}, {Udry}, \& {Zanutta}}]{Pepe2021}
{Pepe}, F., {Cristiani}, S., {Rebolo}, R., {et~al.} 2021, \aap, 645, A96

\bibitem[{{Petigura} {et~al.}(2013){Petigura}, {Howard}, \& {Marcy}}]{Petigura2013}
{Petigura}, E.~A., {Howard}, A.~W., \& {Marcy}, G.~W. 2013, Proceedings of the National Academy of Science, 110, 19273

\bibitem[{{Plotnykov} \& {Valencia}(2020)}]{Plotnykov_2020}
{Plotnykov}, M. \& {Valencia}, D. 2020, \mnras, 499, 932

\bibitem[{{Plotnykov} \& {Valencia}(2024)}]{Plotnykov2024}
{Plotnykov}, M. \& {Valencia}, D. 2024, \mnras, 530, 3488

\bibitem[{{Pojmanski}(1997)}]{Pojmanski1997}
{Pojmanski}, G. 1997, \actaa, 47, 467

\bibitem[{{Rasio} \& {Ford}(1996)}]{Rasio1996}
{Rasio}, F.~A. \& {Ford}, E.~B. 1996, Science, 274, 954

\bibitem[{{Rauer} {et~al.}(2014){Rauer}, {Catala}, {Aerts}, {Appourchaux}, {Benz}, {Brandeker}, {Christensen-Dalsgaard}, {Deleuil}, {Gizon}, {Goupil}, {G{\"u}del}, {Janot-Pacheco}, {Mas-Hesse}, {Pagano}, {Piotto}, {Pollacco}, {Santos}, {Smith}, {Su{\'a}rez}, {Szab{\'o}}, {Udry}, {Adibekyan}, {Alibert}, {Almenara}, {Amaro-Seoane}, {Eiff}, {Asplund}, {Antonello}, {Barnes}, {Baudin}, {Belkacem}, {Bergemann}, {Bihain}, {Birch}, {Bonfils}, {Boisse}, {Bonomo}, {Borsa}, {Brand{\~a}o}, {Brocato}, {Brun}, {Burleigh}, {Burston}, {Cabrera}, {Cassisi}, {Chaplin}, {Charpinet}, {Chiappini}, {Church}, {Csizmadia}, {Cunha}, {Damasso}, {Davies}, {Deeg}, {D{\'\i}az}, {Dreizler}, {Dreyer}, {Eggenberger}, {Ehrenreich}, {Eigm{\"u}ller}, {Erikson}, {Farmer}, {Feltzing}, {de Oliveira Fialho}, {Figueira}, {Forveille}, {Fridlund}, {Garc{\'\i}a}, {Giommi}, {Giuffrida}, {Godolt}, {Gomes da Silva}, {Granzer}, {Grenfell}, {Grotsch-Noels}, {G{\"u}nther}, {Haswell}, {Hatzes}, {H{\'e}brard}, {Hekker}, {Helled}, {Heng}, {Jenkins},
  {Johansen}, {Khodachenko}, {Kislyakova}, {Kley}, {Kolb}, {Krivova}, {Kupka}, {Lammer}, {Lanza}, {Lebreton}, {Magrin}, {Marcos-Arenal}, {Marrese}, {Marques}, {Martins}, {Mathis}, {Mathur}, {Messina}, {Miglio}, {Montalban}, {Montalto}, {Monteiro}, {Moradi}, {Moravveji}, {Mordasini}, {Morel}, {Mortier}, {Nascimbeni}, {Nelson}, {Nielsen}, {Noack}, {Norton}, {Ofir}, {Oshagh}, {Ouazzani}, {P{\'a}pics}, {Parro}, {Petit}, {Plez}, {Poretti}, {Quirrenbach}, {Ragazzoni}, {Raimondo}, {Rainer}, {Reese}, {Redmer}, {Reffert}, {Rojas-Ayala}, {Roxburgh}, {Salmon}, {Santerne}, {Schneider}, {Schou}, {Schuh}, {Schunker}, {Silva-Valio}, {Silvotti}, {Skillen}, {Snellen}, {Sohl}, {Sousa}, {Sozzetti}, {Stello}, {Strassmeier}, {{\v{S}}vanda}, {Szab{\'o}}, {Tkachenko}, {Valencia}, {Van Grootel}, {Vauclair}, {Ventura}, {Wagner}, {Walton}, {Weingrill}, {Werner}, {Wheatley}, \& {Zwintz}}]{Rauer2014}
{Rauer}, H., {Catala}, C., {Aerts}, C., {et~al.} 2014, Experimental Astronomy, 38, 249

\bibitem[{{Raymond} \& {Izidoro}(2017)}]{Raymond2017}
{Raymond}, S.~N. \& {Izidoro}, A. 2017, \icarus, 297, 134

\bibitem[{{Reyl{\'e}} {et~al.}(2021){Reyl{\'e}}, {Jardine}, {Fouqu{\'e}}, {Caballero}, {Smart}, \& {Sozzetti}}]{Reyle2021}
{Reyl{\'e}}, C., {Jardine}, K., {Fouqu{\'e}}, P., {et~al.} 2021, \aap, 650, A201

\bibitem[{{Ribas} {et~al.}(2005){Ribas}, {Guinan}, {G{\"u}del}, \& {Audard}}]{Ribas2005}
{Ribas}, I., {Guinan}, E.~F., {G{\"u}del}, M., \& {Audard}, M. 2005, \apj, 622, 680

\bibitem[{{Ricker} {et~al.}(2014){Ricker}, {Winn}, {Vanderspek}, {Latham}, {Bakos}, {Bean}, {Berta-Thompson}, {Brown}, {Buchhave}, {Butler}, {Butler}, {Chaplin}, {Charbonneau}, {Christensen-Dalsgaard}, {Clampin}, {Deming}, {Doty}, {De Lee}, {Dressing}, {Dunham}, {Endl}, {Fressin}, {Ge}, {Henning}, {Holman}, {Howard}, {Ida}, {Jenkins}, {Jernigan}, {Johnson}, {Kaltenegger}, {Kawai}, {Kjeldsen}, {Laughlin}, {Levine}, {Lin}, {Lissauer}, {MacQueen}, {Marcy}, {McCullough}, {Morton}, {Narita}, {Paegert}, {Palle}, {Pepe}, {Pepper}, {Quirrenbach}, {Rinehart}, {Sasselov}, {Sato}, {Seager}, {Sozzetti}, {Stassun}, {Sullivan}, {Szentgyorgyi}, {Torres}, {Udry}, \& {Villasenor}}]{Ricker2014}
{Ricker}, G.~R., {Winn}, J.~N., {Vanderspek}, R., {et~al.} 2014, in Society of Photo-Optical Instrumentation Engineers (SPIE) Conference Series, Vol. 9143, Space Telescopes and Instrumentation 2014: Optical, Infrared, and Millimeter Wave, ed. J.~M. {Oschmann}, Jr., M.~{Clampin}, G.~G. {Fazio}, \& H.~A. {MacEwen}, 914320

\bibitem[{{Rossiter}(1924)}]{Rossiter1924}
{Rossiter}, R.~A. 1924, \apj, 60, 15

\bibitem[{{Rousset} {et~al.}(2003){Rousset}, {Lacombe}, {Puget}, {Hubin}, {Gendron}, {Fusco}, {Arsenault}, {Charton}, {Feautrier}, {Gigan}, {Kern}, {Lagrange}, {Madec}, {Mouillet}, {Rabaud}, {Rabou}, {Stadler}, \& {Zins}}]{Rousset2003}
{Rousset}, G., {Lacombe}, F., {Puget}, P., {et~al.} 2003, in Society of Photo-Optical Instrumentation Engineers (SPIE) Conference Series, Vol. 4839, Adaptive Optical System Technologies II, ed. P.~L. {Wizinowich} \& D.~{Bonaccini}, 140--149

\bibitem[{{Saumon} {et~al.}(1995){Saumon}, {Chabrier}, \& {van Horn}}]{Saumon_1995}
{Saumon}, D., {Chabrier}, G., \& {van Horn}, H.~M. 1995, \apjs, 99, 713

\bibitem[{{Schlecker} {et~al.}(2021){Schlecker}, {Mordasini}, {Emsenhuber}, {Klahr}, {Henning}, {Burn}, {Alibert}, \& {Benz}}]{Schlecker2021}
{Schlecker}, M., {Mordasini}, C., {Emsenhuber}, A., {et~al.} 2021, \aap, 656, A71

\bibitem[{{Sch{\"o}nrich} {et~al.}(2010){Sch{\"o}nrich}, {Binney}, \& {Dehnen}}]{Schronrich2010}
{Sch{\"o}nrich}, R., {Binney}, J., \& {Dehnen}, W. 2010, \mnras, 403, 1829

\bibitem[{{Schweitzer} {et~al.}(2019){Schweitzer}, {Passegger}, {Cifuentes}, {B{\'e}jar}, {Cort{\'e}s-Contreras}, {Caballero}, {del Burgo}, {Czesla}, {K{\"u}rster}, {Montes}, {Zapatero Osorio}, {Ribas}, {Reiners}, {Quirrenbach}, {Amado}, {Aceituno}, {Anglada-Escud{\'e}}, {Bauer}, {Dreizler}, {Jeffers}, {Guenther}, {Henning}, {Kaminski}, {Lafarga}, {Marfil}, {Morales}, {Schmitt}, {Seifert}, {Solano}, {Tabernero}, \& {Zechmeister}}]{Schweitzer2019}
{Schweitzer}, A., {Passegger}, V.~M., {Cifuentes}, C., {et~al.} 2019, \aap, 625, A68

\bibitem[{{Skrutskie} {et~al.}(2006){Skrutskie}, {Cutri}, {Stiening}, {Weinberg}, {Schneider}, {Carpenter}, {Beichman}, {Capps}, {Chester}, {Elias}, {Huchra}, {Liebert}, {Lonsdale}, {Monet}, {Price}, {Seitzer}, {Jarrett}, {Kirkpatrick}, {Gizis}, {Howard}, {Evans}, {Fowler}, {Fullmer}, {Hurt}, {Light}, {Kopan}, {Marsh}, {McCallon}, {Tam}, {Van Dyk}, \& {Wheelock}}]{2MASS2006}
{Skrutskie}, M.~F., {Cutri}, R.~M., {Stiening}, R., {et~al.} 2006, \aj, 131, 1163

\bibitem[{{Smith} {et~al.}(2012){Smith}, {Stumpe}, {Van Cleve}, {Jenkins}, {Barclay}, {Fanelli}, {Girouard}, {Kolodziejczak}, {McCauliff}, {Morris}, \& {Twicken}}]{Smith2012}
{Smith}, J.~C., {Stumpe}, M.~C., {Van Cleve}, J.~E., {et~al.} 2012, \pasp, 124, 1000

\bibitem[{{Sousa} {et~al.}(2021){Sousa}, {Adibekyan}, {Delgado-Mena}, {Santos}, {Rojas-Ayala}, {Soares}, {Legoinha}, {Ulmer-Moll}, {Camacho}, {Barros}, {Demangeon}, {Hoyer}, {Israelian}, {Mortier}, {Tsantaki}, \& {Monteiro}}]{Sousa2021}
{Sousa}, S.~G., {Adibekyan}, V., {Delgado-Mena}, E., {et~al.} 2021, \aap, 656, A53

\bibitem[{{Speagle}(2020)}]{Speagle2020_dynesty}
{Speagle}, J.~S. 2020, \mnras, 493, 3132

\bibitem[{{Stassun} {et~al.}(2019){Stassun}, {Oelkers}, {Paegert}, {Torres}, {Pepper}, {De Lee}, {Collins}, {Latham}, {Muirhead}, {Chittidi}, {Rojas-Ayala}, {Fleming}, {Rose}, {Tenenbaum}, {Ting}, {Kane}, {Barclay}, {Bean}, {Brassuer}, {Charbonneau}, {Ge}, {Lissauer}, {Mann}, {McLean}, {Mullally}, {Narita}, {Plavchan}, {Ricker}, {Sasselov}, {Seager}, {Sharma}, {Shiao}, {Sozzetti}, {Stello}, {Vanderspek}, {Wallace}, \& {Winn}}]{Stassun2019}
{Stassun}, K.~G., {Oelkers}, R.~J., {Paegert}, M., {et~al.} 2019, \aj, 158, 138

\bibitem[{{Stumpe} {et~al.}(2014){Stumpe}, {Smith}, {Catanzarite}, {Van Cleve}, {Jenkins}, {Twicken}, \& {Girouard}}]{Stumpe2014}
{Stumpe}, M.~C., {Smith}, J.~C., {Catanzarite}, J.~H., {et~al.} 2014, \pasp, 126, 100

\bibitem[{{Stumpe} {et~al.}(2012){Stumpe}, {Smith}, {Van Cleve}, {Twicken}, {Barclay}, {Fanelli}, {Girouard}, {Jenkins}, {Kolodziejczak}, {McCauliff}, \& {Morris}}]{Stumpe2012}
{Stumpe}, M.~C., {Smith}, J.~C., {Van Cleve}, J.~E., {et~al.} 2012, \pasp, 124, 985

\bibitem[{{Su{\'a}rez Mascare{\~n}o} {et~al.}(2025){Su{\'a}rez Mascare{\~n}o}, {Artigau}, {Mignon}, {Delfosse}, \& {Cook}}]{Suarez2025}
{Su{\'a}rez Mascare{\~n}o}, A., {Artigau}, {\'E}., {Mignon}, L., {Delfosse}, X., \& {Cook}, N.~J. 2025, \aap

\bibitem[{{Su{\'a}rez Mascare{\~n}o} {et~al.}(2016){Su{\'a}rez Mascare{\~n}o}, {Rebolo}, \& {Gonz{\'a}lez Hern{\'a}ndez}}]{Suarez2016}
{Su{\'a}rez Mascare{\~n}o}, A., {Rebolo}, R., \& {Gonz{\'a}lez Hern{\'a}ndez}, J.~I. 2016, \aap, 595, A12

\bibitem[{{Su{\'a}rez Mascare{\~n}o} {et~al.}(2015){Su{\'a}rez Mascare{\~n}o}, {Rebolo}, {Gonz{\'a}lez Hern{\'a}ndez}, \& {Esposito}}]{Suarez2015}
{Su{\'a}rez Mascare{\~n}o}, A., {Rebolo}, R., {Gonz{\'a}lez Hern{\'a}ndez}, J.~I., \& {Esposito}, M. 2015, \mnras, 452, 2745

\bibitem[{{Sullivan} {et~al.}(2024){Sullivan}, {Kraus}, {Berger}, {Dupuy}, {Evans}, {Gaidos}, {Huber}, {Ireland}, {Mann}, {Petigura}, {Thao}, {Wood}, \& {Zhang}}]{Sullivan2024}
{Sullivan}, K., {Kraus}, A.~L., {Berger}, T.~A., {et~al.} 2024, \aj, 168, 129

\bibitem[{{Sullivan} {et~al.}(2023){Sullivan}, {Kraus}, {Huber}, {Petigura}, {Evans}, {Dupuy}, {Zhang}, {Berger}, {Gaidos}, \& {Mann}}]{Sullivan2023}
{Sullivan}, K., {Kraus}, A.~L., {Huber}, D., {et~al.} 2023, \aj, 165, 177

\bibitem[{{Tian} {et~al.}(2005){Tian}, {Toon}, {Pavlov}, \& {De Sterck}}]{Tian2005}
{Tian}, F., {Toon}, O.~B., {Pavlov}, A.~A., \& {De Sterck}, H. 2005, \apj, 621, 1049

\bibitem[{{Tokovinin}(2018)}]{Tokovinin2018}
{Tokovinin}, A. 2018, \pasp, 130, 035002

\bibitem[{{Torres}(1999)}]{Torres1999}
{Torres}, G. 1999, \pasp, 111, 169

\bibitem[{{Twicken} {et~al.}(2018){Twicken}, {Catanzarite}, {Clarke}, {Girouard}, {Jenkins}, {Klaus}, {Li}, {McCauliff}, {Seader}, {Tenenbaum}, {Wohler}, {Bryson}, {Burke}, {Caldwell}, {Haas}, {Henze}, \& {Sanderfer}}]{Twicken2018}
{Twicken}, J.~D., {Catanzarite}, J.~H., {Clarke}, B.~D., {et~al.} 2018, \pasp, 130, 064502

\bibitem[{{Twicken} {et~al.}(2010){Twicken}, {Clarke}, {Bryson}, {Tenenbaum}, {Wu}, {Jenkins}, {Girouard}, \& {Klaus}}]{Twicken2010}
{Twicken}, J.~D., {Clarke}, B.~D., {Bryson}, S.~T., {et~al.} 2010, in Society of Photo-Optical Instrumentation Engineers (SPIE) Conference Series, Vol. 7740, Software and Cyberinfrastructure for Astronomy, ed. N.~M. {Radziwill} \& A.~{Bridger}, 774023

\bibitem[{{Valencia} {et~al.}(2007){Valencia}, {Sasselov}, \& {O'Connell}}]{Valencia_2007}
{Valencia}, D., {Sasselov}, D.~D., \& {O'Connell}, R.~J. 2007, \apj, 656, 545

\bibitem[{{Venturini} {et~al.}(2020){Venturini}, {Guilera}, {Haldemann}, {Ronco}, \& {Mordasini}}]{Venturini2020}
{Venturini}, J., {Guilera}, O.~M., {Haldemann}, J., {Ronco}, M.~P., \& {Mordasini}, C. 2020, \aap, 643, L1

\bibitem[{{Venturini} {et~al.}(2024){Venturini}, {Ronco}, {Guilera}, {Haldemann}, {Mordasini}, \& {Miller Bertolami}}]{Venturini2024}
{Venturini}, J., {Ronco}, M.~P., {Guilera}, O.~M., {et~al.} 2024, \aap, 686, L9

\bibitem[{{Wallace} {et~al.}(2025){Wallace}, {Casey}, {Brown}, \& {Castro-Ginard}}]{Wallace2025}
{Wallace}, A.~L., {Casey}, A.~R., {Brown}, A.~G.~A., \& {Castro-Ginard}, A. 2025, \mnras, 536, 2485

\bibitem[{{Weiss} {et~al.}(2024){Weiss}, {Isaacson}, {Howard}, {Fulton}, {Petigura}, {Fabrycky}, {Jontof-Hutter}, {Steffen}, {Schlichting}, {Wright}, {Beard}, {Brinkman}, {Chontos}, {Giacalone}, {Hill}, {Kosiarek}, {MacDougall}, {Mo{\v{c}}nik}, {Polanski}, {Turtelboom}, {Tyler}, \& {Van Zandt}}]{Weiss2024}
{Weiss}, L.~M., {Isaacson}, H., {Howard}, A.~W., {et~al.} 2024, \apjs, 270, 8

\bibitem[{Wilcoxon(1945)}]{Wilcoxon1945}
Wilcoxon, F. 1945, Biometrics Bulletin, 1, 80

\bibitem[{{Wilson} {et~al.}(2022){Wilson}, {Goffo}, {Alibert}, {Gandolfi}, {Bonfanti}, {Persson}, {Collier Cameron}, {Fridlund}, {Fossati}, {Korth}, {Benz}, {Deline}, {Flor{\'e}n}, {Guterman}, {Adibekyan}, {Hooton}, {Hoyer}, {Leleu}, {Mustill}, {Salmon}, {Sousa}, {Suarez}, {Abe}, {Agabi}, {Alonso}, {Anglada}, {Asquier}, {B{\'a}rczy}, {Barrado Navascues}, {Barros}, {Baumjohann}, {Beck}, {Beck}, {Billot}, {Bonfils}, {Brandeker}, {Broeg}, {Bryant}, {Burleigh}, {Buttu}, {Cabrera}, {Charnoz}, {Ciardi}, {Cloutier}, {Cochran}, {Collins}, {Col{\'o}n}, {Crouzet}, {Csizmadia}, {Davies}, {Deleuil}, {Delrez}, {Demangeon}, {Demory}, {Dragomir}, {Dransfield}, {Ehrenreich}, {Erikson}, {Fortier}, {Gan}, {Gill}, {Gillon}, {Gnilka}, {Grieves}, {Grziwa}, {G{\"u}del}, {Guillot}, {Haldemann}, {Heng}, {Horne}, {Howell}, {Isaak}, {Jenkins}, {Jensen}, {Kiss}, {Lacedelli}, {Lam}, {Laskar}, {Latham}, {Lecavelier des Etangs}, {Lendl}, {Lester}, {Levine}, {Livingston}, {Lovis}, {Luque}, {Magrin}, {Marie-Sainte}, {Maxted}, {Mayo},
  {McLean}, {Mecina}, {M{\'e}karnia}, {Nascimbeni}, {Nielsen}, {Olofsson}, {Osborn}, {Osborne}, {Ottensamer}, {Pagano}, {Pall{\'e}}, {Peter}, {Piotto}, {Pollacco}, {Queloz}, {Ragazzoni}, {Rando}, {Rauer}, {Redfield}, {Ribas}, {Ricker}, {Rieder}, {Santos}, {Scandariato}, {Schmider}, {Schwarz}, {Scott}, {Seager}, {S{\'e}gransan}, {Serrano}, {Simon}, {Smith}, {Steller}, {Stockdale}, {Szab{\'o}}, {Thomas}, {Ting}, {Triaud}, {Udry}, {Van Eylen}, {Van Grootel}, {Vanderspek}, {Viotto}, {Walton}, \& {Winn}}]{Wilson2022}
{Wilson}, T.~G., {Goffo}, E., {Alibert}, Y., {et~al.} 2022, \mnras, 511, 1043

\bibitem[{{Winters} {et~al.}(2015){Winters}, {Henry}, {Lurie}, {Hambly}, {Jao}, {Bartlett}, {Boyd}, {Dieterich}, {Finch}, {Hosey}, {Ianna}, {Riedel}, {Slatten}, \& {Subasavage}}]{Winters2015}
{Winters}, J.~G., {Henry}, T.~J., {Lurie}, J.~C., {et~al.} 2015, \aj, 149, 5

\bibitem[{{Wright} {et~al.}(2010){Wright}, {Eisenhardt}, {Mainzer}, {Ressler}, {Cutri}, {Jarrett}, {Kirkpatrick}, {Padgett}, {McMillan}, {Skrutskie}, {Stanford}, {Cohen}, {Walker}, {Mather}, {Leisawitz}, {Gautier}, {McLean}, {Benford}, {Lonsdale}, {Blain}, {Mendez}, {Irace}, {Duval}, {Liu}, {Royer}, {Heinrichsen}, {Howard}, {Shannon}, {Kendall}, {Walsh}, {Larsen}, {Cardon}, {Schick}, {Schwalm}, {Abid}, {Fabinsky}, {Naes}, \& {Tsai}}]{WISE2010}
{Wright}, E.~L., {Eisenhardt}, P. R.~M., {Mainzer}, A.~K., {et~al.} 2010, \aj, 140, 1868

\bibitem[{{Wroblewski} \& {Torres}(1991)}]{Wroblewski1991}
{Wroblewski}, H. \& {Torres}, C. 1991, \aaps, 91, 129

\bibitem[{{Wu} \& {Lithwick}(2011)}]{Wu2011}
{Wu}, Y. \& {Lithwick}, Y. 2011, \apj, 735, 109

\bibitem[{{Wu} \& {Murray}(2003)}]{Wu2003}
{Wu}, Y. \& {Murray}, N. 2003, \apj, 589, 605

\bibitem[{{Yelle}(2004)}]{Yelle2004}
{Yelle}, R.~V. 2004, \icarus, 170, 167

\bibitem[{{Zeng} {et~al.}(2019){Zeng}, {Jacobsen}, {Sasselov}, {Petaev}, {Vanderburg}, {Lopez-Morales}, {Perez-Mercader}, {Mattsson}, {Li}, {Heising}, {Bonomo}, {Damasso}, {Berger}, {Cao}, {Levi}, \& {Wordsworth}}]{Zeng2019}
{Zeng}, L., {Jacobsen}, S.~B., {Sasselov}, D.~D., {et~al.} 2019, Proceedings of the National Academy of Science, 116, 9723

\bibitem[{{Zeng} {et~al.}(2016){Zeng}, {Sasselov}, \& {Jacobsen}}]{Zeng2016}
{Zeng}, L., {Sasselov}, D.~D., \& {Jacobsen}, S.~B. 2016, \apj, 819, 127

\bibitem[{{Ziegler} {et~al.}(2020){Ziegler}, {Tokovinin}, {Brice{\~n}o}, {Mang}, {Law}, \& {Mann}}]{Ziegler2020}
{Ziegler}, C., {Tokovinin}, A., {Brice{\~n}o}, C., {et~al.} 2020, \aj, 159, 19

\end{thebibliography}

\begin{appendix}

\onecolumn
\section{TESS pixel file plots}
In this appendix, we show the TESS pixel file plots for all the sectors observed of TOI-756 (except Sector 10 shown in Fig.~\ref{fig:TPF}).
 
\begin{figure*}[h]
  \centering
    \includegraphics[width=0.45\textwidth]{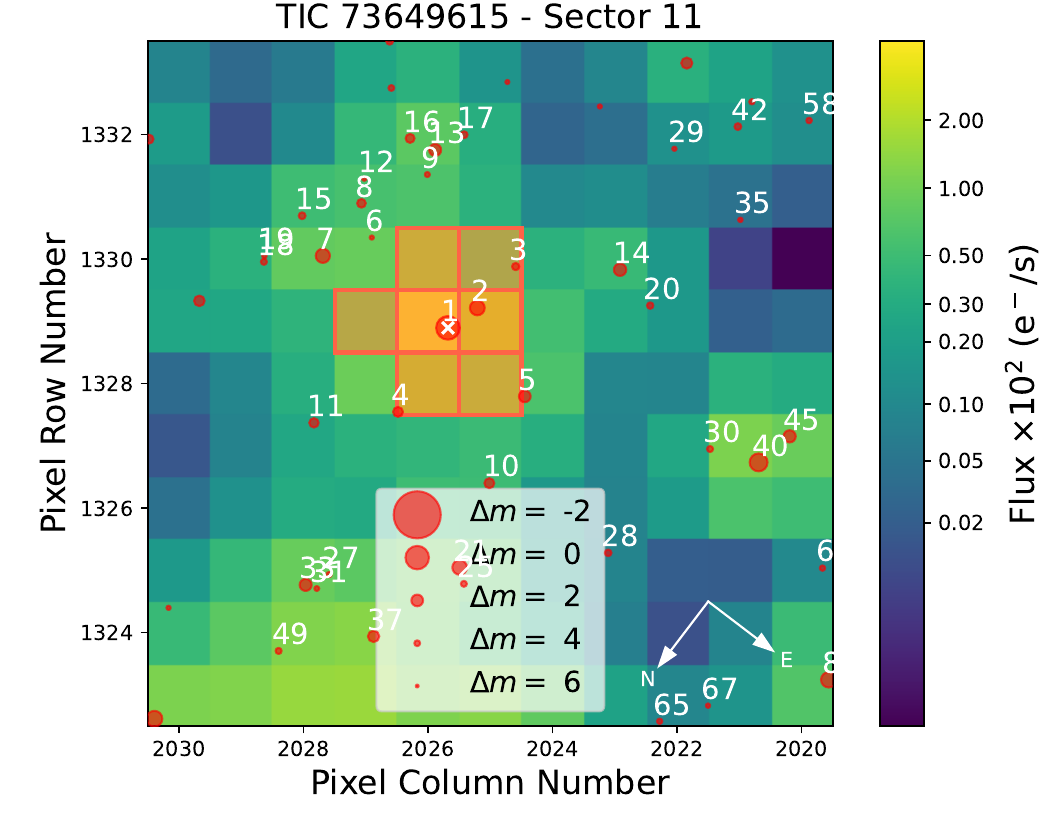}
    \includegraphics[width=0.45\textwidth]{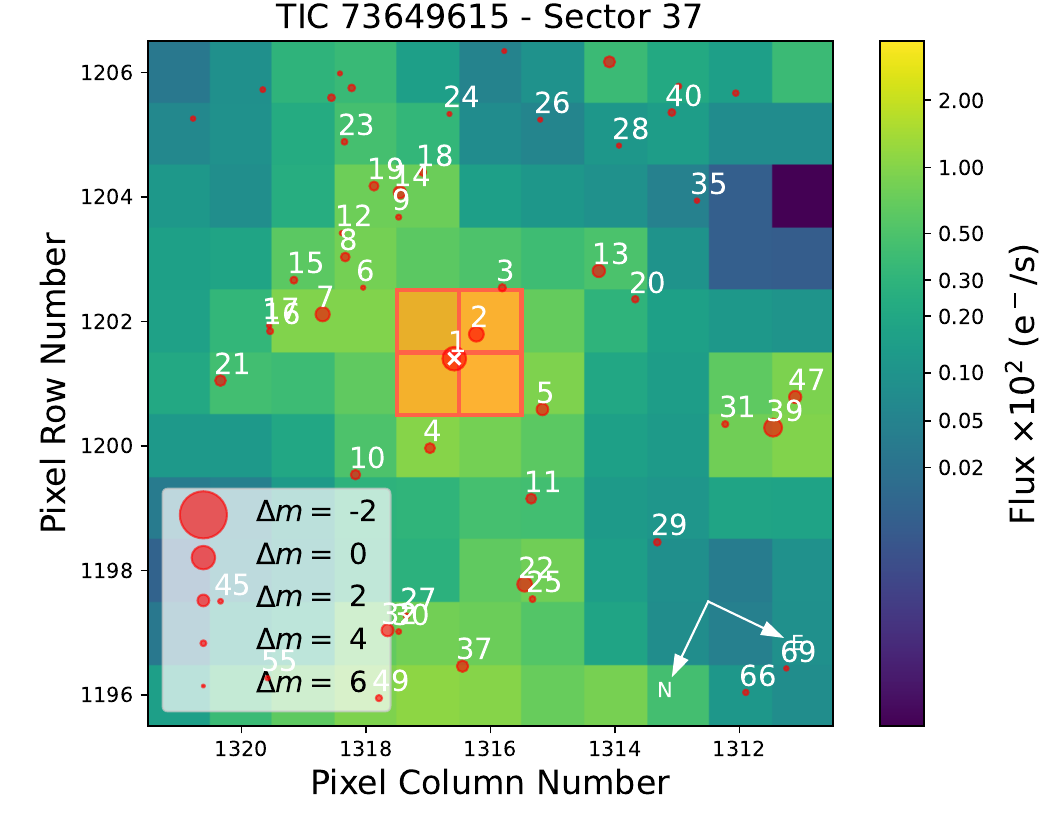}
    \includegraphics[width=0.45\textwidth]{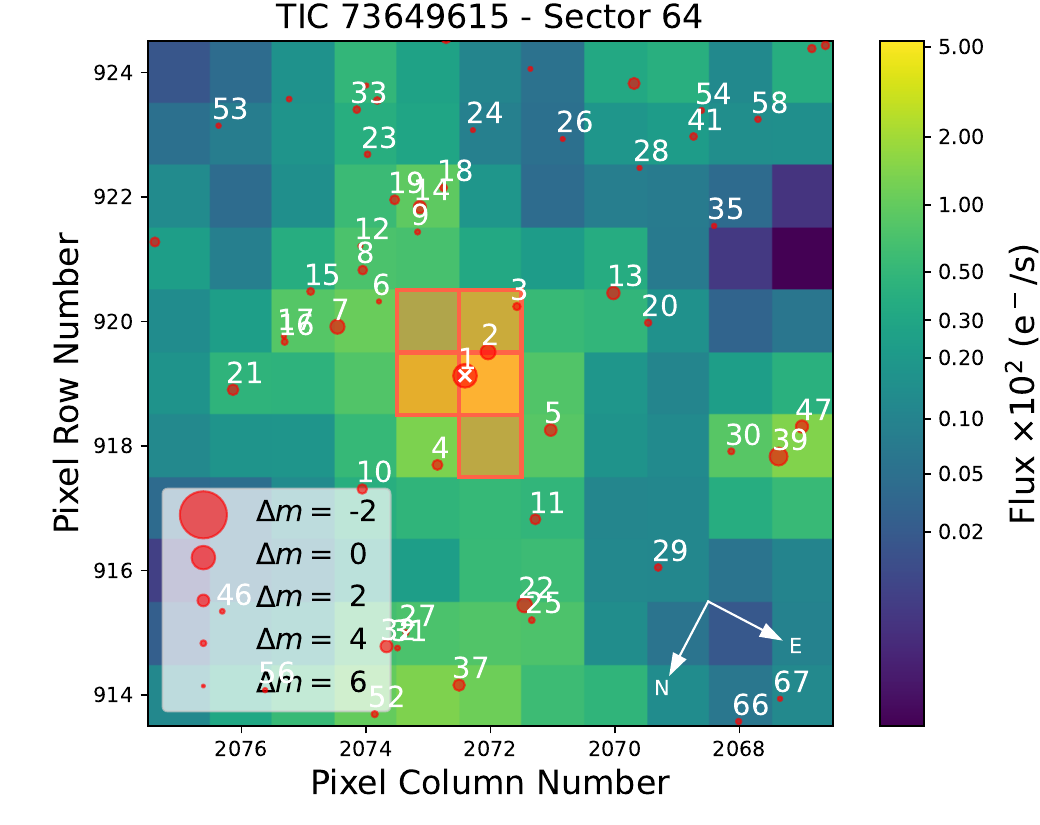}
  \caption{TESS TPF of TOI-756 created with \texttt{tpfplotter} (\citealt{Aller2020}). The orange pixels define the aperture mask used for extracting the photometry. Additionally, the red circles indicate neighboring objects from the Gaia DR3 catalog, with the circle size corresponding to the brightness difference compared to the target (as indicated in the legend). Our target is marked with a white cross. Pixel scale is 21$\arcsec$/pixel. The co-moving companion of TOI-756 corresponds to the star labeled "2".} 
  \label{appendix:TPF}
\end{figure*}

\newpage
\section{High-contrast imaging observations}

\begin{figure*}[h]
\centering
    \includegraphics[width=7cm]{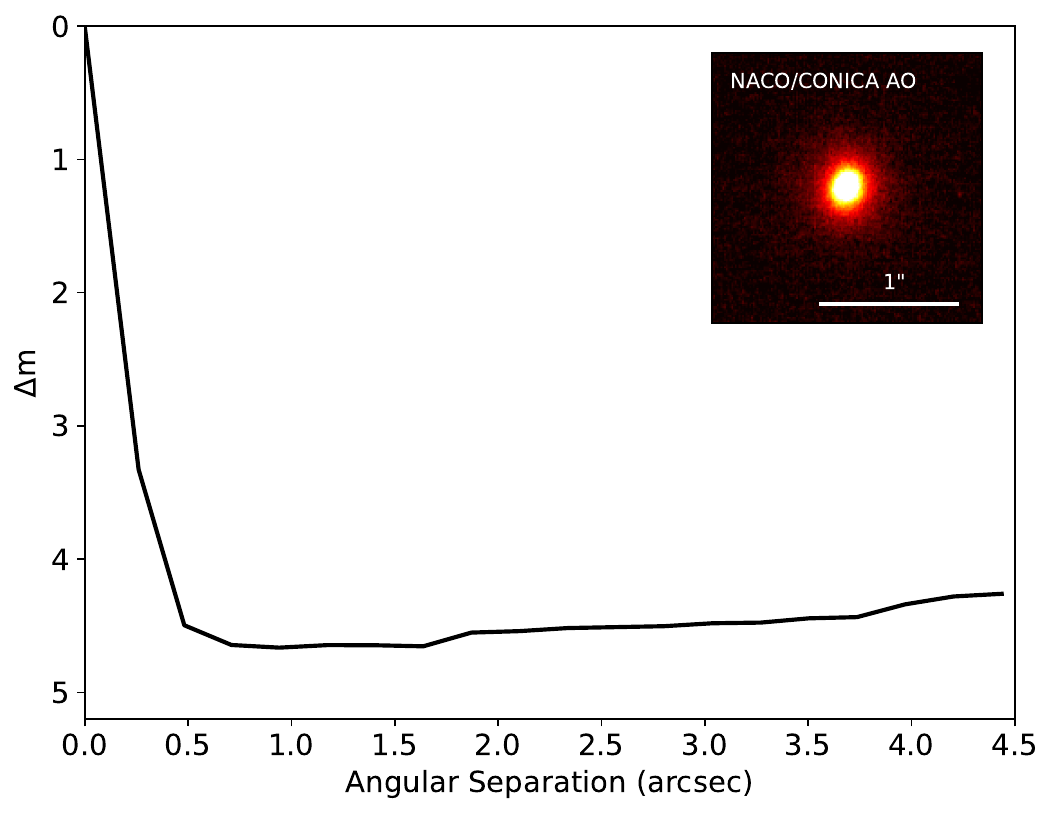}
   \includegraphics[width=7cm]{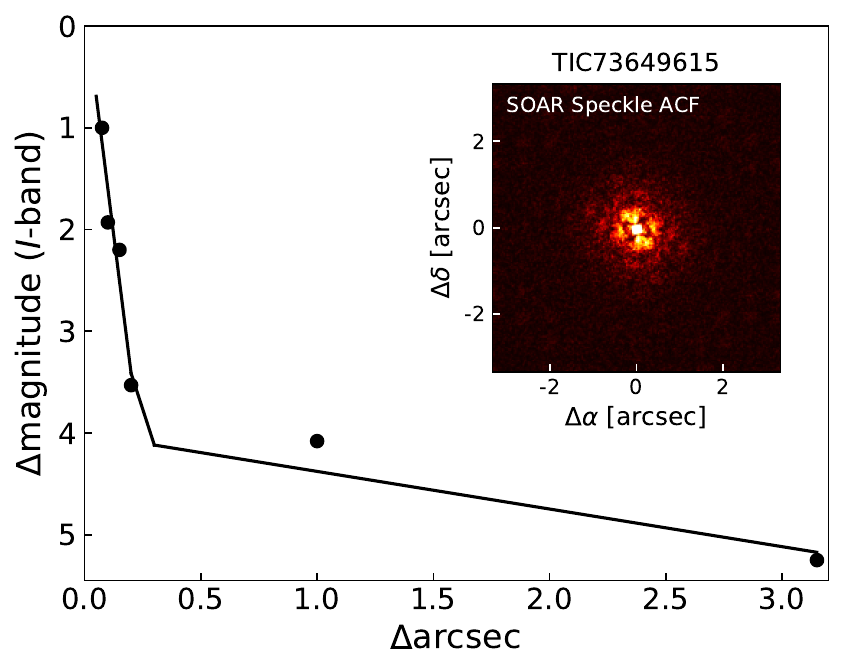}
   \includegraphics[width=7cm]{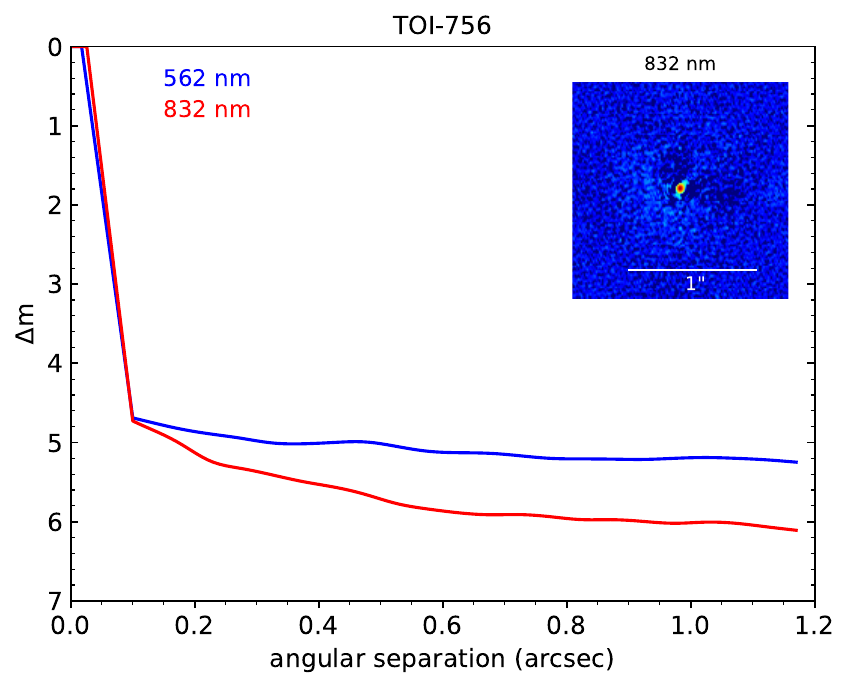}
   \includegraphics[width=7cm]{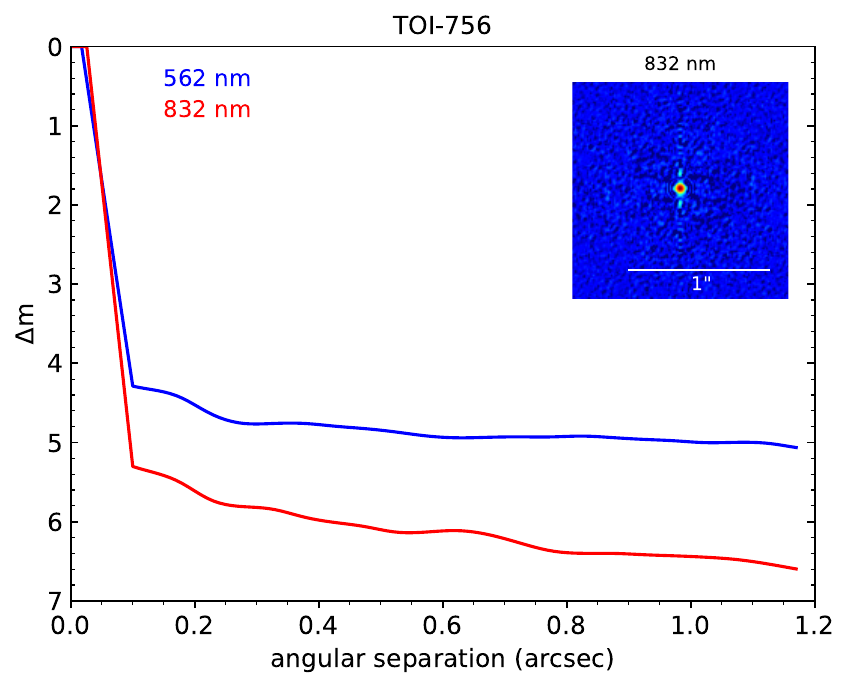}
     \caption{Adaptive optics and speckle imaging plots for TOI-756 showing magnitude contrast in function of angular separation. Top left: NACO/CONICA@VLT. Top right: HRCam@SOAR. Bottom: Zorro@Gemini-South (left: March 12, 2020 ; right:  July 05, 2023). For VLT and Gemini, the inset image is the primary target showing no additional close-in companions. For SOAR, the inset image shows the auto-correlation function.}
     \label{fig:HighRes}
\end{figure*}

\section{Photometric and radial velocity analysis}\label{appendix:tess_fit}

\begin{table*}[h]
\tiny
\caption{Median values and 68$\%$ confidence intervals of the posterior distributions of the TESS-only fit.}
\centering
\renewcommand{\arraystretch}{1.1}
\setlength{\tabcolsep}{38pt}
\begin{center}
\begin{tabular}{lcc}
\hline\hline
\textbf{Parameter} & \textbf{Prior} & \textbf{Value} \\
\hline
\textbf{Stellar parameters} & & \\
Stellar density, $\rho_*$ (\denssol)\dotfill  & $\mathcal{N}(5350, 230)$ & $5291^{+156} _{-163}$ \\
\textbf{Parameters for TOI-756 b} & & \\
Orbital period, $P$ (days)\dotfill  & $\mathcal{N}(1.23926, 0.1)$ & $1.239250\pm0.000001$\\
Semi-major axis, $a$ (AU)\dotfill  & - & $0.0182\pm0.0003$\\
Transit epoch, $T_{0}$ (BJD)\dotfill & $\mathcal{N}(2458570.65, 0.1)$ & $58570.65187^{+0.00067} _{-0.00073}$\\
Scaled planetary radius, $R_{P}$/$R_{*}$\dotfill & $\mathcal{U}(0, 1)$ & $0.0486\pm0.0013$ \\
Impact parameter, $b$\dotfill  & $\mathcal{U}(0, 1)$ & $0.541^{+0.037} _{-0.047}$ \\
Inclination, $\textit{i}$ (deg)\dotfill & - & $85.92^{+0.39} _{-0.31}$ \\
Eccentricity, $e$\dotfill & Fixed & $0.0$ \\
Argument of periastron, $\omega$ (deg)\dotfill & Fixed & $90.0$ \\
\textbf{Limb darkening parameters} & & \\
Limb darkening parameter, $q_{1,\text{TESS}}$\dotfill & $\mathcal{N}(0.792, 0.029)$ & $0.795\pm0.029$ \\
Limb darkening parameter, $q_{2,\text{TESS}}$\dotfill & $\mathcal{N}(0.453, 0.022)$ & $0.455\pm0.021$\\
\hline
\end{tabular}
\begin{tablenotes}
\item
\textbf{Notes:} $\mathcal{N}(\mu, \sigma^{2})$ indicates a normal distribution with mean $\mu$ and variance $\sigma^{2}$, $\mathcal{U}(a, b)$ a uniform distribution between $a$ and $b$.
\end{tablenotes}
\label{tab:posteriors_tess}
\end{center}
\end{table*}

\begin{figure*}[h]
\centering
    \includegraphics[width=1.\textwidth]{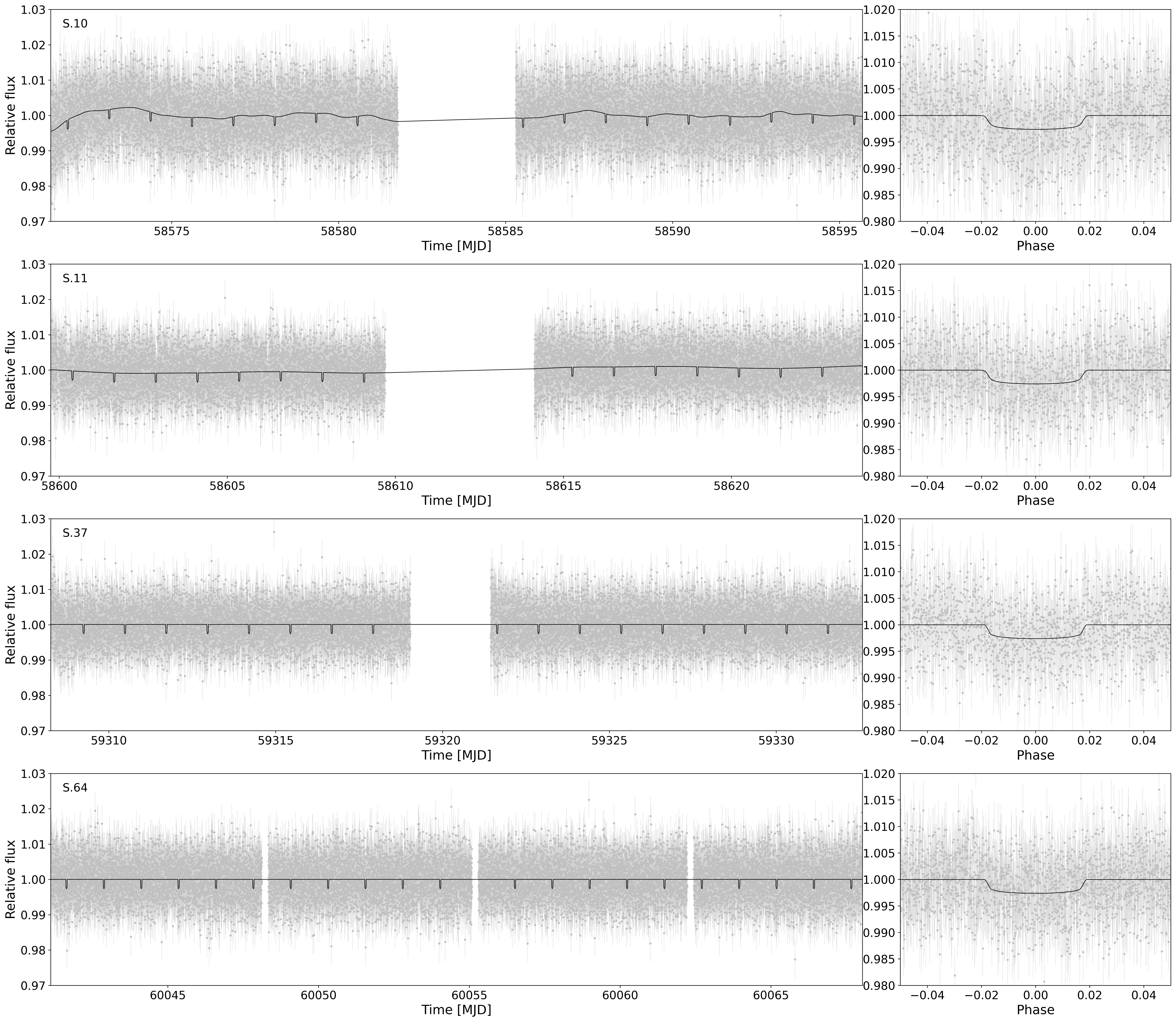}
     \caption{TESS PDCSAP flux light curves of the four different sectors with the best-fit \texttt{juliet} model shown as a black line (see Sect.~\ref{sect:joint_fit_juliet} for details on the modeling).}
     \label{fig:TESS_fit}
\end{figure*}

\begin{figure}[h]
\centering
    \includegraphics[width=0.45\textwidth]{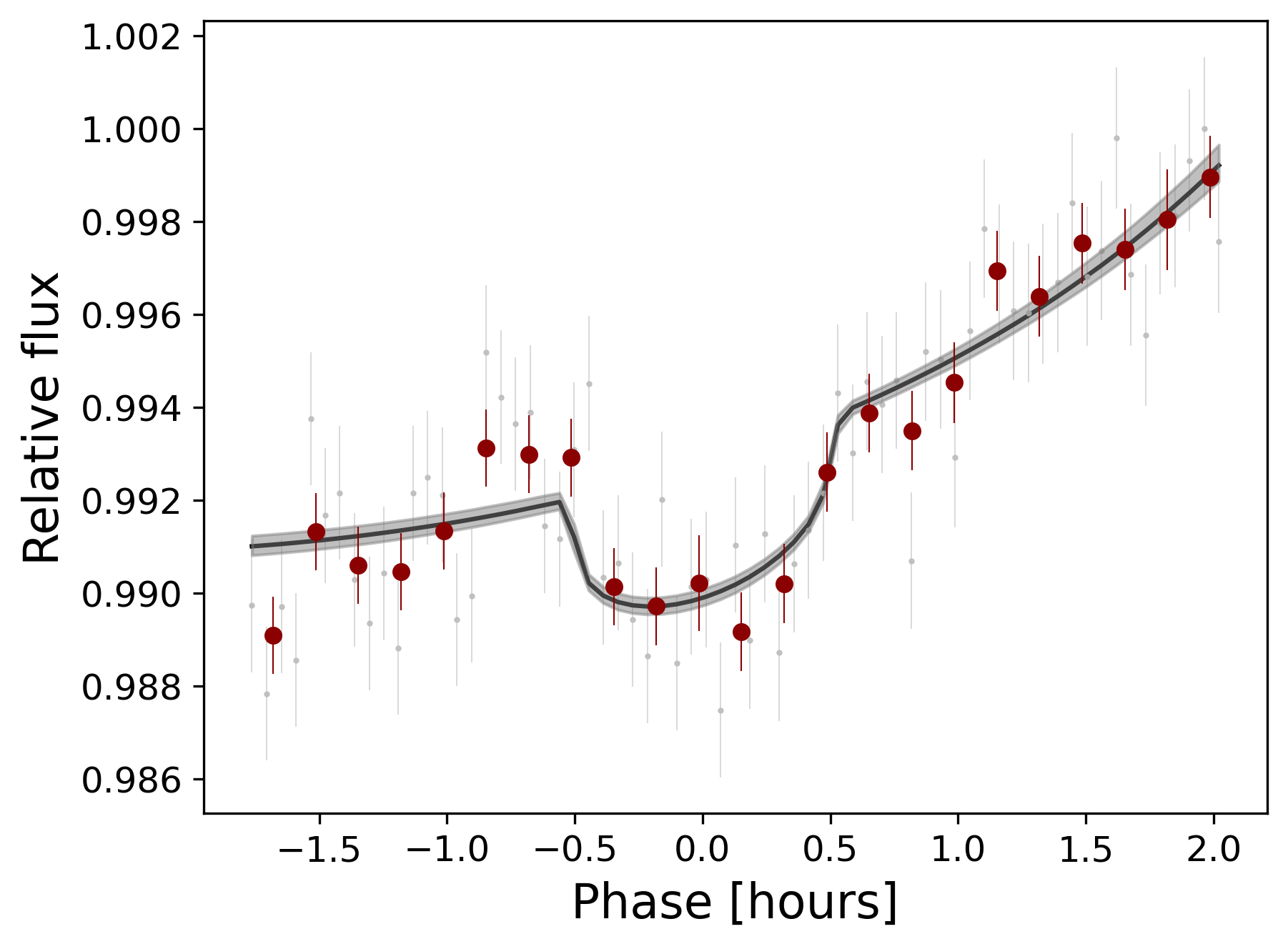}
     \caption{LCO-CTIO light curve from the $g'$-band transit with the best-fit \texttt{juliet} model shown as a black line and model errors as gray area. Dark red circles are data binned to 10 min (see Sect.~\ref{sect:joint_fit_juliet} for details on the modeling). We do not present the $i'$-band transit here, as no detrending was applied, making it identical to the one shown in Fig.~\ref{fig:Photometry}.}
     \label{fig:LCOT2_fit}
\end{figure}

\begin{figure*}[h]
\centering
    \includegraphics[width=1.\textwidth]{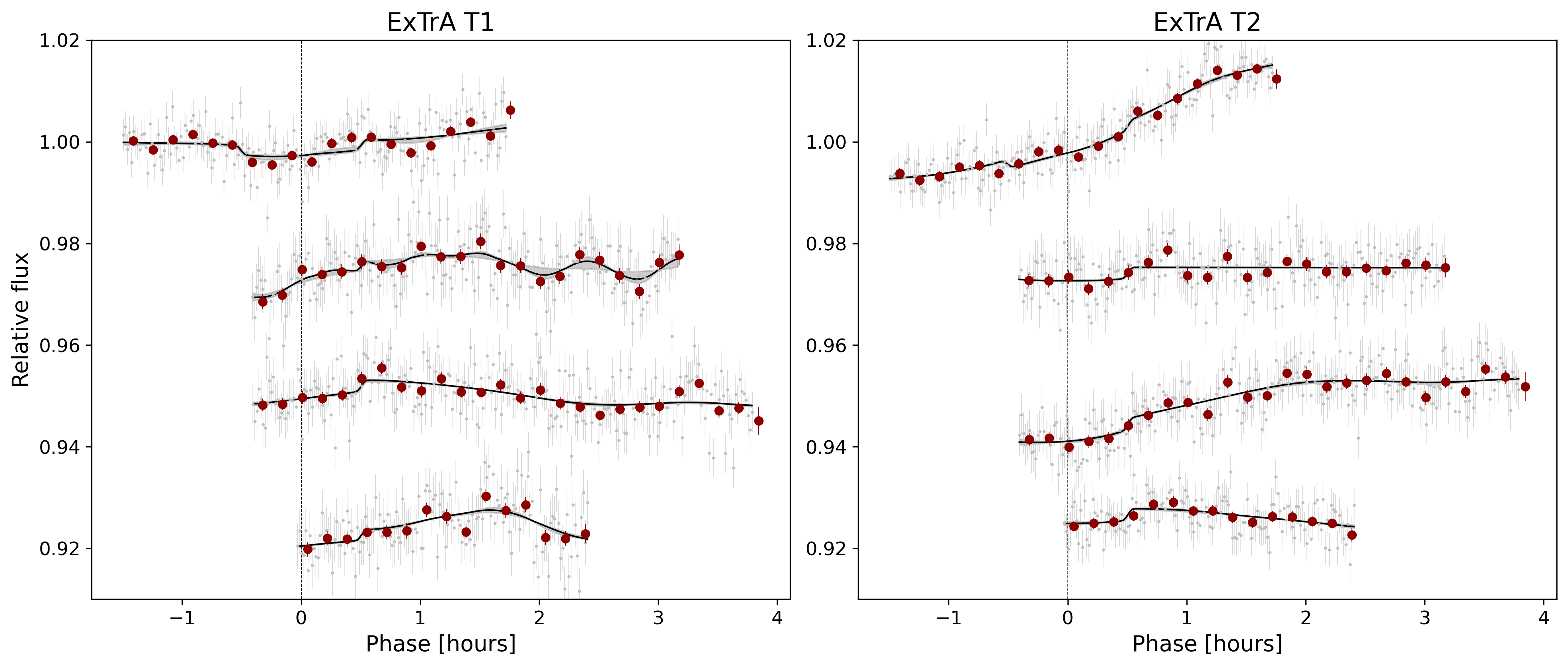}
     \caption{ExTrA light curves of the four different transits, observed with two telescopes (left panel: first telescope; right panel: second telescope). From top to bottom, the first transit is complete, while the remaining three show only the egress. The best-fit \texttt{juliet} models are shown as black lines, with 1$\sigma$ model uncertainties indicated by the gray shaded regions. Dark red circles represent the data binned to 10 minutes. The dashed vertical line define the transit midpoint. An arbitrary offset has been added between the transits for clarity. See Sect.~\ref{sect:joint_fit_juliet} for details on the modeling. }
     \label{fig:ExTrA_fit}
\end{figure*}

\onecolumn
\renewcommand{\arraystretch}{1.1}
\setlength{\tabcolsep}{28pt}
\begin{longtable}[t]{l c c}

\caption{Median values and 68$\%$ confidence intervals of the posterior distributions of the joint fit.\label{tab:posteriors_jointfit}}\\
\hline\hline
\textbf{Parameter} & \textbf{Prior} & \textbf{Value} \\
\hline
\\
\endfirsthead
\caption{continued.}\\
\hline\hline
\textbf{Parameter} & \textbf{Prior} & \textbf{Value} \\
\hline
\\
\endhead

\\
\hline
\endfoot

\\
\hline\hline
\endlastfoot

\textbf{Stellar parameters} & & \\
\\

Stellar density, $\rho_*$ (\denssol)\dotfill  & $\mathcal{N}(5350, 230)$ & $5298^{+164} _{-169}$ \\

\\
\textbf{Fitted parameters for TOI-756 b} & & \\
\\

Orbital period, $P_b$ (days)\dotfill  & $\mathcal{N}(1.23925, 0.00001)$ & $1.23924949^{+0.00000068} _{-0.00000063}$\\
Transit epoch, $T_{0,b}$ (BJD)\dotfill & $\mathcal{N}(2458570.652, 0.001)$ & $2458570.65234^{+0.00035} _{-0.00037}$\\
$r_1$ \dotfill & $\mathcal{N}(0.70,0.1)$ & $0.726^{+0.012} _{-0.014}$ \\
Scaled planetary radius, $R_{P}$/$R_{*} = r_2$\dotfill & $\mathcal{N}(0.049, 0.01)$ & $0.05113^{+0.00082} _{-0.00089}$ \\
RV semi-amplitude, $K_b$ (m/s) \dotfill & $\mathcal{U}(0, 30)$ & $9.22^{+1.70} _{-1.49}$ \\
Eccentricity, $e_b$\dotfill & Fixed & $0.0$ (adopted, 3$\sigma < 0.51$) \\
Argument of periastron, $\omega_b$ (deg)\dotfill & Fixed & $90.0$ \\

\\
\textbf{Fitted parameters for TOI-756 c} & & \\
\\

Orbital period, $P$ (days)\dotfill  & $\mathcal{U}$(10, 300) & $149.40^{+0.16}_{-0.17}$\\
Transit epoch, $T_{0}$ (BJD)\dotfill & $\mathcal{U}(2460000,2460500)$ & $2460498.82^{+0.57} _{-0.52}$\\
RV semi amplitude, $K_c$ (m/s) \dotfill & $\mathcal{U}(100, 500)$ & $273.29^{+2.56}_{-2.60}$\\
$\sqrt{e}_csin(\omega_c)$\dotfill & $\mathcal{U}(-1, 1)$ & $-0.141\pm0.011$ \\
$\sqrt{e}_ccos(\omega_c)$\dotfill & $\mathcal{U}(-1, 1)$ & $-0.652\pm0.006$ \\

\\
\textbf{Fitted parameters for the linear trend} & & \\
\\

RV slope (m/s/day)\dotfill & $\mathcal{U}(-10, 10)$ & $0.399\pm+0.014$ \\
RV intercept (m/s)\dotfill& $\mathcal{U}(-300, 300)$ & $-181.58^{+105.59}_{-78.22}$ \\

\\
\textbf{Instrumental photometric parameters} & & \\
\\

Offset relative flux, $M_{\mathrm{TESS}_{S10}}$ ($\times 10^{-4}$) \dotfill &$\mathcal{N}(0,300)$ & $-0.1\pm1.6$\\
Jitter, $\sigma_{w,\mathrm{TESS}_{S10}}$ (ppm)\dotfill & log$\mathcal{U}(0.01,300)$ & $12.2 ^{+58.3} _{-11.2}$ \\  
Offset relative flux, $M_{\mathrm{TESS}_{S11}}$ ($\times 10^{-4}$) \dotfill &$\mathcal{N}(0,300)$ & $3.7 ^{+17.6} _{-13.1}$\\
Jitter, $\sigma_{w,\mathrm{TESS}_{S11}}$ (ppm)\dotfill & log$\mathcal{U}(0.01,300)$ & $17.4 ^{+77.3} _{-15.8}$ \\ 
Offset relative flux, $M_{\mathrm{TESS}_{S37}}$ ($\times 10^{-4}$) \dotfill &$\mathcal{N}(0,300)$ & $-2.15 ^{+0.65} _{-0.59}$\\
Jitter, $\sigma_{w,\mathrm{TESS}_{S37}}$ (ppm)\dotfill & log$\mathcal{U}(0.01,300)$ & $1.13 ^{+13.13} _{-0.98}$ \\ 
Offset relative flux, $M_{\mathrm{TESS}_{S64}}$ ($\times 10^{-4}$) \dotfill &$\mathcal{N}(0,300)$ & $-15.59 ^{+0.63} _{-0.64}$\\
Jitter, $\sigma_{w,\mathrm{TESS}_{S64}}$ (ppm)\dotfill & log$\mathcal{U}(0.01,300)$ & $0.73 ^{+10.62} _{-0.66}$ \\ 

Offset relative flux, $M_{\mathrm{ExTrA}_{1T1}}$ \dotfill & $\mathcal{N}(0,2)$ & $0.16^{+0.46}_{-0.18}$ \\
Jitter, $\sigma_{w,\mathrm{ExTrA}_{1T1}}$ (ppm) \dotfill& $\log\mathcal{U}(0.01,5000)$ & $2146^{+308}_{-348}$ \\
Offset relative flux, $M_{\mathrm{ExTrA}_{2T1}}$ \dotfill& $\mathcal{N}(0,2)$ & $0.005^{+0.24}_{-0.004}$ \\
Jitter, $\sigma_{w,\mathrm{ExTrA}_{2T1}}$ (ppm) \dotfill& $\log\mathcal{U}(0.01,5000)$ & $1.97^{+45.03}_{-1.80}$ \\
Offset relative flux, $M_{\mathrm{ExTrA}_{3T1}}$ \dotfill& $\mathcal{N}(0,2)$ & $0.05^{+0.20}_{-0.12}$ \\
Jitter, $\sigma_{w,\mathrm{ExTrA}_{3T1}}$ (ppm) \dotfill& $\log\mathcal{U}(0.01,5000)$ & $67.39^{+1286.35}_{-62.43}$ \\
Offset relative flux, $M_{\mathrm{ExTrA}_{4T1}}$ \dotfill& $\mathcal{N}(0,2)$ & $0.04^{+0.20}_{-0.05}$ \\
Jitter, $\sigma_{w,\mathrm{ExTrA}_{4T1}}$ (ppm) \dotfill& $\log\mathcal{U}(0.01,5000)$ & $34.90^{+318.35}_{-32.80}$ \\
Offset relative flux, $M_{\mathrm{ExTrA}_{1T2}}$ \dotfill& $\mathcal{N}(0,2)$ & $0.32^{+0.38}_{-0.30}$ \\
Jitter, $\sigma_{w,\mathrm{ExTrA}_{1T2}}$ (ppm) \dotfill& $\log\mathcal{U}(0.01,5000)$ & $0.94^{+8.48}_{-0.85}$ \\
Offset relative flux, $M_{\mathrm{ExTrA}_{2T2}}$ \dotfill& $\mathcal{N}(0,2)$ & $0.028^{+0.20}_{-0.06}$ \\
Jitter, $\sigma_{w,\mathrm{ExTrA}_{2T2}}$ (ppm) \dotfill& $\log\mathcal{U}(0.01,5000)$ & $3.4^{+39.5}_{-3.1}$ \\
Offset relative flux, $M_{\mathrm{ExTrA}_{3T2}}$ \dotfill& $\mathcal{N}(0,2)$ & $0.002^{+0.098}_{-0.065}$ \\
Jitter, $\sigma_{w,\mathrm{ExTrA}_{3T2}}$ (ppm) \dotfill& $\log\mathcal{U}(0.01,5000)$ & $1.84^{+19.27}_{-1.65}$ \\
Offset relative flux, $M_{\mathrm{ExTrA}_{4T2}}$ \dotfill& $\mathcal{N}(0,2)$ & $-0.002^{+0.011}_{-0.020}$ \\
Jitter, $\sigma_{w,\mathrm{ExTrA}_{4T2}}$ (ppm) \dotfill& $\log\mathcal{U}(0.01,5000)$ & $1.66^{+34.90}_{-1.53}$ \\

Offset relative flux, $M_{\mathrm{LCO-\textit{i'}}}$  ($\times 10^{-4}$) \dotfill & $\mathcal{N}(0,2000)$ & $39.14^{+0.97}_{-0.96}$\\
Jitter, $\sigma_{w,\mathrm{LCO-\textit{i'}}}$ (ppm) \dotfill & log$\mathcal{U}(0.1,1000)$ & $4.2^{+40.4}_{-3.6}$ \\ 
Offset relative flux, $M_{\mathrm{LCO-\textit{g'}}}$  ($\times 10^{-4}$) \dotfill & $\mathcal{N}(0,2000)$ & $230.5^{+10.1}_{-9.7}$\\
Jitter, $\sigma_{w,\mathrm{LCO-\textit{g'}}}$ (ppm) \dotfill & log$\mathcal{U}(0.1,1000)$ & $34.6^{+309.8}_{-31.0}$ \\ 

\\
\textbf{Instrumental RV parameters} & & \\
\\

Systemic RV, $\mu_\mathrm{NIRPS}$ (m/s) \dotfill& $\mathcal{U}(10000,20000)$ & $14783.20^{+79.74}_{-103.06}$\\
Jitter, $\sigma_{w,\mathrm{NIRPS}}$ (m/s) \dotfill & log$\mathcal{U}(0.001,100)$ & $17.64\pm2.16$ \\ 
Systemic RV, $\mu_\mathrm{HARPS}$ (m/s) \dotfill& $\mathcal{U}(10000,20000)$ &$14688.28^{+80.74}_{-102.61}$ \\
Jitter, $\sigma_{w,\mathrm{HARPS}}$ (m/s) \dotfill & log$\mathcal{U}(0.001,100)$ & $13.52^{+1.25}_{-1.14}$ \\ 

\\
\textbf{GP/detrending parameters} & & \\
\\

$\rho_{GP,\mathrm{TESS}_{S10}}$ (days)\dotfill & log$\mathcal{U}(0.001, 50)$ & $0.62 ^{+0.66} _{-0.39}$ \\  
$\sigma_{GP,\mathrm{TESS}_{S10}}$ (10$^{-4}$ relative flux)\dotfill &   log$\mathcal{U}$(10$^{-2}$, 5 $\times$ 10$^{5}$) & $6.63^{+1.49} _{-1.24}$\\  
$\rho_{GP,\mathrm{TESS}_{S11}}$ (days)\dotfill & log$\mathcal{U}(0.001, 50)$ & $19.00 ^{+9.90} _{-6.68}$ \\  
$\sigma_{GP,\mathrm{TESS}_{S11}}$ (10$^{-4}$ relative flux)\dotfill &   log$\mathcal{U}$(10$^{-2}$, 5 $\times$ 10$^{5}$) & $18.8 ^{+12.9} _{+7.0}$\\

$\rho_{GP,\mathrm{ExTrA}_{1T1}}$ (days) \dotfill & log$\mathcal{U}(0.001, 10)$ & $2.05 ^{+1.68} _{-1.01}$ \\  
$\sigma_{GP,\mathrm{ExTrA}_{1T1}}$ ($10^{-2}$ relative flux) \dotfill & log$\mathcal{U}(10^{-4}, 10^{2})$ & $20.2^{+21.6}_{-10.1}$ \\  
$\rho_{GP,\mathrm{ExTrA}_{2T1}}$ (days) \dotfill & log$\mathcal{U}(0.001, 10)$ & $0.06 ^{+0.98} _{-0.04}$ \\  
$\sigma_{GP,\mathrm{ExTrA}_{2T1}}$ ($10^{-2}$ relative flux) \dotfill & log$\mathcal{U}(10^{-4}, 10^{2})$ & $1.2 ^{+27.2} _{-0.9}$ \\  
$\rho_{GP,\mathrm{ExTrA}_{3T1}}$ (days) \dotfill & log$\mathcal{U}(0.001, 10)$ & $1.36 ^{+0.84} _{-0.53}$ \\  
$\sigma_{GP,\mathrm{ExTrA}_{3T1}}$ ($10^{-2}$ relative flux) \dotfill & log$\mathcal{U}(10^{-4}, 10^{2})$ & $18.0 ^{+16.6} _{-9.2}$ \\  
$\rho_{GP,\mathrm{ExTrA}_{4T1}}$ (days) \dotfill & log$\mathcal{U}(0.001, 10)$ & $0.41 ^{+0.54} _{-0.32}$ \\  
$\sigma_{GP,\mathrm{ExTrA}_{4T1}}$ ($10^{-2}$ relative flux) \dotfill & log$\mathcal{U}(10^{-4}, 10^{2})$ & $8.2 ^{+16.2} _{-6.8}$ \\  
$\rho_{GP,\mathrm{ExTrA}_{1T2}}$ (days) \dotfill & log$\mathcal{U}(0.001, 10)$ & $1.19 ^{+0.89} _{-0.61}$ \\  
$\sigma_{GP,\mathrm{ExTrA}_{1T2}}$ ($10^{-2}$ relative flux) \dotfill & log$\mathcal{U}(10^{-4}, 10^{2})$ & $25.7 ^{+27.5} _{16.3}$ \\  
$\rho_{GP,\mathrm{ExTrA}_{2T2}}$ (days) \dotfill & log$\mathcal{U}(0.001, 10)$ & $5.6 ^{+2.3} _{-2.0}$ \\  
$\sigma_{GP,\mathrm{ExTrA}_{2T2}}$ ($10^{-2}$ relative flux) \dotfill & log$\mathcal{U}(10^{-4}, 10^{2})$ & $7.2 ^{+10.3} _{-3.3}$ \\  
$\rho_{GP,\mathrm{ExTrA}_{3T2}}$ (days) \dotfill & log$\mathcal{U}(0.001, 10)$ & $1.02 ^{+1.49} _{-0.74}$ \\  
$\sigma_{GP,\mathrm{ExTrA}_{3T2}}$ ($10^{-2}$ relative flux) \dotfill & log$\mathcal{U}(10^{-4}, 10^{2})$ & $12.5 ^{+22.5} _{-10.2}$ \\  
$\rho_{GP,\mathrm{ExTrA}_{4T2}}$ (days) \dotfill & log$\mathcal{U}(0.001, 10)$ & $0.40 ^{+0.82} _{-0.25}$ \\  
$\sigma_{GP,\mathrm{ExTrA}_{4T2}}$ ($10^{-2}$ relative flux) \dotfill & log$\mathcal{U}(10^{-4}, 10^{2})$ & $1.6 ^{+5.0} _{-1.1}$ \\

$\theta_{0,\mathrm{LCO-}\textit{g'}}$ (10$^{-4}$ relative flux)\dotfill &   $\mathcal{U}$(-10$^{6}$, 10$^{6}$) & $129.3 ^{+7.9} _{-7.4}$\\   
\\

\textbf{Limb darkening parameters} & & \\
\\

$q_{1,\text{TESS}}$\dotfill & $\mathcal{N}(0.792, 0.029)$ & $0.794^{+0.017} _{-0.019}$ \\
$q_{2,\text{TESS}}$\dotfill & $\mathcal{N}(0.453, 0.022)$ & $0.456^{+0.015} _{-0.016}$\\
$q_{1,\text{ExTrA}}$\dotfill & $\mathcal{N}(0.779, 0.074)$ & $0.744^{+0.038} _{-0.044}$ \\
$q_{2,\text{ExTrA}}$\dotfill & $\mathcal{N}(0.284, 0.028)$ & $0.287^{+0.016} _{-0.017}$\\
$q_{1,\text{LCO-\textit{i'}}}$\dotfill & $\mathcal{N}(0.825, 0.023)$ & $0.814\pm0.016$ \\
$q_{2,\text{LCO-\textit{i'}}}$\dotfill & $\mathcal{N}(0.506, 0.030)$ & $0.514^{+0.019} _{-0.017}$\\
$q_{1,\text{LCO-\textit{g'}}}$\dotfill & $\mathcal{N}(0.888, 0.011)$ & $0.887^{+0.007} _{-0.006}$ \\
$q_{2,\text{LCO-\textit{g'}}}$\dotfill & $\mathcal{N}(0.664, 0.015)$ & $0.671\pm0.011$\\

\end{longtable}

\begin{tablenotes}
\item
\textbf{Notes:} $\mathcal{N}(\mu, \sigma^{2})$ indicates a normal distribution with mean $\mu$ and variance $\sigma^{2}$, $\mathcal{U}(a, b)$ a uniform distribution between $a$ and $b$ and log$\mathcal{U}(a, b)$ a log-uniform distribution between $a$ and $b$.
\end{tablenotes}

\newpage
\section{Interior modeling of TOI-756~b}

\begin{figure*}[h]
\centering
    \includegraphics[width=0.56\textwidth]{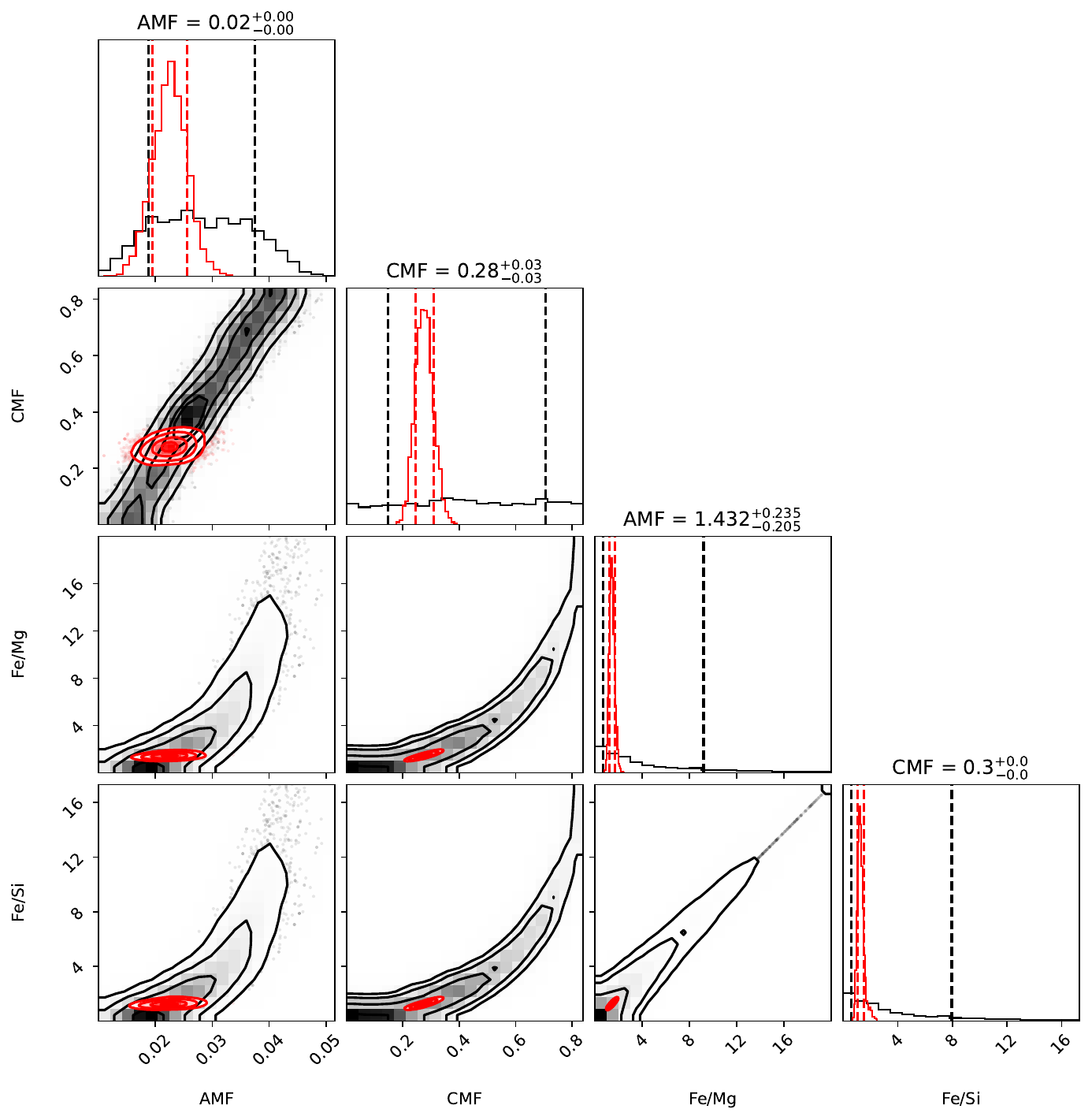}
    \includegraphics[width=0.56\textwidth]{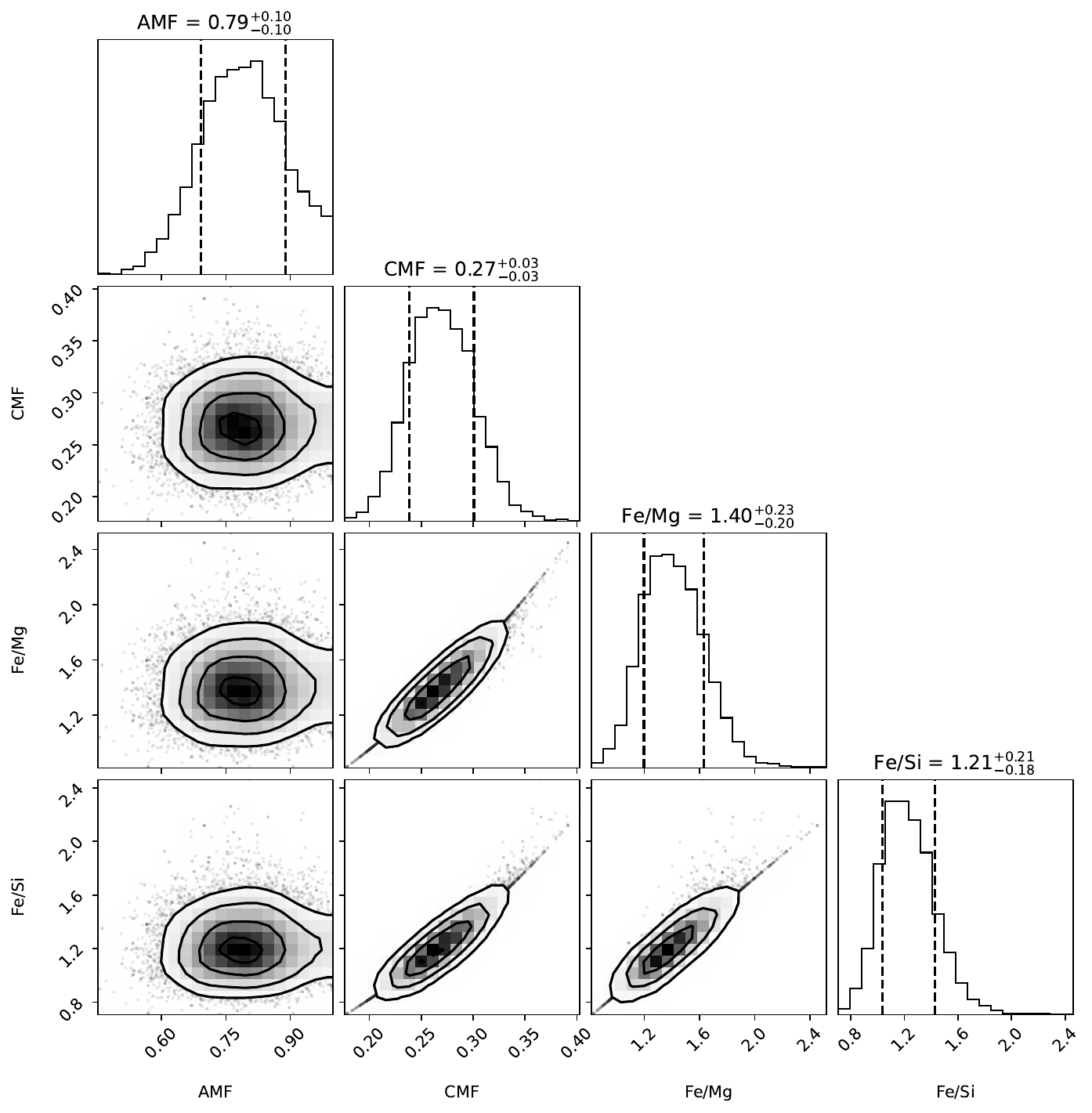}
     \caption{Corner plots from the interior modeling of TOI-756~b (see Sect.\ref{sect:interior}). The top panel shows scenario (1), assuming a H/He envelope: results with uninformative priors are shown in grey, and those using stellar-informed priors based on the host star's refractory abundances are in red. The bottom panel corresponds to scenario (2), assuming a pure H$_2$O envelope with stellar-informed priors.}
     \label{fig:corner_interiors}
\end{figure*}
\end{appendix}

\end{document}